%% file: mass.tex
\documentclass[titlepage,12pt,twoside]{article}

\usepackage{epsfig}
\usepackage{rotating}

\input{texdef}

\begin{document}

\begin{titlepage}

\vspace{1cm}

\begin{center}

\begin{large}
EUROPEAN ORGANISATION FOR NUCLEAR RESEARCH (CERN)
\end{large}

\vglue 0.5cm
\vspace{0.5cm}
\begin{flushright}
{\small CERN-PH-EP/2006-004} \\
{\small 25 April 2006} \\
{\small Journal Version} \\
\end{flushright}
\vspace*{1cm}

\vspace{1.5cm}

\begin{Large}
\begin{bf}
Measurement of the W boson Mass and Width \\
in {\boldmath\Pepem}\ Collisions at LEP \\
\end{bf}
\end{Large}

\vspace{1.0cm}
\begin{large}
The ALEPH Collaboration $^{*)}$ \\
\end{large}

\input{abstract}



\vspace*{2.cm}
\centerline{\it \small {To be published in Eur. Phys. J. C}}
\vspace*{2.cm}
\noindent
--------------------------------------------\hfil\break
{\small {$^*)$ See next pages for the list of authors}}
\newcommand{\prel}{}

\end{center}
\end{titlepage}
\normalsize

\newpage
\pagestyle{plain}
\setcounter{page}{1}
\newpage
\include{authb}

\newpage 

\setcounter{page}{1}

\input{introduction}

\vspace*{-0.5mm}
\input{detector}

\input{montecarlo}

\input{selection}

\input{extraction}

\input{crlimit}

\input{systematics}

\input{rad_returns}

\input{results}

\input{conclusions}

\section*{Acknowledgements}
  It is a pleasure to congratulate our colleagues from the CERN 
  accelerator divisions for the very successful operation of LEP2.  
  We are indebted to the engineers and technicians in all our institutions 
  for their contributions to the excellent performance of ALEPH. 
  Those of us from non-member countries thank CERN for its hospitality.

\input{appendices}


\input{biblio}
\end{document}

%% file: texdef.tex
\def\PL#1#2#3{{ Phys. Lett.}{ B#1 }(#2) #3}
\def\NIM#1#2#3{{ Nucl. Inst. Meth.}{ #1 }(#2) #3}
\newcommand{\PW}{\mbox{$\mathrm{W}$}}
\newcommand{\PWm}{\mbox{$\mathrm{W}^-$}}
\newcommand{\PWp}{\mbox{$\mathrm{W}^+$}}
\newcommand{\PZz}{\mbox{$\mathrm{Z}$}}
\newcommand{\PMW}{\mbox{$m_{\mathrm{W}}$}}
\newcommand{\PMZ}{\mbox{$m_{\mathrm{Z}}$}}
\newcommand{\PGW}{\mbox{$\Gamma_{\mathrm{W}}$}}

\newcommand{\GeV}{\mbox{$\mathrm{GeV}$}}
\newcommand{\MeV}{\mbox{$\mathrm{MeV}$}}
\newcommand{\GeVc}{\mbox{$\mathrm{GeV}/{\it c}$}}

\newcommand{\GeVcsq}{\mbox{$\mathrm{GeV}/{{\it c}^2}$}}
\newcommand{\MeVcsq}{\mbox{$\mathrm{MeV}/{{\it c}^2}$}}
\newcommand{\Pem}{\mbox{$\mathrm{e}^-$}}
\newcommand{\Pep}{\mbox{$\mathrm{e}^+$}}
\newcommand{\Pepem}{\mbox{$\mathrm{e}^+\mathrm{e}^-$}}

\newcommand{\PWW}{\mbox{$\mathrm{W}\mathrm{W}$}}
\newcommand{\CCC}{\mbox{$\mathrm{CC03}$}}

\newcommand{\emis}{\mbox{${\not{\!E}}$}}
\newcommand{\mm}{\mbox{$\mu^+ \mu^-$}}
\newcommand{\toto}{\mbox{$\tau^+ \tau^-$}}
\newcommand{\ipb}{\ensuremath{\mathrm {pb}^{-1}}}
\newcommand{\oalph}{\mbox{$\mathcal{O}\rm(\alpha)$}}

\def\ra{\rightarrow} 
\def\gappeq{\mathrel{ \rlap{\raise.5ex\hbox{$>$}}
                      {\lower.5ex\hbox{$\sim$}}  } }
\def\lappeq{\mathrel{ \rlap{\raise.5ex\hbox{$<$}}
                      {\lower.5ex\hbox{$\sim$}}  } }
   \def\epem{{e^+e^-}}
\def\be{\begin{equation}}
\def\ee{\end{equation}}
\def\bea{\begin{eqnarray}}
\def\eea{\end{eqnarray}}
   \def\ra{\rightarrow}

  \def\eqref#1{(\ref{#1})}

  \def\mf2{m_f^2}

  \def\s0h{{\sigma}^{{\mathrm{ peak,0}}}_{{\mathrm{ had}} }}

\newcommand{\qqbar}{\mbox{$\mathrm{q\bar{q}}$}}
\newcommand{\ppbar}{\mbox{$\mathrm{p\bar{p}}$}}
\newcommand{\lvqq}{\mbox{$\ell\nu \mathrm{q\bar{q}}$}}
\newcommand{\evqq}{\mbox{$\mathrm{e}\nu \mathrm{q\bar{q}}$}}
\newcommand{\mvqq}{\mbox{$\mu\nu \mathrm{q\bar{q}}$}}
\newcommand{\tvqq}{\mbox{$\tau\nu \mathrm{q\bar{q}}$}}



%% file: abstract.tex
\begin{quotation}
\noindent
\begin{center}
{\bf Abstract}
\end{center}
{\normalsize
The mass of the \PW\ boson is determined from the direct reconstruction of 
\PW\ decays in $\PWW\ra\qqbar\qqbar$ and $\PWW\ra\lvqq$ events in e$^+$e$^-$ 
collisions at LEP.
The data sample corresponds to an integrated luminosity of 683~pb$^{-1}$ 
collected with the ALEPH detector at centre-of-mass energies up to 
209 \GeV. To minimise any effect from colour reconnection a new procedure is 
adopted in which low energy particles are not considered in the mass 
determination from the \qqbar\qqbar\ channel. 
The combined result from all channels 
is
\[
m_\mathrm{W} = 80.440 
\pm 0.043{\mathrm{ (stat.)}} 
\pm 0.024{\mathrm{ (syst.)}}
\pm 0.009{\mathrm{ (FSI)}}
\pm 0.009{\mathrm{ (LEP)}}
~\GeVcsq,
\]
where FSI represents the possible effects of final state interactions
in the $\qqbar\qqbar$ channel and LEP indicates the uncertainty in the beam 
energy.
From two-parameter fits to the \PW\ mass and width, the \PW\ 
width is found to be 
\[
\Gamma_\mathrm{W} = 2.14 
\pm 0.09{\mathrm{ (stat.)}} 
\pm 0.04{\mathrm{ (syst.)}}
\pm 0.05{\mathrm{ (FSI)}}
\pm 0.01{\mathrm{ (LEP)}}
~\GeV.
\]
}
\end{quotation}

%% file: authb.tex
\pagestyle{empty}
\newpage
\small
%
%
\newlength{\saveparskip}
\newlength{\savetextheight}
\newlength{\savetopmargin}
\newlength{\savetextwidth}
\newlength{\saveoddsidemargin}
\newlength{\savetopsep}
\setlength{\saveparskip}{\parskip}
\setlength{\savetextheight}{\textheight}
\setlength{\savetopmargin}{\topmargin}
\setlength{\savetextwidth}{\textwidth}
\setlength{\saveoddsidemargin}{\oddsidemargin}
\setlength{\savetopsep}{\topsep}
%
%
\setlength{\parskip}{0.0cm}
\setlength{\textheight}{25.0cm}
\setlength{\topmargin}{-1.5cm}
\setlength{\textwidth}{16 cm}
\setlength{\oddsidemargin}{-0.0cm}
\setlength{\topsep}{1mm}
\pretolerance=10000
\centerline{\large\bf The ALEPH Collaboration}
\footnotesize
\vspace{0.5cm}
{\raggedbottom
\begin{sloppypar}
\samepage\noindent
S.~Schael,
\nopagebreak
\begin{center}
\parbox{15.5cm}{\sl\samepage
Physikalisches Institut das RWTH-Aachen, D-52056 Aachen, Germany}
\end{center}\end{sloppypar}
\vspace{2mm}
\begin{sloppypar}
\noindent
R.~Barate,
R.~Bruneli\`ere,
I.~De~Bonis,
D.~Decamp,
C.~Goy,
S.~J\'ez\'equel,
J.-P.~Lees,
F.~Martin,
E.~Merle,
\mbox{M.-N.~Minard},
B.~Pietrzyk,
B.~Trocm\'e
\nopagebreak
\begin{center}
\parbox{15.5cm}{\sl\samepage
Laboratoire de Physique des Particules (LAPP), IN$^{2}$P$^{3}$-CNRS,
F-74019 Annecy-le-Vieux Cedex, France}
\end{center}\end{sloppypar}
\vspace{2mm}
\begin{sloppypar}
\noindent
S.~Bravo,
M.P.~Casado,
M.~Chmeissani,
J.M.~Crespo,
E.~Fernandez,
M.~Fernandez-Bosman,
Ll.~Garrido,$^{15}$
M.~Martinez,
A.~Pacheco,
H.~Ruiz
\nopagebreak
\begin{center}
\parbox{15.5cm}{\sl\samepage
Institut de F\'{i}sica d'Altes Energies, Universitat Aut\`{o}noma
de Barcelona, E-08193 Bellaterra (Barcelona), Spain$^{7}$}
\end{center}\end{sloppypar}
\vspace{2mm}
\begin{sloppypar}
\noindent
A.~Colaleo,
D.~Creanza,
N.~De~Filippis,
M.~de~Palma,
G.~Iaselli,
G.~Maggi,
M.~Maggi,
S.~Nuzzo,
A.~Ranieri,
G.~Raso,$^{24}$
F.~Ruggieri,
G.~Selvaggi,
L.~Silvestris,
P.~Tempesta,
A.~Tricomi,$^{3}$
G.~Zito
\nopagebreak
\begin{center}
\parbox{15.5cm}{\sl\samepage
Dipartimento di Fisica, INFN Sezione di Bari, I-70126 Bari, Italy}
\end{center}\end{sloppypar}
\vspace{2mm}
\begin{sloppypar}
\noindent
X.~Huang,
J.~Lin,
Q. Ouyang,
T.~Wang,
Y.~Xie,
R.~Xu,
S.~Xue,
J.~Zhang,
L.~Zhang,
W.~Zhao
\nopagebreak
\begin{center}
\parbox{15.5cm}{\sl\samepage
Institute of High Energy Physics, Academia Sinica, Beijing, The People's
Republic of China$^{8}$}
\end{center}\end{sloppypar}
\vspace{2mm}
\begin{sloppypar}
\noindent
D.~Abbaneo,
T.~Barklow,$^{27}$
O.~Buchm\"uller,$^{27}$
M.~Cattaneo,
B.~Clerbaux,$^{23}$
H.~Drevermann,
R.W.~Forty,
M.~Frank,
F.~Gianotti,
J.B.~Hansen,
J.~Harvey,
D.E.~Hutchcroft,$^{31}$,
P.~Janot,
B.~Jost,
M.~Kado,$^{2}$
P.~Mato,
A.~Moutoussi,
F.~Ranjard,
L.~Rolandi,
D.~Schlatter,
F.~Teubert,
A.~Valassi,
I.~Videau
\nopagebreak
\begin{center}
\parbox{15.5cm}{\sl\samepage
European Laboratory for Particle Physics (CERN), CH-1211 Geneva 23,
Switzerland}
\end{center}\end{sloppypar}
\vspace{2mm}
\begin{sloppypar}
\noindent
F.~Badaud,
S.~Dessagne,
A.~Falvard,$^{20}$
D.~Fayolle,
P.~Gay,
J.~Jousset,
B.~Michel,
S.~Monteil,
D.~Pallin,
J.M.~Pascolo,
P.~Perret
\nopagebreak
\begin{center}
\parbox{15.5cm}{\sl\samepage
Laboratoire de Physique Corpusculaire, Universit\'e Blaise Pascal,
IN$^{2}$P$^{3}$-CNRS, Clermont-Ferrand, F-63177 Aubi\`{e}re, France}
\end{center}\end{sloppypar}
\vspace{2mm}
\begin{sloppypar}
\noindent
J.D.~Hansen,
J.R.~Hansen,
P.H.~Hansen,
A.C.~Kraan,
B.S.~Nilsson
\nopagebreak
\begin{center}
\parbox{15.5cm}{\sl\samepage
Niels Bohr Institute, 2100 Copenhagen, DK-Denmark$^{9}$}
\end{center}\end{sloppypar}
\vspace{2mm}
\begin{sloppypar}
\noindent
A.~Kyriakis,
C.~Markou,
E.~Simopoulou,
A.~Vayaki,
K.~Zachariadou
\nopagebreak
\begin{center}
\parbox{15.5cm}{\sl\samepage
Nuclear Research Center Demokritos (NRCD), GR-15310 Attiki, Greece}
\end{center}\end{sloppypar}
\vspace{2mm}
\begin{sloppypar}
\noindent
A.~Blondel,$^{12}$
\mbox{J.-C.~Brient},
F.~Machefert,
A.~Roug\'{e},
H.~Videau
\nopagebreak
\begin{center}
\parbox{15.5cm}{\sl\samepage
Laoratoire Leprince-Ringuet, Ecole
Polytechnique, IN$^{2}$P$^{3}$-CNRS, \mbox{F-91128} Palaiseau Cedex, France}
\end{center}\end{sloppypar}
\vspace{2mm}
\begin{sloppypar}
\noindent
V.~Ciulli,
E.~Focardi,
G.~Parrini
\nopagebreak
\begin{center}
\parbox{15.5cm}{\sl\samepage
Dipartimento di Fisica, Universit\`a di Firenze, INFN Sezione di Firenze,
I-50125 Firenze, Italy}
\end{center}\end{sloppypar}
\vspace{2mm}
\begin{sloppypar}
\noindent
A.~Antonelli,
M.~Antonelli,
G.~Bencivenni,
F.~Bossi,
G.~Capon,
F.~Cerutti,
V.~Chiarella,
P.~Laurelli,
G.~Mannocchi,$^{5}$
G.P.~Murtas,
L.~Passalacqua
\nopagebreak
\begin{center}
\parbox{15.5cm}{\sl\samepage
Laboratori Nazionali dell'INFN (LNF-INFN), I-00044 Frascati, Italy}
\end{center}\end{sloppypar}
\vspace{2mm}
\begin{sloppypar}
\noindent
J.~Kennedy,
J.G.~Lynch,
P.~Negus,
V.~O'Shea,
A.S.~Thompson
\nopagebreak
\begin{center}
\parbox{15.5cm}{\sl\samepage
Department of Physics and Astronomy, University of Glasgow, Glasgow G12
8QQ,United Kingdom$^{10}$}
\end{center}\end{sloppypar}
\vspace{2mm}
\begin{sloppypar}
\noindent
S.~Wasserbaech
\nopagebreak
\begin{center}
\parbox{15.5cm}{\sl\samepage
Utah Valley State College, Orem, UT 84058, U.S.A.}
\end{center}\end{sloppypar}
\vspace{2mm}
\begin{sloppypar}
\noindent
R.~Cavanaugh,$^{4}$
S.~Dhamotharan,$^{21}$
C.~Geweniger,
P.~Hanke,
V.~Hepp,
E.E.~Kluge,
A.~Putzer,
H.~Stenzel,
K.~Tittel,
M.~Wunsch$^{19}$
\nopagebreak
\begin{center}
\parbox{15.5cm}{\sl\samepage
Kirchhoff-Institut f\"ur Physik, Universit\"at Heidelberg, D-69120
Heidelberg, Germany$^{16}$}
\end{center}\end{sloppypar}
\vspace{2mm}
\begin{sloppypar}
\noindent
R.~Beuselinck,
W.~Cameron,
G.~Davies,
P.J.~Dornan,
M.~Girone,$^{1}$
N.~Marinelli,
J.~Nowell,
S.A.~Rutherford,
J.K.~Sedgbeer,
J.C.~Thompson,$^{14}$
R.~White
\nopagebreak
\begin{center}
\parbox{15.5cm}{\sl\samepage
Department of Physics, Imperial College, London SW7 2BZ,
United Kingdom$^{10}$}
\end{center}\end{sloppypar}
\vspace{2mm}
\begin{sloppypar}
\noindent
V.M.~Ghete,
P.~Girtler,
E.~Kneringer,
D.~Kuhn,
G.~Rudolph
\nopagebreak
\begin{center}
\parbox{15.5cm}{\sl\samepage
Institut f\"ur Experimentalphysik, Universit\"at Innsbruck, A-6020
Innsbruck, Austria$^{18}$}
\end{center}\end{sloppypar}
\vspace{2mm}
\begin{sloppypar}
\noindent
E.~Bouhova-Thacker,
C.K.~Bowdery,
D.P.~Clarke,
G.~Ellis,
A.J.~Finch,
F.~Foster,
G.~Hughes,
R.W.L.~Jones,
M.R.~Pearson,
N.A.~Robertson,
T.~Sloan,
M.~Smizanska
\nopagebreak
\begin{center}
\parbox{15.5cm}{\sl\samepage
Department of Physics, University of Lancaster, Lancaster LA1 4YB,
United Kingdom$^{10}$}
\end{center}\end{sloppypar}
\vspace{2mm}
\begin{sloppypar}
\noindent
O.~van~der~Aa,
C.~Delaere,$^{29}$
G.Leibenguth,$^{32}$
V.~Lemaitre$^{30}$
\nopagebreak
\begin{center}
\parbox{15.5cm}{\sl\samepage
Institut de Physique Nucl\'eaire, D\'epartement de Physique, Universit\'e Catholique de Louvain, 1348 Louvain-la-Neuve, Belgium}
\end{center}\end{sloppypar}
\vspace{2mm}
\begin{sloppypar}
\noindent
U.~Blumenschein,
F.~H\"olldorfer,
K.~Jakobs,
F.~Kayser,
A.-S.~M\"uller,
B.~Renk,
H.-G.~Sander,
S.~Schmeling,
H.~Wachsmuth,
C.~Zeitnitz,
T.~Ziegler
\nopagebreak
\begin{center}
\parbox{15.5cm}{\sl\samepage
Institut f\"ur Physik, Universit\"at Mainz, D-55099 Mainz, Germany$^{16}$}
\end{center}\end{sloppypar}
\vspace{2mm}
\begin{sloppypar}
\noindent
A.~Bonissent,
P.~Coyle,
C.~Curtil,
A.~Ealet,
D.~Fouchez,
P.~Payre,
A.~Tilquin
\nopagebreak
\begin{center}
\parbox{15.5cm}{\sl\samepage
Centre de Physique des Particules de Marseille, Univ M\'editerran\'ee,
IN$^{2}$P$^{3}$-CNRS, F-13288 Marseille, France}
\end{center}\end{sloppypar}
\vspace{2mm}
\begin{sloppypar}
\noindent
F.~Ragusa
\nopagebreak
\begin{center}
\parbox{15.5cm}{\sl\samepage
Dipartimento di Fisica, Universit\`a di Milano e INFN Sezione di
Milano, I-20133 Milano, Italy.}
\end{center}\end{sloppypar}
\vspace{2mm}
\begin{sloppypar}
\noindent
A.~David,
H.~Dietl,$^{33}$
G.~Ganis,$^{28}$
K.~H\"uttmann,
G.~L\"utjens,
W.~M\"anner$^{33}$,
\mbox{H.-G.~Moser},
R.~Settles,
M.~Villegas,
G.~Wolf
\nopagebreak
\begin{center}
\parbox{15.5cm}{\sl\samepage
Max-Planck-Institut f\"ur Physik, Werner-Heisenberg-Institut,
D-80805 M\"unchen, Germany\footnotemark[16]}
\end{center}\end{sloppypar}
\vspace{2mm}
\begin{sloppypar}
\noindent
J.~Boucrot,
O.~Callot,
M.~Davier,
L.~Duflot,
\mbox{J.-F.~Grivaz},
Ph.~Heusse,
A.~Jacholkowska,$^{6}$
L.~Serin,
\mbox{J.-J.~Veillet}
\nopagebreak
\begin{center}
\parbox{15.5cm}{\sl\samepage
Laboratoire de l'Acc\'el\'erateur Lin\'eaire, Universit\'e de Paris-Sud,
IN$^{2}$P$^{3}$-CNRS, F-91898 Orsay Cedex, France}
\end{center}\end{sloppypar}
\vspace{2mm}
\begin{sloppypar}
\noindent
P.~Azzurri, 
G.~Bagliesi,
T.~Boccali,
L.~Fo\`a,
A.~Giammanco,
A.~Giassi,
F.~Ligabue,
A.~Messineo,
F.~Palla,
G.~Sanguinetti,
A.~Sciab\`a,
G.~Sguazzoni,
P.~Spagnolo,
R.~Tenchini,
A.~Venturi,
P.G.~Verdini
\samepage
\begin{center}
\parbox{15.5cm}{\sl\samepage
Dipartimento di Fisica dell'Universit\`a, INFN Sezione di Pisa,
e Scuola Normale Superiore, I-56010 Pisa, Italy}
\end{center}\end{sloppypar}
\vspace{2mm}
\begin{sloppypar}
\noindent
O.~Awunor,
G.A.~Blair,
G.~Cowan,
A.~Garcia-Bellido,
M.G.~Green,
T.~Medcalf,$^{25}$
A.~Misiejuk,
J.A.~Strong,
P.~Teixeira-Dias
\nopagebreak
\begin{center}
\parbox{15.5cm}{\sl\samepage
Department of Physics, Royal Holloway \& Bedford New College,
University of London, Egham, Surrey TW20 OEX, United Kingdom$^{10}$}
\end{center}\end{sloppypar}
\vspace{2mm}
\begin{sloppypar}
\noindent
R.W.~Clifft,
T.R.~Edgecock,
P.R.~Norton,
I.R.~Tomalin,
J.J.~Ward
\nopagebreak
\begin{center}
\parbox{15.5cm}{\sl\samepage
Particle Physics Dept., Rutherford Appleton Laboratory,
Chilton, Didcot, Oxon OX11 OQX, United Kingdom$^{10}$}
\end{center}\end{sloppypar}
\vspace{2mm}
\begin{sloppypar}
\noindent
\mbox{B.~Bloch-Devaux},
D.~Boumediene,
P.~Colas,
B.~Fabbro,
E.~Lan\c{c}on,
\mbox{M.-C.~Lemaire},
E.~Locci,
P.~Perez,
J.~Rander,
A.~Trabelsi,$^{25}$
B.~Tuchming,
B.~Vallage
\nopagebreak
\begin{center}
\parbox{15.5cm}{\sl\samepage
CEA, DAPNIA/Service de Physique des Particules,
CE-Saclay, F-91191 Gif-sur-Yvette Cedex, France$^{17}$}
\end{center}\end{sloppypar}
\vspace{2mm}
\begin{sloppypar}
\noindent
A.M.~Litke,
G.~Taylor
\nopagebreak
\begin{center}
\parbox{15.5cm}{\sl\samepage
Institute for Particle Physics, University of California at
Santa Cruz, Santa Cruz, CA 95064, USA$^{22}$}
\end{center}\end{sloppypar}
\vspace{2mm}
\begin{sloppypar}
\noindent
C.N.~Booth,
S.~Cartwright,
F.~Combley,$^{26}$
P.N.~Hodgson,
M.~Lehto,
L.F.~Thompson
\nopagebreak
\begin{center}
\parbox{15.5cm}{\sl\samepage
Department of Physics, University of Sheffield, Sheffield S3 7RH,
United Kingdom$^{10}$}
\end{center}\end{sloppypar}
\vspace{2mm}
\begin{sloppypar}
\noindent
A.~B\"ohrer,
S.~Brandt,
C.~Grupen,
J.~Hess,
A.~Ngac,
G.~Prange
\nopagebreak
\begin{center}
\parbox{15.5cm}{\sl\samepage
Fachbereich Physik, Universit\"at Siegen, D-57068 Siegen, Germany$^{16}$}
\end{center}\end{sloppypar}
\vspace{2mm}
\begin{sloppypar}
\noindent
C.~Borean,
G.~Giannini
\nopagebreak
\begin{center}
\parbox{15.5cm}{\sl\samepage
Dipartimento di Fisica, Universit\`a di Trieste e INFN Sezione di Trieste,
I-34127 Trieste, Italy}
\end{center}\end{sloppypar}
\vspace{2mm}
\begin{sloppypar}
\noindent
H.~He,
J.~Putz,
J.~Rothberg
\nopagebreak
\begin{center}
\parbox{15.5cm}{\sl\samepage
Experimental Elementary Particle Physics, University of Washington, Seattle,
WA 98195 U.S.A.}
\end{center}\end{sloppypar}
\vspace{2mm}
\begin{sloppypar}
\noindent
S.R.~Armstrong,
K.~Berkelman,
K.~Cranmer,
D.P.S.~Ferguson,
Y.~Gao,$^{13}$
S.~Gonz\'{a}lez,
O.J.~Hayes,
H.~Hu,
S.~Jin,
J.~Kile,
P.A.~McNamara III,
J.~Nielsen,
Y.B.~Pan,
\mbox{J.H.~von~Wimmersperg-Toeller}, 
W.~Wiedenmann,
J.~Wu,
Sau~Lan~Wu,
X.~Wu,
G.~Zobernig
\nopagebreak
\begin{center}
\parbox{15.5cm}{\sl\samepage
Department of Physics, University of Wisconsin, Madison, WI 53706,
USA$^{11}$}
\end{center}\end{sloppypar}
\vspace{2mm}
\begin{sloppypar}
\noindent
G.~Dissertori
\nopagebreak
\begin{center}
\parbox{15.5cm}{\sl\samepage
Institute for Particle Physics, ETH H\"onggerberg, 8093 Z\"urich,
Switzerland.}
\end{center}\end{sloppypar}
}
\footnotetext[1]{Also at CERN, 1211 Geneva 23, Switzerland.}
\footnotetext[2]{Now at Fermilab, PO Box 500, MS 352, Batavia, IL 60510, USA}
\footnotetext[3]{Also at Dipartimento di Fisica di Catania and INFN Sezione di
 Catania, 95129 Catania, Italy.}
\footnotetext[4]{Now at University of Florida, Department of Physics, Gainesville, Florida 32611-8440, USA}
\footnotetext[5]{Also IFSI sezione di Torino, INAF, Italy.}
\footnotetext[6]{Also at Groupe d'Astroparticules de Montpellier, Universit\'{e} de Montpellier II, 34095, Montpellier, France.}
\footnotetext[7]{Supported by CICYT, Spain.}
\footnotetext[8]{Supported by the National Science Foundation of China.}
\footnotetext[9]{Supported by the Danish Natural Science Research Council.}
\footnotetext[10]{Supported by the UK Particle Physics and Astronomy Research
Council.}
\footnotetext[11]{Supported by the US Department of Energy, grant
DE-FG0295-ER40896.}
\footnotetext[12]{Now at Departement de Physique Corpusculaire, Universit\'e de
Gen\`eve, 1211 Gen\`eve 4, Switzerland.}
\footnotetext[13]{Also at Department of Physics, Tsinghua University, Beijing, The People's Republic of China.}
\footnotetext[14]{Supported by the Leverhulme Trust.}
\footnotetext[15]{Permanent address: Universitat de Barcelona, 08208 Barcelona,
Spain.}
\footnotetext[16]{Supported by Bundesministerium f\"ur Bildung
und Forschung, Germany.}
\footnotetext[17]{Supported by the Direction des Sciences de la
Mati\`ere, C.E.A.}
\footnotetext[18]{Supported by the Austrian Ministry for Science and Transport.}
\footnotetext[19]{Now at SAP AG, 69185 Walldorf, Germany}
\footnotetext[20]{Now at Groupe d' Astroparticules de Montpellier, Universit\'e de Montpellier II, 34095 Montpellier, France.}
\footnotetext[21]{Now at BNP Paribas, 60325 Frankfurt am Mainz, Germany}
\footnotetext[22]{Supported by the US Department of Energy,
grant DE-FG03-92ER40689.}
\footnotetext[23]{Now at Institut Inter-universitaire des hautes Energies (IIHE), CP 230, Universit\'{e} Libre de Bruxelles, 1050 Bruxelles, Belgique}
\footnotetext[24]{Now at Dipartimento di Fisica e Tecnologie Relative, Universit\`a di Palermo, Palermo, Italy.}
\footnotetext[25]{Now at Faculté des Sciences de Tunis, 2092, Campus Universitaire, Tunisia.}
\footnotetext[26]{Deceased.}
\footnotetext[27]{Now at SLAC, Stanford, CA 94309, U.S.A}
\footnotetext[28]{Now at CERN, 1211 Geneva 23, Switzerland}
\footnotetext[29]{Research Fellow of the Belgium FNRS}
\footnotetext[30]{Research Associate of the Belgium FNRS} 
\footnotetext[31]{Now at Liverpool University, Liverpool L69 7ZE, United Kingdom} 
\footnotetext[32]{Supported by the Federal Office for Scientific, Technical and Cultural Affairs through
the Interuniversity Attraction Pole P5/27} 
\footnotetext[33]{Now at Henryk Niewodnicznski Institute of Nuclear Physics, Polish Academy of Sciences, Cracow, Poland}   
\setlength{\parskip}{\saveparskip}
\setlength{\textheight}{\savetextheight}
\setlength{\topmargin}{\savetopmargin}
\setlength{\textwidth}{\savetextwidth}
\setlength{\oddsidemargin}{\saveoddsidemargin}
\setlength{\topsep}{\savetopsep}
\normalsize
\newpage
\pagestyle{plain}
\setcounter{page}{1}

%% file: introduction.tex
\section{Introduction}
\label{Intro}
\vspace*{-0.5mm}
The electroweak Standard Model (SM) successfully 
describes all interactions of quarks and leptons at the \PZz\ resonance 
provided that quantum radiative corrections are included~\cite{lepew2004}. 
In this model, the mass of the \PW\ boson (\PMW) can be calculated as
follows:  
\begin{center}
\vspace{-0.5cm}
\begin{eqnarray*}
\PMW^2(1-\frac{\PMW^2}{\PMZ^2}) = \frac{\pi\alpha}{\sqrt2 G_{\mu}}(1+\Delta r)
\end{eqnarray*}
\end{center}
where \PMZ\ (\PZz\ mass), $G_{\mu}$ (Fermi coupling constant) and $\alpha$ (fine 
structure constant) are measured with high precision.
In this equation, $\Delta r$ parametrises the loop corrections which lead to a 
quadratic dependence on the top quark mass, ($m_{\rm {top}}$), 
and a weaker logarithmic dependence on the Higgs boson mass. A global 
fit of electroweak observables measured at the \PZz\ resonance together with the 
measured $m_{\rm {top}}$~\cite{mtop} yields a \PW\ mass of 
$80.373 \pm 0.023~\GeVcsq$~\cite{lepew2004} in the SM. 

The comparison of a direct measurement of \PMW\ with this prediction was
a primary goal of LEP, enabling a stringent test of the Standard Model 
to be made. 
This paper describes the final measurement of the \PW\ mass and width (\PGW) 
from ALEPH. They are determined from the direct reconstruction 
of the invariant mass of its decay products in both the $\PWW\ra\qqbar\qqbar$  
hadronic and $\PWW\rightarrow \ell (\ell =\rm e,\mu,\tau)\nu\qqbar$ semileptonic 
channels. Measurements were published previously using 
the data collected at centre-of-mass (CM) energies of 172, 183 and 189 
~\GeV~\cite{mass_172,mass_183,mass_189}. The most recent ALEPH publication 
included a weighted average result obtained from the combination of all these 
measurements as well as those obtained earlier from the total \PW\ pair cross 
sections at 161 and 172 \GeV~\cite{ALEPH-THRESHOLD,wwxsec_172}. The 
statistical precision for the mass was 61 \MeVcsq\ with a 
systematic uncertainty of 47 \MeVcsq.  In the last two years up to the closure of 
LEP in 2000, more data were collected at CM energies up to 209 \GeV\ 
increasing the total sample by a factor of three. 
 
All these data, except for the small sample at 172 \GeV, are included in the 
analysis corresponding to an integrated luminosity of 683 \ipb. 
The data were sub-divided into eight 
samples labelled as 183, 189, 192, 196, 200, 202, 205 and 207 \GeV\ according to 
their CM energies. This sub-division is the same as that used in the 
measurement of the \PWW\ cross section~\cite{xsec_final}.

A constrained kinematic fit conserving energy and momentum is applied to each 
selected event in data and Monte Carlo (MC) simulation. As in 
previous analyses, the simulated mass spectra are fitted to the 
data using a reweighting technique to extract the \PW\ mass and width. 
Very large MC productions ($>10^6$ signal events per CM energy) enable 
multi-dimensional fits to be used with significant gains in precision. The signal 
events are weighted to account for the effect of 
${\cal{O}}(\alpha)$ corrections~\cite{YFSWW} in \PMW\ and \PGW. 

Since the statistical error on \PMW\ is now comparable with the previously 
published systematic uncertainties, a more detailed evaluation of all important 
uncertainties has been performed. 
In the previous analysis~\cite{mass_189}, the dominant systematic uncertainty in 
the \qqbar\qqbar\ channel was due to colour reconnection (CR). This affects the 
topological distribution of lower energy particles in an event. 
Two new analysis procedures have been adopted. In one of these, PCUT, low energy 
particles are not considered in the reconstruction of jets, whilst in the other, 
CONE, only particles close to the jet axes are used. These significantly reduce 
the difficult-to-estimate CR uncertainty at the expense of some statistical power 
in this channel. 
The effect of these modified reconstructions has been checked using di-jets in 
the \lvqq\ channels, where no final state interactions are present between the 
\PW's.
 
The paper is organised as follows. In Section~2, the important properties
of the ALEPH detector are described. In Section~3, the event 
reconstruction procedures and calibrations are recalled and detailed studies of 
the detector simulation reported. Section~4 contains a full description of the 
event samples generated for the signal and background processes involved. 
Section~5 describes the event selection 
and kinematic reconstruction procedures in the 
different channels highlighting, where appropriate, the modifications and 
improvements applied since the earlier analyses at 183 and 
189~\GeV~\cite{mass_183,mass_189}.
Section~6 describes the extraction of \PMW\ and \PGW. 
Section~7 describes the specific studies made to set a limit on colour 
reconnection from the data using event reconstructions where low momentum 
particles or particles between jets are excluded.
Section~8 describes all studies of systematic uncertainties.
The analysis of radiative returns to the \PZz\ peak is reported in Section~9, 
providing a cross check on the \PW\ mass measurement. 
The measurements in all channels are combined in 
Section~10, taking into account common sources of systematic uncertainties. The 
\PW\ masses measured in the purely hadronic and combined semileptonic 
channels are compared in this section.    
Final conclusions are given in Section~11.


%% file: detector.tex
\section{The ALEPH detector}
\label{sec:detector}
\vspace*{-0.5mm}

A detailed description of the ALEPH detector can be found in 
Ref.~\cite{detector} and of its performance in Ref.~\cite{perf}. 
The tracking detectors include a silicon vertex detector (VDET), a cylindrical 
drift chamber and a large time projection chamber (TPC)
which measures up to 31 coordinates along the charged particle trajectories.
A 1.5~T axial magnetic field, provided by a superconducting solenoidal coil, 
yields a resolution of $\delta p_{\rm T}/p_{\rm T}=6 \times 10^{-4} p_{\rm T} 
\oplus 0.005$ ($p_{\rm T}$ in \GeVc). Charged particle tracks reconstructed 
with at least four hits in the TPC and  originating
from within a cylinder of 2~cm radius and 20~cm length, 
centred on the nominal interaction point 
and parallel to the beam axis, are called {\it good tracks}. In addition to its 
r\^ole as a tracking device, the TPC also measures the specific energy loss by 
ionisation ${\mathrm d}E/{\mathrm d}x$. 

Electrons and photons are identified in the electromagnetic calorimeter (ECAL)
by their characteristic longitudinal and transverse shower
development. The calorimeter is a lead/wire-plane sampling detector 
with fine readout segmentation. Each tower element is projective, subtending an 
angle of $\sim 1^{\circ}$ in both $\theta$ and $\phi$, and segmented longitudinally 
into three {\em stacks}. It provides a relative energy resolution 
of \mbox{$0.180/\sqrt{E} + 0.009$ ($E$ in \GeV)} for isolated electrons and photons.
 The three-dimensional fine segmentation allows a good spatial resolution to be 
achieved for photons and $\pi^0$'s in jets. 
Such deposits are separately identified and their energies evaluated by a fine 
clustering algorithm~\cite{perf}. 
Muons are identified by their penetration pattern in the
hadron calorimeter (HCAL), a 1.2~m thick iron yoke instrumented with 23 layers of 
streamer tubes, together with two surrounding double layers of muon chambers. 
The hadron calorimeter also provides a measurement of the energies of charged and
neutral hadrons with a relative resolution of \mbox{$0.85/\sqrt{E}$ 
($E$ in \GeV)}. At low polar angles, electromagnetic energy deposits are detected in 
the luminosity calorimeters (LCAL and SiCAL~\cite{perf}) down to 34 mrad with 
respect to the beam axis. 

\section{Event Reconstruction}
\label{sec:evrecons}
\subsection{Simulation of electromagnetic deposits}
\label{ECAL:FULLSIM}
Aiming for a precise measurement of \PMW\ to 1 part in 2000 imposes a level of 
understanding of this detector and its simulation not required in previous 
analyses of ALEPH data. The fine granularity and longitudinal 
segmentation of the ECAL detector elements~\cite{detector} allow nearby 
energy deposits to be identified. The treatment of these deposits has been 
revised following a detailed simulation. 

The normal simulation of the lateral and longitudinal development of 
electromagnetic energy deposits in the ECAL tower elements is based upon a 
parametrisation of showers measured in a test beam.  This parametrisation was  
employed in the generation of all reference events used in the analysis 
($\sim 10^8$ events). It provides a good description of 
the individual shower cores but fails to simulate the correlated 
fluctuations in their development through the sampling layers of ECAL, which 
can lead to the production of objects separate from the main deposit, called 
{\em satellites}. Mostly below 1 \GeV\ and confined to one stack, the observed rate 
of such objects significantly exceeds expectation.   
To understand the origin of this discrepancy, a more complete simulation 
(FULLSIM) of the response of ECAL to electromagnetic showers was 
developed using {\tt GEANT3}~\cite{GEANT}.  The effect of correlated fluctuations 
is included. As expected, there is better agreement in the reproduction of low 
energy satellites. However, since FULLSIM was restricted to an average medium for 
the ECAL sampling layers, it does not describe the lateral shower shape as well as 
the parametrisation. Consequently, its use was confined to the study of 
calorimeter systematic effects for which samples of $\sim 10^6$ events were 
generated.  

Specific studies with 45 \GeV\ Bhabha electrons 
show an excess in the data of objects formed entirely from connected elements 
from within the same stack. Similar effects are seen in the close neighbourhood 
of particles in jets. Not identified as electromagnetic, all single stack 
objects are removed from both data and simulated events unless 
related to a good track or a HCAL energy deposit. After this ECAL `cleaning' 
process, the multiplicity of single stack objects in ECAL matches the prediction 
from FULLSIM. The multiplicity of identified photons is unaffected by this 
procedure.  
The ECAL cleaning removes $\sim$3\% from the total energy of a hadronic jet both in 
the data and FULLSIM. 
 
\subsection{Energy flow}
\label{sec:eflow}
The total visible energy and momentum per event and thus the missing energy and 
momentum, are evaluated by an energy flow reconstruction algorithm~\cite{perf} 
which combines all measurements from calorimeters and tracking devices.     
The algorithm also provides a list of charged and neutral
reconstructed particles, called {\it energy flow objects}, from which jets
are reconstructed.
The four-momentum of a jet is defined as the sum of the four-momenta
of all particles in the jet, assuming the pion mass for all charged hadrons.
The typical jet angular resolution is 30~mrad in space. 
The jet energy resolution is approximately
$\sigma_{E_\mathrm{jet}}=(0.6\sqrt{E_\mathrm{jet}}+0.6)~\GeV$ where  
$E_\mathrm{jet}$ (in \GeV) is the jet energy. 

In order to bring better agreement between data and simulation, all energy flow 
objects in data and simulated events found to 
subtend angles less than 15$^{\circ}$ to the beams are rejected. 
All photonic and hadronic objects identified only in the ECAL are rejected if their 
energies are less than 1.5 \GeV. 
Hadronic objects identified in the third stack of the ECAL combined 
spatially with an HCAL deposit are rejected if their energies are less 
than 2 \GeV.  Objects with energies below these thresholds are not perfectly  
described by the simulation of the detectors. 

\subsection{Calibrations} 
\label{sec:calibrations}
Large samples of \PZz\ decays were collected at 91.2 \GeV\ CM energy at the start 
and end of LEP2 running each year. Di-lepton and di-jet events were used to monitor 
the performance of the detector and to 
compare reconstructed particle and jet four-momenta with the predictions of the
simulation. The following subsections describe the corrections applied where 
significant discrepancies between data and simulation were found. 

\subsubsection{Charged particles}
\label{sec:sagitta-corr}
For charged particles, small sagitta corrections are applied in data as 
determined using di-muons. They are proportional to momentum and opposite in 
sign for positively and negatively charged particles reaching a relative 
difference of 2\% for 45.6~\GeVc\ tracks at the smallest polar angles.
 
\subsubsection{Momentum of isolated leptons in 
e,\boldmath $\mu\nu\mathrm{q\bar{q}}$ events}
\label{sec:emumom}
Electron candidates from semileptonic \PW\ decays are corrected for energy losses 
due to bremsstrahlung in the detector material by combining their four-momenta
with those of any detected photons that are consistent with this
hypothesis. These photons can appear either as an excess of energy in
the ECAL electron cluster or as a separate deposit topologically consistent with 
bremsstrahlung. This correction is not applied when the electron is accompanied by 
other charged particles with summed momenta greater than 5 \GeVc\ within 
$6^\circ$ of the electron track. 
In addition, for muon and electron candidates, a search is made for isolated final 
state (FSR) photons associated with the lepton. Such a photon must be closer to the 
good lepton track than to any other object or the beam axis and at least $40^\circ$ 
away from any other good track. Their four-momenta are then combined.

In addition to the treatment of sagitta distortion described in 
subsection~\ref{sec:sagitta-corr}, the simulation of electrons from Bhabha events at 
91.2, 130-136 and 183-209 \GeV\ was compared with data. Small systematic biases as 
a function of polar angle $\theta$ and electron energy $E_{\rm e}$ were found 
arising from an imperfect simulation of saturation and leakage effects in ECAL. The 
main effect is a global relative shift 
of 0.45\%  parametrised as 
$\Delta E_{\rm e}/E_{\rm e}(\%) = 0.45(0.04)-4(8)\cdot 10^{-4}
[E_{\rm e}-45.6~\GeV]$ (errors in brackets).
This is applied as a correction to the simulated electrons to match the data.
A similar study for muons yields a small miscalibration of the momentum at 
45 \GeV\ (0.08$\pm$0.03\%) with no significant dependence on momentum or polar 
angle. In this case, no corrections are applied.

\subsubsection{Identified photons}
Using the energy flow algorithm, photons are identified in ECAL both in isolation 
and from within clusters of overlapping objects. Any bias in the 
photon energies from the simulated events relative to data was investigated by 
comparing $\pi^0$ mass distributions made from tau pairs at the 
\PZz. In addition, directly measured single photons from $\mu\mu\gamma$ events were 
compared event-by-event with the corresponding kinematically reconstructed 
values. Small biases are corrected to match the data, parametrised 
separately for the barrel, endcaps and the `overlap' region in between. 

\subsubsection{Jets}
\label{sec:jetcorr} 
\begin{figure}[htp]
  \begin{center}
    \mbox
    {\epsfig{file=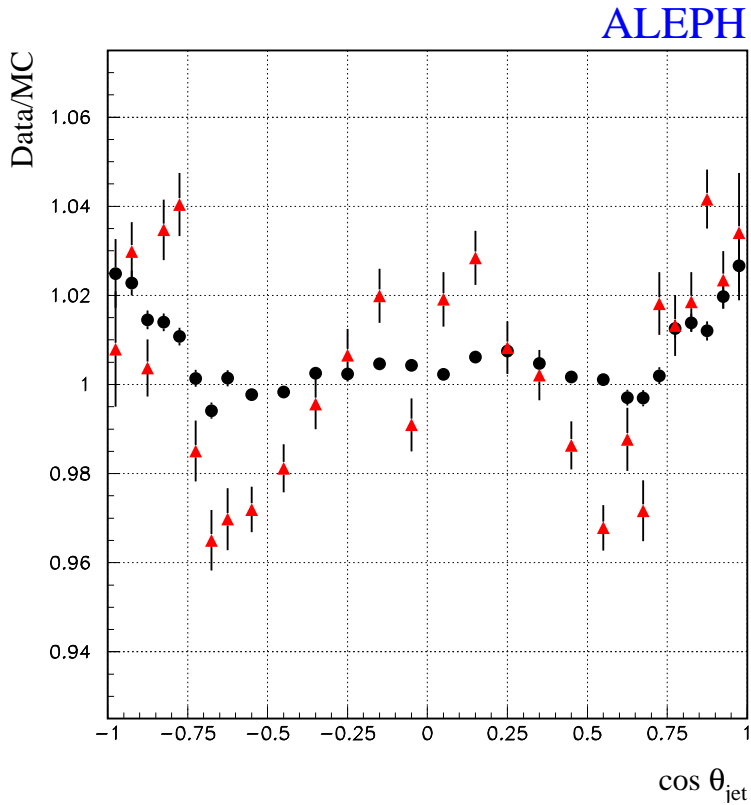,width=12.0cm}}
    \caption{\footnotesize 
The ratios of jet energies (circles) and resolutions (triangles) at 
the \PZz\ peak between data taken at the \PZz\ and corresponding 
simulation as a function of $\cos \theta_{\rm {jet}}$.
\label{fig:jetecorr}
     }
\end{center}
\end{figure}

Following these corrections to the charged hadrons and photons within jets,      
simulated hadronic events at the \PZz\ with energies of 45 \GeV\ were compared with 
the data.  The hadrons are clustered into two jets using the 
{\tt DURHAM-PE} algorithm~\cite{durham}.
Only \qqbar\ events with thrust values in the range 0.8 to 0.9925 are used, to 
suppress three-jet configurations and tau pairs. 
Using all \PZz\ calibration data collected during the LEP2 data taking periods,
a statistical precision of about 0.2\% on jet energies is obtained. 
Figure~\ref{fig:jetecorr} shows the ratio of jet energies in data to simulation, 
determined from the mean values in each bin, as a function of jet polar 
angle $\cos\theta_\mathrm{jet}$. The relative biases 
in the barrel region do not exceed 0.5\% and reach a maximum of 2.5\% for 
$|\cos\theta_\mathrm{jet} |>0.95$.  The Monte Carlo reconstructed 
jet energies are corrected bin-by-bin for these biases as a function of 
$\cos\theta_\mathrm{jet}$ before event kinematic fits are applied. 
Figure~\ref{fig:jetecorr} also shows the relative jet energy resolutions as 
determined from the RMS values of the distributions in each 
$\cos\theta_\mathrm{jet}$ bin. 
The simulation agrees with the data to within 1\% for the barrel and 4\% for the 
endcaps; no correction is applied.
 
\begin{figure}[htp]
  \begin{center}
    \mbox{
    {\epsfig{file=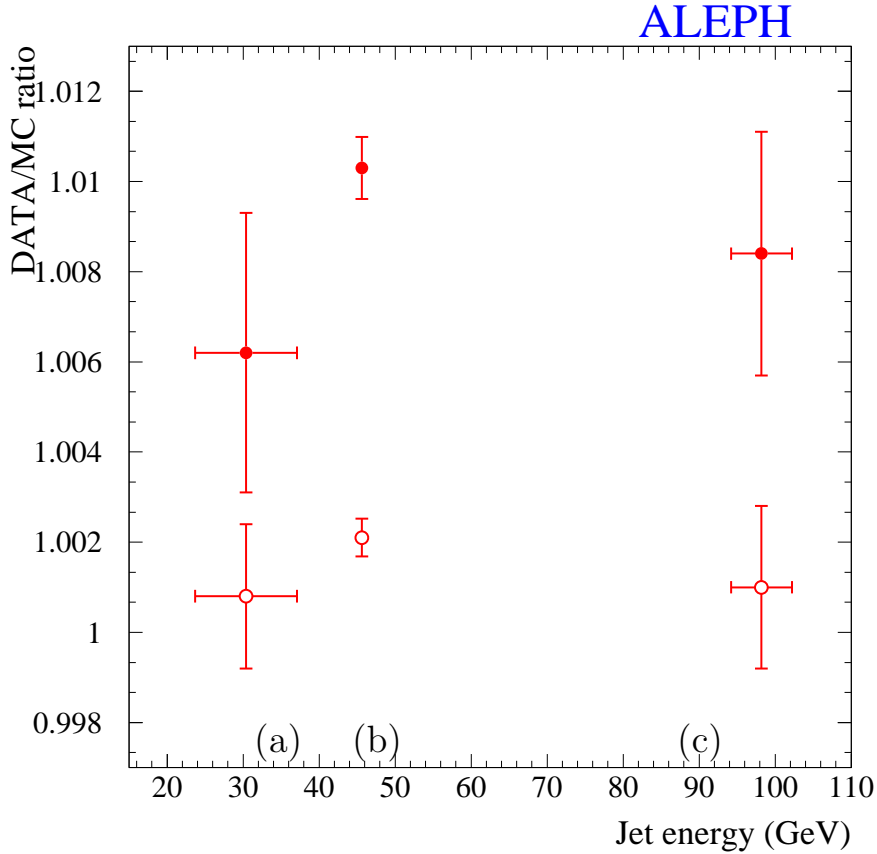,width=12.0cm}}
         }
    \caption{\footnotesize 
Data/MC ratio of average $E_{\rm {jet}}/E_{\rm {beam}}$ for $E_{\rm {jet}}$ 
energies of (a) 3-jet events at the \PZz, (b) di-jet events at the \PZz\ and (c) high
 energy di-jet events. Closed points: $|\cos\theta_{\rm {jet}}|>0.7$, open points: 
$|\cos\theta_{\rm {jet}}|<0.7$.   
             }
    \label{fig:jetlin}
  \end{center}
  \vspace{-4.1cm}
  \begin{center}
  \hspace*{-1cm}
  \begin{picture}(400,10)
  \put(140,10){\large (a)}
  \put(177,10){\large (b)}
  \put(300,10){\large (c)}
  \end{picture}
  \end{center}
  \vspace*{3.0cm}

\end{figure}

Variations in the corrections as a function of jet energy have been 
studied above 45 \GeV\ by comparing di-jet event samples from data and simulation at 
CM energies from 130 to 209 \GeV. Below 45 \GeV, a large sample of jets 
with energies centred at 30$\pm$7 \GeV\ were obtained from three-jet events at 
the \PZz\ peak. 
In this way, the full range of jet energies from \PWW\ decays is 
covered. Figure~\ref{fig:jetlin} shows the average ratios of measured jet 
energies in data to simulation for the barrel and endcap regions separately 
for three values of jet energies from 30 to 98 \GeV. No significant deviations 
are observed from the ratios found for 45 \GeV\ jets.  


%% file: montecarlo.tex
\newcommand{\nnbar}{\mbox{$\mathrm{\nu\bar{\nu}}$}}
\newcommand{\GEVcc}{\mbox{$\mathrm{GeV}/{{\it c}^2}$}}
\newcommand{\GEV}{\mbox{$\mathrm{GeV}$}}
\section{Simulation of Physics Processes}
\label{sec:MCsamples}

The {\tt KORALW} event generator, version 1.51~\cite{KORALW},
is used to produce \PW\ pair events. These events are weighted by the \oalph\ 
correction to the doubly resonant W-pair process using the {\tt YFSWW3} 
program version 1.16~\cite{YFSWW}.
Within {\tt KORALW} 
all four-fermion (4f) diagrams producing \PWW-like final states 
are computed, including Cabbibo suppressed decay modes,
using the fixed-width scheme for \PW\ and \PZz\ propagators.
The {\tt JETSET~7.4}~\cite{JETSET} or {\tt PYTHIA 6.1}~\cite{PYTHIA} packages
are used for the hadronisation of quarks in the final states.
Their parameters are tuned at the \PZz\ from a selection of \qqbar\ events 
with anti-b tagging.
Colour reconnection and Bose-Einstein 
final state interactions are not included. 
A sample of 10$^6$ 4f events
was generated with {\tt KORALW} at each of eight CM energies 
ranging from 182.7 to 206.5 \GeV~\cite{xsec_final}. 
The \PW\ mass was set to 
80.35 \GeVcsq\ and the width taken from Standard Model (SM) predictions to be 
2.094 \GeV. These samples are used 
as reference samples for fitting to the data in the reweighting 
procedure (Sec.~\ref{sec:extraction}), 
as well as for the study of detector systematic errors.
Additional samples of 200k events 
were generated with \PW\ masses up to 0.5 \GeVcsq\  
and \PW\ widths up to 0.6 \GeV\ different from the reference sample, for 
checking the stability of the results. 
Also, an independent sample of 500k \PW\ pair events was generated 
at each CM energy 
with {\tt KORALW} restricted 
to the doubly resonant \CCC\ diagrams~\cite{xsec_final}. 
This sample is used to train neural networks 
and parametrise the corrections used in the kinematic fitting.

For studies of the systematic errors from fragmentation in \PW\ decays, 
$10^6$ \PW\ pair events generated with {\tt KORALW} 
were hadronised using {\tt JETSET}, {\tt HERWIG~6.2}~\cite{HERWIG} and 
{\tt ARIADNE 4.10}~\cite{ARIADNE} and processed through the 
detector simulation. To suppress statistical fluctuations 
in the comparison between these models, the same events at the 
parton level were used.
Similarly, simulated samples of 100k to 500k events,
generated with {\tt KORALW}, were hadronised with modified versions
of {\tt JETSET}~\cite{CRSK,GAL}, {\tt HERWIG} and {\tt ARIADNE}~\cite{2-step}
containing various implementations of colour reconnection, to assess the 
influence of final state interactions between \PW\ decay products on the mass 
and width. The {\tt KORALW} events
were also hadronised with a version of {\tt JETSET} 
that includes Bose Einstein correlations~\cite{LoSj,Lonnblad},
to determine their influence on the \PW\ mass and width measurements.     

Simulated samples of events of at least hundred times the data luminosity
were generated for all background processes at each CM energy. 
The $\Pepem\rightarrow \qqbar(\gamma)$ events were generated with 
{\tt KK} version 4.14~\cite{KKMC} with hadronisation performed by 
{\tt PYTHIA} and including final state photon radiation (FSR) in the parton 
shower step.
Interference between initial and final state was not taken into account.
Events from ZZ-like final states were generated using {\tt PYTHIA} (NC08 
diagrams), but particular care was taken to avoid double counting of ZZ 
events already included in the signal generation as WW-like events 
(i.e. u$\bar{\rm u}$d$\bar{\rm d}$, \mm \nnbar,..). The same applies to Zee 
final states, generated with a 12 \GEVcc\ minimum mass for the Z system, for 
what concerns double counting of $\Pep\Pem\nnbar$ events.
Two-photon ($\gamma\gamma$) reactions into leptons and hadrons 
were simulated with the {\tt PHOT02}~\cite{PHOT02}, {\tt PYTHIA} and 
{\tt HERWIG} 
generators but no events survived the selection cuts in the \qqbar\qqbar\ and 
\lvqq\ channels. 
Di-lepton final states were simulated
using {\tt KK} for $\tau\tau(\gamma)$ and $\mu\mu(\gamma)$ 
and {\tt BHWIDE 1.01}~\cite{BHWIDE} for ee$(\gamma)$ events.
Table~\ref{tab:mcprod} lists the number of simulated events passed through 
{\tt GEANT}, generated for each 
process at each CM energy with corresponding integrated luminosities for the 
data.

\begin{table} [ht]
\caption [kk] {\label{tab:mcprod} {\footnotesize
Overview of the numbers of simulated events generated (in units of 1000 events) for 
each process type at eight average centre-of-mass energies and corresponding data 
integrated luminosities.
Events generated from the same 4f (2f) samples are shown in {\em italics}.
}}
\begin{center}
\begin{tabular} {@{\extracolsep{-1.5mm}} l   r  r  r r r r r r}
\hline
\hline
 year       & 1997    & 1998    & 1999    & 1999 & 1999 & 1999 & 2000 & 2000 \\
 Energy (\GEV) & 182.65 & 188.63 & 191.58 & 195.52 & 199.52 & 201.62 &  204.86 & 206.53 \\
Luminosity and & 56.81 & 174.21 & 28.93 & 79.86 & 86.28 & 41.89 & 81.41 & 133.21 \\
total error (\ipb) &$\pm 0.31$ &$\pm0.77$ &$\pm 0.14$&$\pm 0.36$ &$\pm 0.38$ &$\pm 0.20$ &$\pm 0.38$ &$\pm 0.60$ \\
\hline
$4f$- signal  & 1200 & 1400 & 1200 & 1200 & 1200 & 1200 & 1200 & 1400 \\
$4f$-JETSET   &      & {\em1000} & & & & & & {\em1000} \\
$4f$-HERWIG   &      & {\em1000} & & & & & & {\em1000} \\
$4f$-ARIADNE  &      & {\em1000} & & & & & & {\em1000} \\
\hline
ZZ (NC08)   &  200 &  200 &  200 &  200 &  200 &  200 &  200 &  200 \\
Zee ($>12~\GEVcc$)& 200 &  200 &  200 &  200 &  200 &  200 &  200 &  200 \\
\Pepem      & 1000 & 3000 & 1000 & 3000 & 3000 & 1000 & 3000 & 3200 \\
\mm         &  300 &  300 &  300 &  350 &  300 &  300 &  300 &  300 \\
\toto       &  100 &  160 &  100 &  100 &  100 &  100 &  100 &  100 \\
\qqbar      & 2000 & 1200 & 2000 & 2000 & 2000 & 2000 & 2000 & 1000 \\
\qqbar-JETSET &    & {\em1000} &     &      &     &      &       & {\em1000} \\
\qqbar-HERWIG &    & {\em1000} &     &      &     &      &       & {\em1000} \\
\qqbar-ARIADNE &   & {\em 750} &     &      &     &      &       & {\em 750} \\
\qqbar-BE32   & {\em 150} & {\em 150} & {\em 150} & {\em 150} & {\em 150} & {\em 150} & {\em 150} & {\em 150} \\
\hline
$\gamma\gamma\ra \Pepem$  &  700 & 1200 & 700 & 900 & 900 & 1100 & 900 & 1100 \\
$\gamma\gamma\ra \mm$     &  600 & 1200 & 700 & 900 & 900 & 800 & 900 & 1100 \\
$\gamma\gamma\ra \toto$   &  800 & 1200 & 700 & 900 & 900 & 800 & 900 & 1100 \\
$\gamma\gamma\ra$ hadrons &      &     &     &      &      &     &     &     \\
   un-tagged & 1000 &3000 & 500 & 1500 & 1500 & 500 & 1500 & 2500 \\
   tagged & 500 & 1000 & 500 & 1000 & 1000 & 500 & 1000 & 1000 \\
\hline
CC03-JETSET  &{\em500}&{\em500}&{\em500}&{\em500}&{\em500}&{\em500}&{\em500}&{\em500}\\
CR model SKI & {\em500}  &{\em500}  &{\em500}  & {\em500} &{\em500}&{\em500}&{\em500}& {\em500}\\
CR model SKII  &   &{\em500}&  &  &  &  &  & {\em500}\\
CR model SKII$^\prime$ &   &{\em500}&  &  &  &  &  & {\em500}\\
CR model GAL &{\em250}  &{\em250}  &{\em250}  &{\em250}  &{\em250}  &{\em250}  &{\em250}  &{\em250} \\
BE32 2 models & {\em300}  &{\em300}  &{\em200}  &{\em200}  &{\em300}  &{\em200}  &{\em200}  &{\em300} \\
\hline
CC03-ARIADNE & {\em500}& {\em500}& {\em500}& {\em500}& {\em500}& {\em500}& {\em500}&{\em500}\\
CR model AR2& {\em500}& {\em500}& {\em500}& {\em500}& {\em500}& {\em500}& {\em500}& {\em500}\\
\hline
CC03-HERWIG & {\em500}& {\em500}& {\em500}& {\em500}& {\em500}& {\em500}& {\em500}& {\em500}\\
CR model $11\%$& {\em500}& {\em500}& {\em500}& {\em500}& {\em500}& {\em500}& {\em500}& {\em500}\\
\hline
\hline
\end{tabular}
\end{center}
\end{table}

%% file: selection.tex
\section{Event selections and kinematic reconstruction}
\label{sec:selection}

In the following subsections, the event selections and kinematic 
reconstruction procedures for the mass extraction are described for the 
following four classes of \PWW\ events \qqbar\qqbar, \evqq, \mvqq\ and 
\tvqq. The selections are those required for the 
\PWW\ cross section measurement~\cite{xsec_final}. 
For the \qqbar\qqbar, \evqq, and \mvqq\ channels, the cuts developed earlier at 
189 \GeV~\cite{mass_189} for the leptons and jets are used followed by 
re-optimised neural networks for the higher CM energies. 
A new selection has been developed for the \tvqq\ channel. 
All selections are mutually exclusive.

\boldmath
\subsection{$\PW\PW\ra\qqbar\qqbar$ selection}
\unboldmath

A first preselection step aims at removing events with an energetic undetected 
initial state radiation (ISR) photon from radiative returns to the \PZz\ by 
requiring that the 
absolute value of the total longitudinal momentum be less than 
$1.5(M_{\rm vis}-M_{\rm Z})$ where $M_{\rm vis}$ is the observed visible mass.
All accepted particles are then forced to form 
four jets using the {\tt DURHAM-PE} algorithm~\cite{durham}. Only events where 
the  jet resolution parameter, $y_{34}$, is larger than 0.001 are 
kept. To reject $\qqbar$ events with a visible ISR photon, 
none of the four jets can have more than 95\% of electromagnetic energy 
in a $1^\circ$ cone around any particle included in the jet. 
Four-fermion final states in which one of the fermions is a charged lepton are 
rejected by requiring that the leading charged  particle of each jet carries 
less than 90\% of the jet energy.

The same neural network (NN) as in Ref.~\cite{xsec_final}, trained at five CM 
energies (189, 196, 200, 205 and 207 \GeV) on Monte Carlo \CCC\ and background 
events, is used to tag the preselected events.
There are 14 input variables based on global event properties, heavy quark 
flavour tagging,  reconstructed jet properties and \PWW\ kinematics.  
The signal is well separated from the $\qqbar(\gamma)$ background with 90\% 
efficiency and 80\% purity by requiring a NN output in excess of 
0.3~\cite{xsec_final}.

According to the simulation, a significant fraction ($\sim$6\%) of the 
accepted events are accompanied by an ISR photon 
that can be detected in the calorimeters separately from the hadronic jets. 
Such photons can be removed from the jet clustering process, thus improving 
the invariant mass resolution for \PW\ pairs.
Studies at 189 \GeV\ show that such photons with energies above 3 \GeV\ are 
identified 
in SiCAL or LCAL and above 5 \GeV\ in ECAL with an overall efficiency of 63\% 
and purity of 72\% if an isolation criterion 
based on a minimum angular separation from the closest energy flow object is 
applied. The minimum separation applied is 8$^\circ$ in SiCAL or LCAL and 
18$^\circ$ in ECAL for all CM energies. 
These events are treated differently in the subsequent kinematic fit.    

\boldmath
\subsection{$\PW\PW\ra\lvqq$ selection}
\unboldmath

A preselection common to the three lepton topologies requires at least seven
tracks in the event. Background from \qqbar\ events is reduced by requiring 
the estimated sum of missing energy and missing momentum to be greater than 
35~\GEV. 
The $\mathrm{Z}\gamma$ events in which the photon is undetected are rejected by
requiring the missing longitudinal momentum to be smaller than 
\[\rm{Max}((s-M_{\rm Z}^2)/2\sqrt{s}-27.5 \ \GeV, 
(\sqrt{s}-M_{\rm Z}^2/\sqrt{s}-\sqrt{\emis^2-{\not{\!p_T}}^2}-6 \ \GeV) \]
where ${\not{\!p_{\rm T}}}$ is the transverse missing momentum
and $\emis$ is the missing energy. 

Following the identification of the lepton and associated objects, the 
remaining particles are clustered into two jets using the {\tt DURHAM-PE} 
algorithm as in the \qqbar\qqbar\ channel.
    
\boldmath
\subsubsection{\evqq\ and \mvqq\ selection}
\unboldmath
In addition to the common preselection, a tighter cut is used on the total 
visible energy and visible longitudinal momentum to further reject 
$\mathrm{Z}\gamma$ events:
\[E^{\rm {vis}} (s-M_{\rm Z}^2)/(s+M_{\rm Z}^2) - P_{z}^{\rm {vis}} > 
5 \ \GeV\ \]
where $E^{\rm {vis}}$ and $P_{z}^{\rm {vis}}$ are the visible energy and 
longitudinal momentum, respectively.

The lepton candidate is chosen as the good track with the largest 
P $\sin{(\theta_{\rm {lj}}/2)}$ where P 
is the track momentum and $\theta_{\rm {lj}}$ is the angle from the track
to the closest jet clustered from the remaining tracks using the 
{\tt DURHAM-PE} algorithm ($y_{\rm cut}=0.0003$).
Events are further considered if this lepton candidate satisfies the 
electron or muon criteria defined in Ref.~\cite{xsec_final} and if the sum of the
lepton and missing energies is greater than 30~\GEV. 

Two different NN's have been trained to select and classify 
\evqq\ and \mvqq\ signal events~\cite{xsec_final}.
Both use three discriminant variables, the event transverse momentum, 
the lepton energy and the lepton isolation. 
The event is classified as \evqq\ or \mvqq\ if the corresponding NN 
output value is larger than 0.6~\cite{xsec_final}. The efficiency and 
purity of the \evqq\ selection are 82\% and 93\% respectively. The corresponding 
values for the \mvqq\ channel are 89\% and 98\%.    

Detailed studies of neutral objects not already classified as 
bremsstrahlung within $2.5^\circ$ of the electron track impact point on ECAL 
show a higher multiplicity than expected even after the removal of single 
stack objects (Sec.~\ref{ECAL:FULLSIM}). 
The reference simulation fails to reproduce the data for angles up to 
8$^{\circ}$. Further studies show that a smaller but still significant excess 
of charged objects are present in the data for both \evqq\ and \mvqq\ events. 
Although the summed energy of these objects near the isolated lepton is 
small, their impact on the closest jet is significant, especially for the 
\evqq\ channel. 
Therefore, all these objects up to 8$^{\circ}$ from the lepton are removed 
from the jet reconstruction. Also, they are not included in the calculation of 
the lepton four-momentum.

\boldmath
\subsubsection{\tvqq\ selection}
\unboldmath

A new selection has been designed~\cite{xsec_final}, based on an improved tau 
reconstruction~\cite{djamel}.
Leptonic tau decays are searched for by examining those events with e or 
$\mu$ candidates which fail the \evqq\ or \mvqq\ selection. 
These events are subjected to a similar three variable NN but trained 
on leptonic tau decays. 
Events with the NN output greater than 0.4 are kept~\cite{xsec_final}.

After removing the events which have satisfied any of the three variable NN
selections for \evqq, \mvqq\ or \tvqq, the remaining events 
are further examined for additional \tvqq\ final states. 
Use is made of the fact that one-prong tau decays are characterised by a 
low visible mass with a mean about 0.75~\GeVcsq. The first step is to perform a 
jet clustering using the JADE algorithm~\cite{jade} with a low 
{\it $y_{\rm cut}= (0.75 / E_{\rm vis})^2$} ($E_{\rm vis}$ in \GeV).
The tau candidate is defined as the jet which maximises 
$p_{\rm j} \ (1-\cos\theta_{\rm j}$), 
where $\theta_{\rm j}$ is the smallest angle with respect to other jets and 
$p_{\rm j}$ is the jet momentum. The event is then subjected to additional cuts,
in particular the invariant mass of the hadronic recoil system to the tau 
candidate must be in the range 60 to 105~\GeVcsq.
For those events which fail, the procedure is repeated with increasingly higher
values of {\it $y_{\rm cut}$}. When this exceeds 
$(5.0 / E_{\rm vis})^2$ the iterations are stopped and the event is kept 
requiring only that the recoiling invariant mass is larger than 
20 \GeVcsq~\cite{xsec_final}.

If a $\tau$-jet candidate is found, the event is subjected to further cuts 
to remove the main backgrounds. Most of the $\gamma\gamma$ interactions are
rejected by requiring the visible mass of the event to be larger than 
50 \GeVcsq\ and the missing transverse momentum greater than 10~\GeVc. 
The event is divided into two hemispheres with respect to a plane perpendicular
to the thrust axis. The acollinearity angle between the two hemispheres is 
required to be less than $175^\circ$  to reject most of the \qqbar\ background.
About  80\% of the events with a tau candidate satisfy these
cuts but significant background remains, mainly from \qqbar\ events. These 
events are then subjected to a 15 variable neural network. 
The event is selected if the result is greater than 0.4.  The efficiency and 
purity of the \tvqq\ selection are 65\% and 86\% respectively~\cite{xsec_final}. 

\subsection{Kinematic fit}
\label{subsec:kf}

The biases and resolutions used in the kinematic fits for the jet energies and
directions are determined from an independent \CCC\ simulated sample. The
distributions of the differences between the reconstructed jet energies
and angles and those of the jets built directly from the generated
particles are binned in jet energy and polar angle. Each of these
distributions is fitted to a Gaussian and the mean values and sigmas
are fed to the fitting algorithms.

Except for the \tvqq\ channel, \PW\ pair events are treated as four body final 
states with either four jets or two jets, a charged lepton and neutrino to which 
the measured missing momentum is assigned.
For each selected event, two invariant masses
are computed from the \PW\ decay products. In order to improve 
resolution, kinematic fits are made with the constraint 
of event four momentum conservation and fixing the velocities 
($p/E$) of the jets to their measured values. 
Imposing energy and momentum conservation alone
corresponds to a four-constraint (4C) fit in the case of fully hadronic events,
and a one-constraint (1C) fit in the case of semileptonic events, giving two
different fitted masses per event. An equal mass constraint for the two bosons 
corresponds, respectively, to a five (5C) or two-constraint (2C) fit. 
In the \tvqq\ channel, since the tau energy is largely unknown due to
neutrinos in the tau decay, only the hadronic side of the event is used with
the sole constraint of the beam energy.

The average raw resolution of 12\% on the total jet momentum improves by a
factor 2 and by a factor up to 5 for polar angles down to 20 degrees, due 
largely to the kinematic fitting.

For all classes of events the fits converge successfully 
producing flat $\chi^2$ probability distributions for $P(\chi^2)>0.05$.
The peak below $P(\chi^2)=0.05$ is populated by 
events that do not fully satisfy the fitting hypothesis.
Monte Carlo studies show that approximately half of these events 
have ISR energies greater than 0.5 \GeV, 
leading to a significant positive bias in the reconstructed di-jet masses.
However, these events are not removed since the simulation adequately 
describes the observed $\chi^2$ probability distributions in all channels. 

In the \qqbar\qqbar\ channel
for those events with an identified ISR photon in the detector, the 
procedure of event clustering and fitting is modified~\cite{mass_189}. 
In this case, the energy flow objects from which the ISR photon has been removed 
are forced into four jets. The fit is performed taking into account the modified 
constraints
\[
  \left[ \sum_{{\rm i}=1}^4 (E_{\rm i}, \vec{p}_{\rm i}) = 
  ( \sqrt s, \vec{0}) \right] \,\,
  \rightarrow \,\,
  \left[\sum_{{\rm i}=1}^4 (E_{\rm i}, \vec{p}_{\rm i}) = 
  ( \sqrt{s}\!-\!E_{\gamma}, -\vec{p}_{\gamma})\right].
\]
Of the 4861 data events selected after all cuts, 220 are treated in this way. 
Monte Carlo studies at 189 \GeV\ show that the invariant mass resolution for 
these events improves from 4.1 to 2.9~\GeVcsq\ and the mean displacement of 
the masses from their true values is zero within error.
The improvement in the expected error on \PMW\ for all
selected events is $\sim$2\%.

\subsubsection*{Jet pairing in the \boldmath\qqbar\qqbar\ channel}
\label{sec:invm4q}
At most one of the three possible jet pairings is chosen, based on the
the \CCC\ matrix element 
$|{\cal M}({p}_{{\rm f}_1},{p}_{\bar{{\rm f}_2}},{p}_{{\rm f}_3},
{p}_{\bar{{\rm f}_4}},m_\mathrm{W}^\mathrm{ref})|^2$,
where the ${p}_{{\rm f}_{\rm j}}$'s denote the kinematically fitted four-momenta 
of the respective jets and $m_\mathrm{W}^{\rm ref}$ the reference \PW\ mass, 
taken to be 80.35 \GeVcsq.  The combination with the largest value of 
$|{\cal M}|^2$
is in general selected (in 90\% of the cases), provided that
(a) it does not have the smallest sum of jet-jet angles and
(b) both fitted masses lie in the [60,110] \GeVcsq\ window.
Otherwise (in 10\% of the cases) if it satisfies the same criteria, the 
combination with the next-to-largest value of $|{\cal M}|^2$ is chosen. If
the pairings with the two largest values of $|{\cal M}|^2$ are not accepted, the
event is rejected. At 189 \GeV~\cite{mass_189} for example, the fraction of 
kinematically fitted signal events surviving these criteria is 80\%. 
Of these events, 90\% are found to have the correct 
combination of di-jets when comparing their directions to those of the  
original $\PW\ra\qqbar$ decays. The bias 
from the choice of reference mass is found to be negligible. 
In addition, the combinatorial and physical backgrounds do not show particular 
structure in the defined mass window.

%% file: extraction.tex
\section{Extraction of the W mass and width} 
\label{sec:extraction}

The \PW\ boson mass and width are extracted by fitting simulated 
invariant mass spectra to the observed distributions. 
As in previous analyses~\cite{mass_172, mass_183, mass_189} 
an unbinned maximum likelihood procedure is employed to find the best fits,
using probability density functions obtained from the binned distributions 
of reference event samples, reweighting the Monte Carlo signal 
events with the \CCC\ matrix elements corresponding to various values of \PMW\ 
and \PGW. 
Two types of fits are performed for all four channels individually.
In the first, a one-parameter fit for \PMW\ is made,
where \PGW\ varies with \PMW\ according to the Standard Model
as $\PGW=2.094\;\GeV\times(\PMW/(80.35\;\GeVcsq))^3$.
These results provide the most precise value of \PMW.
In the second, two-parameter fits are performed allowing \PMW\ and \PGW\ to 
vary as two independent parameters.
Although the shape of the invariant mass spectra are dominated by experimental 
resolutions, these fits are used to test the validity of the SM prediction 
for \PGW\ and check for any correlation between the two fitted parameters.
Technically, the matrix element calculation 
assumes the Standard Model value for \PGW\ at a given \PW\ mass, 
for the coupling of electrons and their neutrinos to \PW\ bosons and allows 
the width to vary freely only in the \PW\ propagator.

At LEP1, the \PZz\ mass was defined using a running-width scheme in the 
Breit-Wigner propagator. However, a fixed-width scheme has been employed in 
generating all \PWW\ events with {\tt KORALW}. As a result, to make both mass 
measurements consistent with each other, a positive shift 
of 27 \MeVcsq\ is applied to the extracted \PW\ mass~\cite{conv}. 
The corresponding shift to the fitted width of
0.7 \MeV\ is not significant. 

The statistical error on \PMW\ and \PGW\ is computed from the fits to the 
data distributions.  Also, a large number of subsamples are 
studied, each with the same number of events observed in the data, 
to evaluate the expected errors.

The selection efficiency is found to be independent of the \PW\ mass.
The variation of the total signal cross section with \PMW\ affects the purity 
of the selected events and is taken into account, whereas its dependence on 
\PGW\ is assumed to be negligible.

The reweighting procedure was tested at 189 and 207 \GeV\ by comparing 
the fitted with the input mass in each channel individually for
four independent 4f Monte Carlo samples 
generated with \PMW\ values of 79.850, 80.100, 80.600 and 80.850 \GeVcsq. 
The relationship between the fitted and true masses was found to be 
linear for all channels over this range.
The best straight line fits through the points are consistent 
with calibration curves of unit slope and zero bias, within the statistical 
precision of the test. Small deviations are observed in the \evqq\ channel 
from which a systematic uncertainty is derived (Sec.~\ref{sec:CCCME}). 
   
Table~\ref{tab:selec-xsec} gives the expected and observed numbers of events 
from all contributing processes for each channel which satisfy the kinematic 
fitting criteria after all window cuts are applied.
The numbers of expected \PWW\ events are calculated with \oalph\ corrections 
using the standard 4f reference samples generated at 
\PMW\ = 80.35 \GeVcsq.
\begin{table}[tbhp]
\caption{
\protect\footnotesize
Expected numbers of events corresponding to the whole data sample
(183-209 \GeV) for signal and background processes after all selection, 
quality and window cuts for the four categories of events used in the 
extraction of \PMW\ and \PGW. 
All \PWW\ events are regarded as signal in the calculation of 
the quoted purities per channel.
The signal cross sections are determined with \PMW=80.35 \GeVcsq\ and 
\PGW=2.094 \GeV\ and the \oalph\ correction is applied.
}
\begin{center}
\begin{tabular}{|l|c|c|c|c|}
\hline
Process  & $\mathrm{4q}$ & $\evqq$ & $\mvqq$ & $\tvqq$  \\
\hline
\hline
$\PWW\ra\qqbar\qqbar$    & 4264 & 0.1   & 0.0  & 5.0 \\
$\PWW\ra\evqq$           & 2.1  & 1217  & 0.1  & 120.4 \\
$\PWW\ra\mvqq$           & 1.9  & 0.5   & 1295 & 41.7 \\
$\PWW\ra\tvqq$           & 10.2 & 41.5  & 41.6 & 959.4 \\ 
\hline
$\qqbar(\gamma)$         & 591  & 17.9  & 0.6  & 35.4 \\
$\PZz\PZz$               &  95  & 2.2   & 4.3  & 23.8 \\
$\PZz\mathrm{ee}$        & 2.2  & 7.4   & 0.0  & 16.3 \\
$\PZz\nu\nu$             & 0.0  & 0.0   & 0.0  & 0.7  \\
$\tau\tau$               & -    & 0.2   & -    & 0.4 \\
$\gamma\gamma\ra\tau\tau$& -    & 0.0   & -    & 0.1 \\
$\gamma\gamma\ra$hadrons & -    & 0.4   & -    & 0.2 \\
\hline
\hline
Predicted events        & 4966  & 1288  & 1342  & 1203   \\
Observed events         & 4861  & 1259  & 1371  & 1226  \\
Purity (\%)             & 86.1  & 97.8  & 99.6  & 93.6  \\
\hline
\end{tabular}
\end{center}
\label{tab:selec-xsec}
\end{table}

\subsection{The \boldmath\qqbar\qqbar\ channel}

The two-dimensional reweighting fits used in the previously published analyses 
at 183 and 189 \GeV~\cite{mass_183, mass_189} are replaced by 
three-dimensional (3-D) fits which better exploit the available information 
from each event. The following three estimators were selected: (i) the 5C fitted 
mass, $M_\mathrm{5C}$, (ii) a random choice of one of the 4C di-jet unrescaled 
masses, $M_\mathrm{4C}$ and (iii) the kinematic fit error on the 5C mass, 
$\sigma_{M_{\rm 5C}}$. Using a binned 3-D probability density function, a 
maximum likelihood fit is performed to the data within the following 
acceptance windows: \quad $70 < M_\mathrm{5C} < 90$ \GeVcsq, 
\quad $0 < \sigma_{M_\mathrm{5C}} < 4$ \GeVcsq\  
\quad and $60 < M_\mathrm{4C} < 110$ \GeVcsq\  
for both the one and two-parameter fits. The allowed fit range for \PGW\ is 
loosely constrained to \quad $1.1 < \PGW\ < 4.1$ \GeV.
Bin sizes in the probability density distribution of the 5C and 4C masses are 
chosen for signal and summed backgrounds separately such that the 
number of events of each type per bin is approximately constant. The third 
dimension is subdivided into four bins chosen dynamically to equalise the 
number of signal events in each bin. This binning is kept for the summed 
background. The fitted mass is extracted in each of these bins in the third 
dimension and the likelihoods combined to determine the final mass and error. 
To avoid any bias, the minimum number of signal Monte Carlo events per 3-D bin 
is 200. 

 Fig.~\ref{fig:massdists}(a) shows the mass distribution from the 5C kinematic 
fits to the data before the window cuts between 70 and 
90~\GeVcsq\ are applied. For comparison the mass distribution predicted from the 
simulation, reweighted to the fitted W mass in data, is superimposed.

\subsection{The {\boldmath \evqq} and {\boldmath \mvqq} channels}
\label{sec-MLemu}
The following variables are used to form a three-dimensional (3-D)
probability density function:  
the 2C mass, $M_\mathrm{2C}$, where the leptonic and hadronic masses are 
constrained to be equal, the kinematic fit uncertainty on the 2C mass, 
$\sigma_{M_\mathrm{2C}}$ 
and the 1C hadronic mass, $M_\mathrm{1C}^{\mathrm{q\bar{q}}}$.  
  The event-by-event correlation between $M_\mathrm{1C}^{\mathrm{q\bar{q}}}$ 
  and $M_\mathrm{2C}$ was found to be 43\% at 189 \GeV.
  By construction, the 3-D probability 
  density function from the simulation takes into account all correlations 
  amongst the three variables and leads to an improvement 
  in statistical precision compared with a 1-D fit.
 Using a binned 3-D
probability density function, a maximum likelihood fit is performed 
to the data within the following acceptance windows:
\quad $70 < M_\mathrm{2C} < 90$ \GeVcsq, 
\quad $0 < \sigma_{M_\mathrm{2C}} < 10$ \GeVcsq, 
\quad $60 < M_\mathrm{1C}^{\mathrm{q\bar{q}}} < 110$ \GeVcsq\ and with the 
fitted \PGW\ being constrained in the range \quad $1.1 < \PGW\ < 4.1$ \GeV. 
The bin sizes for the Monte Carlo events are chosen using the same criteria as 
for the \qqbar\qqbar\ channel. The binning of the 3-D probability density 
function has 3 intervals along the event-by-event error axis.    
A stable mass value and statistical error are obtained when the minimum number 
of Monte Carlo events in any bin is 200 or greater. 

Figs.~\ref{fig:massdists}(b) and (c) display the mass distributions for data 
resulting from the 2C kinematic fits to these semileptonic final states together 
with the predictions from the simulation.  

\subsection{The {\boldmath \tvqq} channel}
\label{sec:tauextn}
For \tvqq\ candidates, a 2-D reweighting fit uses the 1-C hadronic mass, 
$M_\mathrm{1C}^{\mathrm{q\bar{q}}}$ and its uncertainty, 
$\sigma_{M_\mathrm{1C}^{\mathrm{q\bar{q}}}}$, from the kinematic fit.
The events must be within the following mass and error acceptance windows:
\quad $70 < M_\mathrm{1C}^{\mathrm{q\bar{q}}} < 90$ \GeVcsq\ and  
\quad $0 < \sigma_{M_\mathrm{1C}^{\mathrm{q\bar{q}}}} < 10$ \GeVcsq. 
In this channel, the allowed fit range for \PGW\ is 
\quad $0.9 < \PGW\ < 4.3$ \GeV. 
The binning of the 2-D probability density function 
has four intervals along the event-by-event error axis  
and 60 intervals of varying size along the 1C mass axis.

\begin{figure*}[bth!]
\begin{flushleft}
\hspace*{-0.5cm}
\begin{tabular}[t]{@{}ll@{}}
\mbox{\epsfig{file=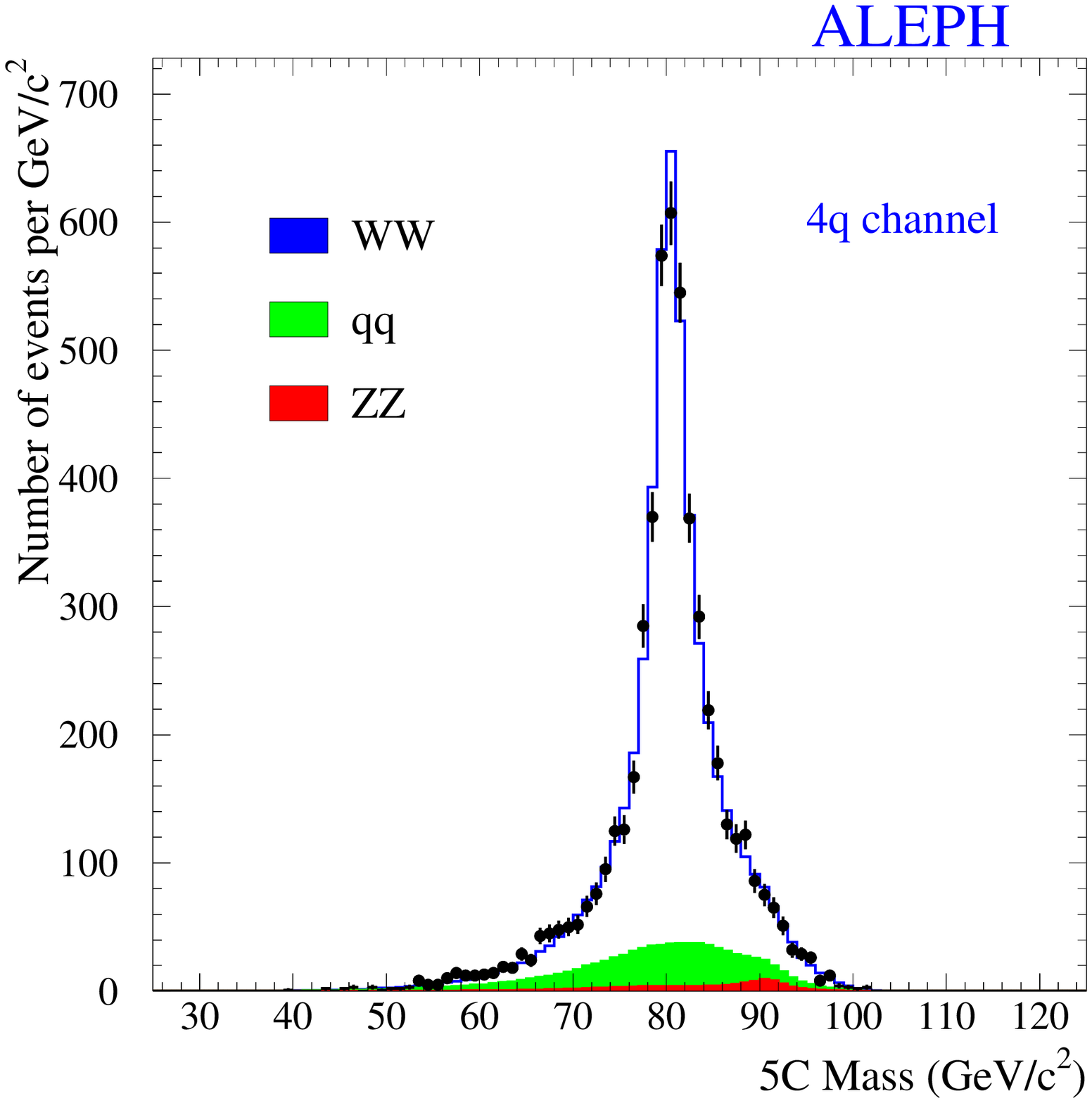,height=8cm}}
&
\mbox{\hspace*{-0.0cm}
      \epsfig{file=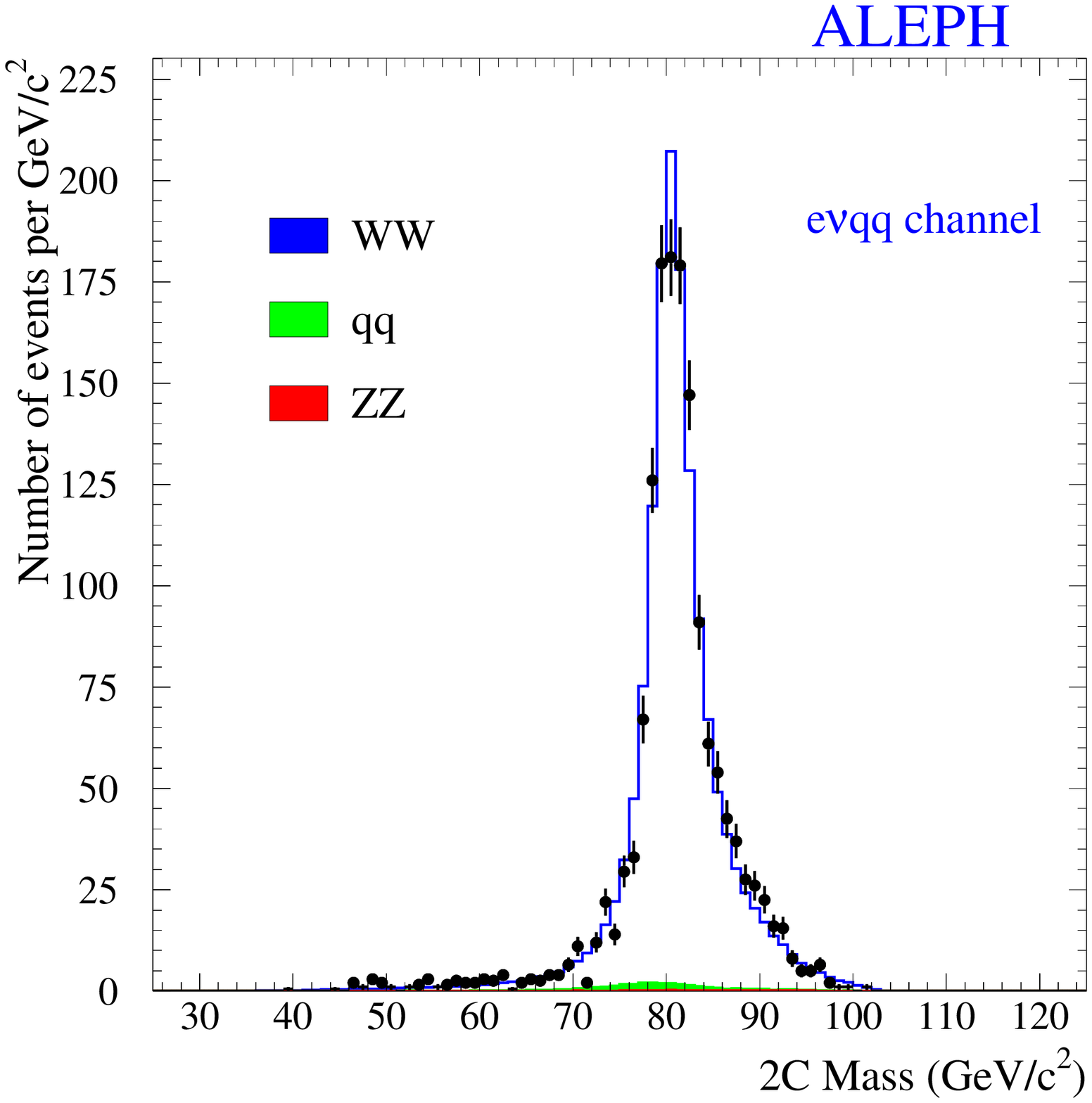,height=8cm}}
\end{tabular}
\end{flushleft}
\vspace{-3.5cm}
\begin{center}
\hspace*{-1cm}
\begin{picture}(400,10)
\put(030,5){\large (a)}
\put(275,5){\large (b)}
\end{picture}
\end{center}
\vspace{1.6cm}
\begin{flushleft}
\hspace*{-0.5cm}
\begin{tabular}[t]{@{}ll@{}}
\mbox{\epsfig{file=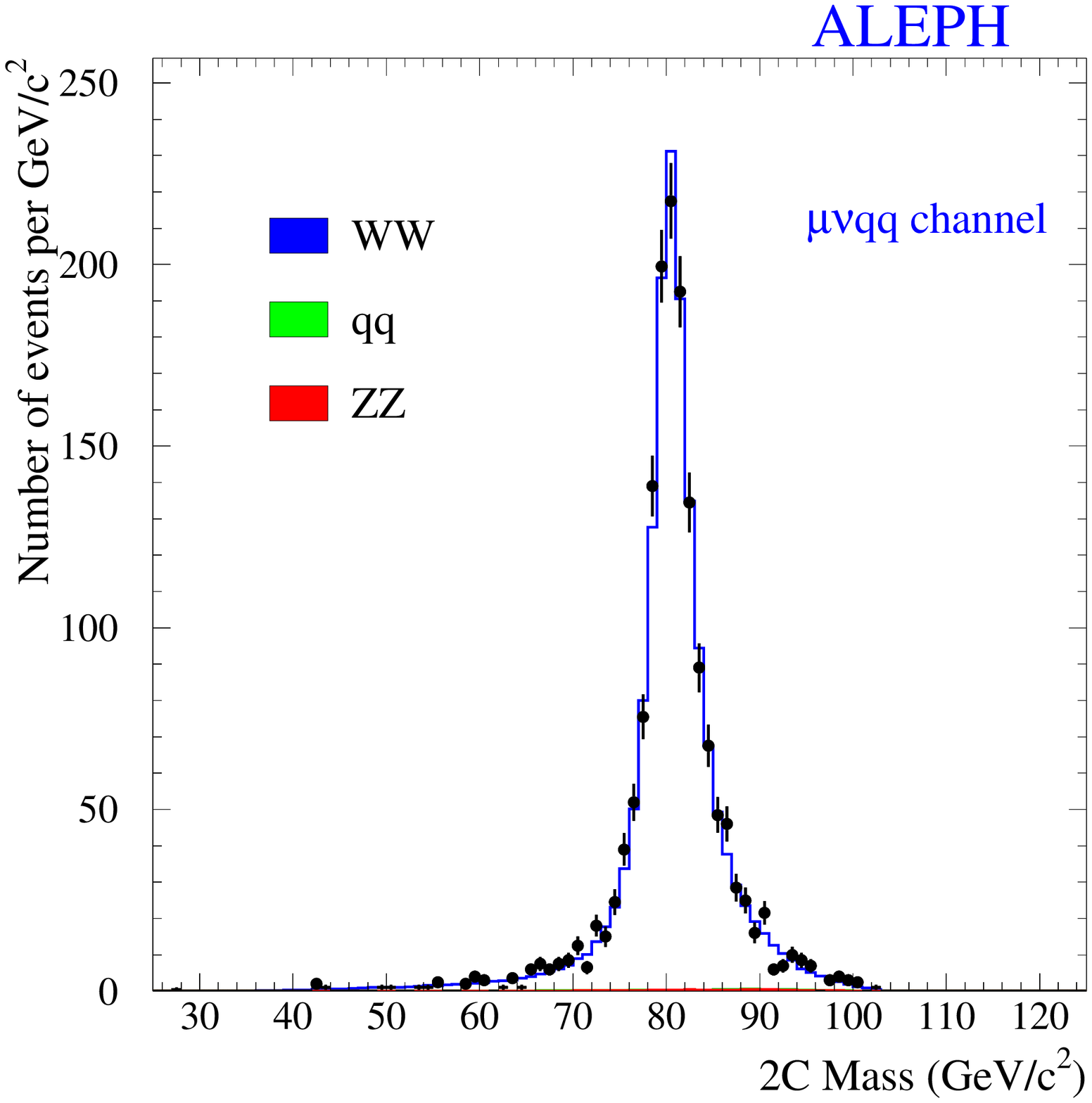,height=8cm}}
&
\mbox{\hspace*{-0.0cm}
      \epsfig{file=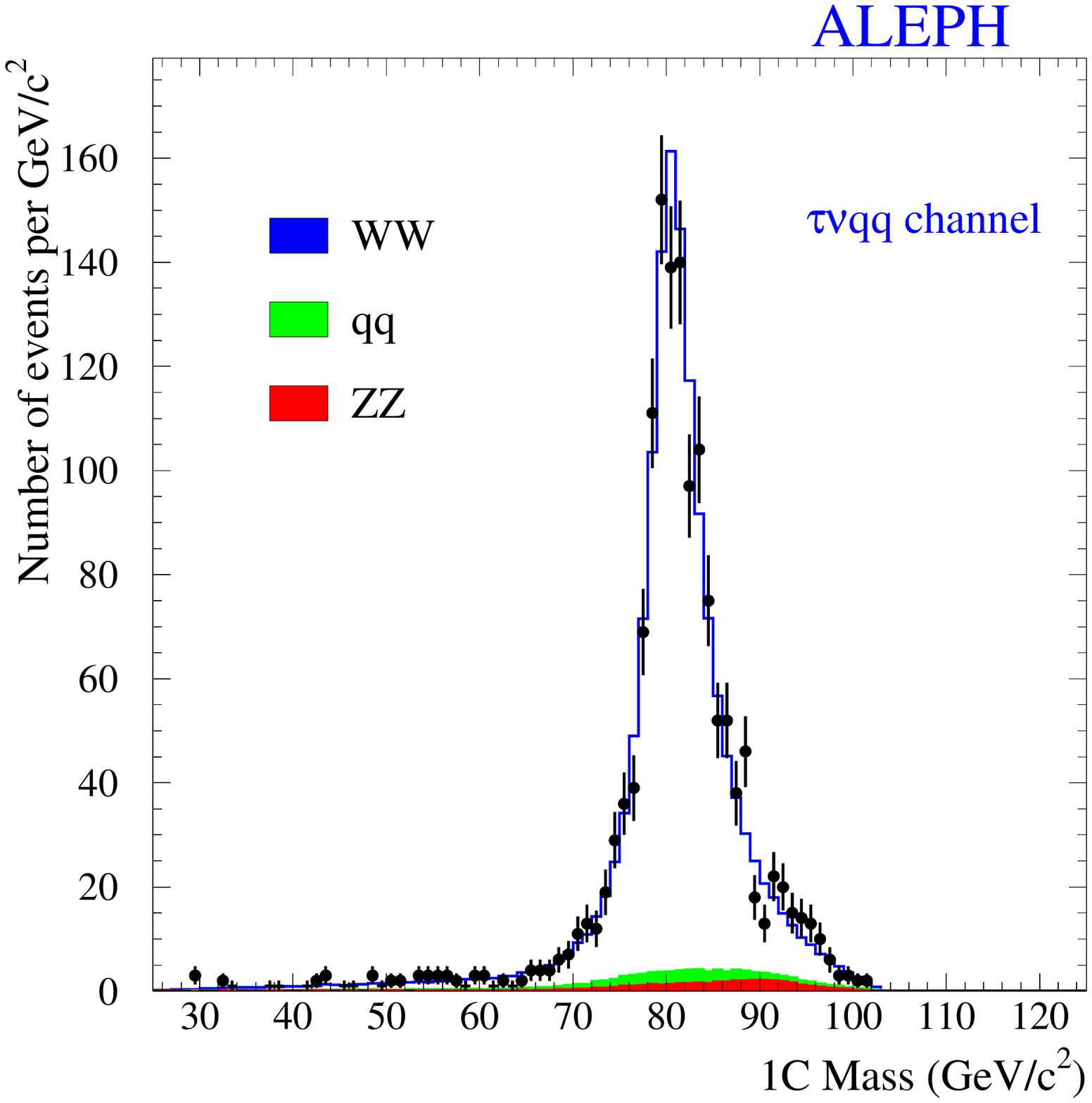,height=8cm}}
\end{tabular}
\end{flushleft}
\vspace{-3.5cm}
\begin{center}
\hspace*{-1cm}
\begin{picture}(400,10)
\put(030,5){\large (c)}
\put(275,5){\large (d)}
\end{picture}
\end{center}
\vspace{1.3cm}
\caption[]
{\protect\footnotesize
Mass distributions in the: (a) $\mathrm{4q}$, (b) \evqq, (c) \mvqq\ and (d)
\tvqq\ channels for data (points with error bars), 
non-\PWW\ background (shaded area) and 
signal+background Monte Carlo with \PMW\ values set 
to those fitted from each individual channel (solid line histogram).
For the $\mathrm{4q}$ channel, the distribution shows the 5C kinematically 
fitted 
dijet masses before window cuts are applied. For the \lvqq\ channels, the 
distributions show the 2C (or 1C) kinematic fits before window cuts.  
}
\label{fig:massdists}
\end{figure*}

Figs.~\ref{fig:massdists}(d) displays the mass distribution resulting 
from the 1C kinematic fits to the data events together with the prediction from 
the simulation.  

%% file: crlimit.tex
\clearpage
\section{Studies on colour reconnection}
\label{sec:crlimit}

The W bosons decay at a short distance from each other 
($1/\Gamma \approx 0.1$ fm), so that in the \qqbar\qqbar\ channel their decay 
products hadronise closely in space time at the typical hadronic scale of 
$\approx 1$ fm. 
An interaction between the partons from different W decays may then occur.

At the perturbative level, the shift in the reconstructed \PW\ mass due to single 
gluon exchange is 
suppressed by the square of the number of colours and by an additional factor of 
\PGW/\PMW. The mass shift is of the order of a few \MeVcsq~\cite{CRSK}. However, 
when the scale of gluon exchange is not large compared with \PGW, 
non-perturbative colour reconnections (CR) in the parton cascades may lead to 
much larger \PMW\ shifts.

At energies well above the pair production threshold, as in the
present data set, the final state QED interconnection between the W's
induces a shift in \PMW\ of order $\alpha_{em} \Gamma_W / \pi$,
which is a few \MeVcsq~\cite{mwshiftQED} and insignificant compared with the 
uncertainties from non-perturbative QCD.

\subsection{Monte Carlo models}
\label{CR_models}
At the non-perturbative level, all phenomenological implementations of CR 
within existing hadronisation models predict that 
the particle flow distributions per event are modified with the low 
momentum particles in the inter-jet regions being most affected. Any 
effect on high momentum particles would occur only when a jet from a \PWm\ is 
aligned with another from a \PWp. Such a topology would not survive 
the 4-jet selection.   
The effect of CR on the fitted \PMW\ is studied using the 
following variants of the parton evolution schemes: \\ 
\indent (a) SKI, SKII, SKII$^{\prime}$~\cite{CRSK} and GAL~\cite{GAL} in 
{\tt JETSET},\\ 
\indent (b) 2-step variants AR2 and AR20~\cite{2-step} in {\tt ARIADNE} and \\
\indent (c) HWCR in {\tt HERWIG}~\cite{HERWIG}. 

As formulated, the SK versions in {\tt JETSET} predict no effect at the \PZz\ and 
therefore, unlike the other variants, cannot be calibrated with \PZz\ data. 
The probability of an event to be reconnected depends on the string overlap 
between partons from the two \PW\ decays. In SKI, this is governed by a freely 
adjustable `string' overlap parameter, $k_{\rm i}$, whereas the predictions of the 
SKII and SKII$^{\prime}$ variants are fixed once the string parameters are fitted 
in {\tt JETSET}. 
When $k_{\rm i}$ is set to 0.65, the fraction of reconnected events is similar to 
SKII~(29.2\%) and SKII$^{\prime}$ (26.7\%). However, SKI($k_{\rm i}$=0.65) 
predicts a larger shift in \PMW\ than the other SK versions. The authors 
state that all SK models are equally valid~\cite{CRSKcomm}. 

The GAL implementation within {\tt JETSET} allows string rearrangements to occur 
by colour exchange with the probability for reconnection depending upon the 
reduction in total string area between the old and new configurations.
After tuning at the \PZz\ on global event properties, the fitted value of a 
non-perturbative free `strength' parameter, $R_0$, is found to be 0.04 correlated 
with the shower cut-off, $Q_0$, of 1.57 \GeVc. The author recommends a larger 
value for $R_0$ of 0.1 from fits to deep-inelastic scattering data which would 
lead to a correspondingly larger \PMW\ shift ($\sim$100 \MeVcsq).  
      
For AR2, both intra-W and inter-W reconnections are allowed between all dipoles 
with the same colour indices formed from emitted gluons with energy $E_{\rm g}$. 
The parton cascade is performed in two steps (i) allowing only intra-W 
reconnections with $E_{\rm g} > \PGW$ and (ii) allowing also inter-W reconnections 
but only for $E_{\rm g} < \PGW$. For AR20, no CR is applied either between or 
within the \PW's. In principle, the predicted net shift in \PMW\ 
due to inter-W reconnections is determined from the difference found between 
AR2 and the corresponding variant, AR21, where only 
intra-W reconnections are allowed. However, in practice it is found that the 
difference between AR20  and AR21 when tuned at the \PZz\ is not 
significant. Thus, the \PMW\ shift is taken from the comparison of AR2 and 
AR20 fitted events. 
   
In HWCR, the criterion for allowing colour reconnections is based on the reduction 
in space-time distances within the colour singlet clusters at the end of the 
parton shower. The reconnection probability is set to $1/9$ for allowed 
re-arrangements. 
The parameter VMIN2, the minimum squared virtuality of partons, is set to 
0.1 $(\GeVcsq)^2$.

Table~\ref{tab:mwshifts_mod} gives the predicted mass shifts 
$\delta\PMW = \PMW(\rm {CR}) - \PMW(\rm {no CR})$ 
from these models averaged over CM energies from 183 to 209 \GeV. Details of the 
parameter settings used in the models are given in Appendix A.
\begin{table}[htbp]
\begin{center}
\caption{\protect\footnotesize Predicted \PW\ mass shifts from various models 
averaged over all CM energies.
\label{tab:mwshifts_mod}
   } 
\begin{tabular}[h]{|l||c|}
\hline
Model             & $\delta\PMW$ (\MeVcsq) \\ \hline

SKI ($k_{\rm i}$=0.65)  & +39$\pm$2   \\
SKI ($k_{\rm i}$=1.0)   & +56$\pm$2   \\
SKII              & +6$\pm$8   \\  
SKII$^{\prime}$   & +4$\pm$8   \\
AR2-AR20          & +54$\pm$5   \\ 
HWCR              & +39$\pm$4  \\
GAL ($R_0$=0.04)  & +44$\pm$8  \\   
\hline
\end{tabular}
\end{center}
\end{table} 

\noindent The predicted mass shifts from the models tunable at the \PZz: GAL, AR2  
and HWCR, range from 40 to 55 \MeVcsq, suggesting for 
consistency that the value of $k_{\rm i}$ in the SKI model should be of order 0.8. 

To examine the validity of some of the tunable models of CR at the \PZz,  
the particle distributions in selected three-jet events were compared 
specifically with the predictions of AR2 and GAL~\cite{cr_ygap_lep}.  
If it can be assumed that the behaviour of colour rearrangements in the 
parton cascades of \PZz\ decays is the same as for \PWW, these observations 
suggest that the two models overestimate the effects on 
\PMW\ from CR. 

\subsection{Data Analysis}
\label{sec:CRmeas}
Keeping the originally reconstructed jets in each selected event, the W mass 
analysis is repeated twice, 
either removing all low momentum particles (PCUT analysis)~\cite{djamel}, 
or rejecting particles outside cones directed along the four jet axes 
(CONE analysis)~\cite{Hugo_thesis}. The difference from the mass
measured without these additional cuts, called the {\em standard} analysis, is a 
sensitive observable of the CR effect according to all the above models.

For each of five values of the particle momentum cut off from 1 to 3 \GeVc\
in the PCUT analysis, each jet energy and angle is recomputed.
In the CONE analysis, each jet energy is kept
unchanged, whilst its three-momentum is recomputed from the vector sum
of its remaining participating particles, rescaled by the ratio of the
original jet energy to the energy of the particles inside the cone. 
Seven values of the cone opening angle R are used from 0.4 to
0.9 radians. Studies show that fragmentation uncertainties increase rapidly for 
momentum cut-offs beyond 3 \GeVc\ or cone angles smaller than 0.4 radians. 
These values were found to provide optimal balance between statistical and 
systematic uncertainties on \PMW. 

\begin{figure}[htp]
  \begin{center}
    \mbox{
    {\epsfig{file=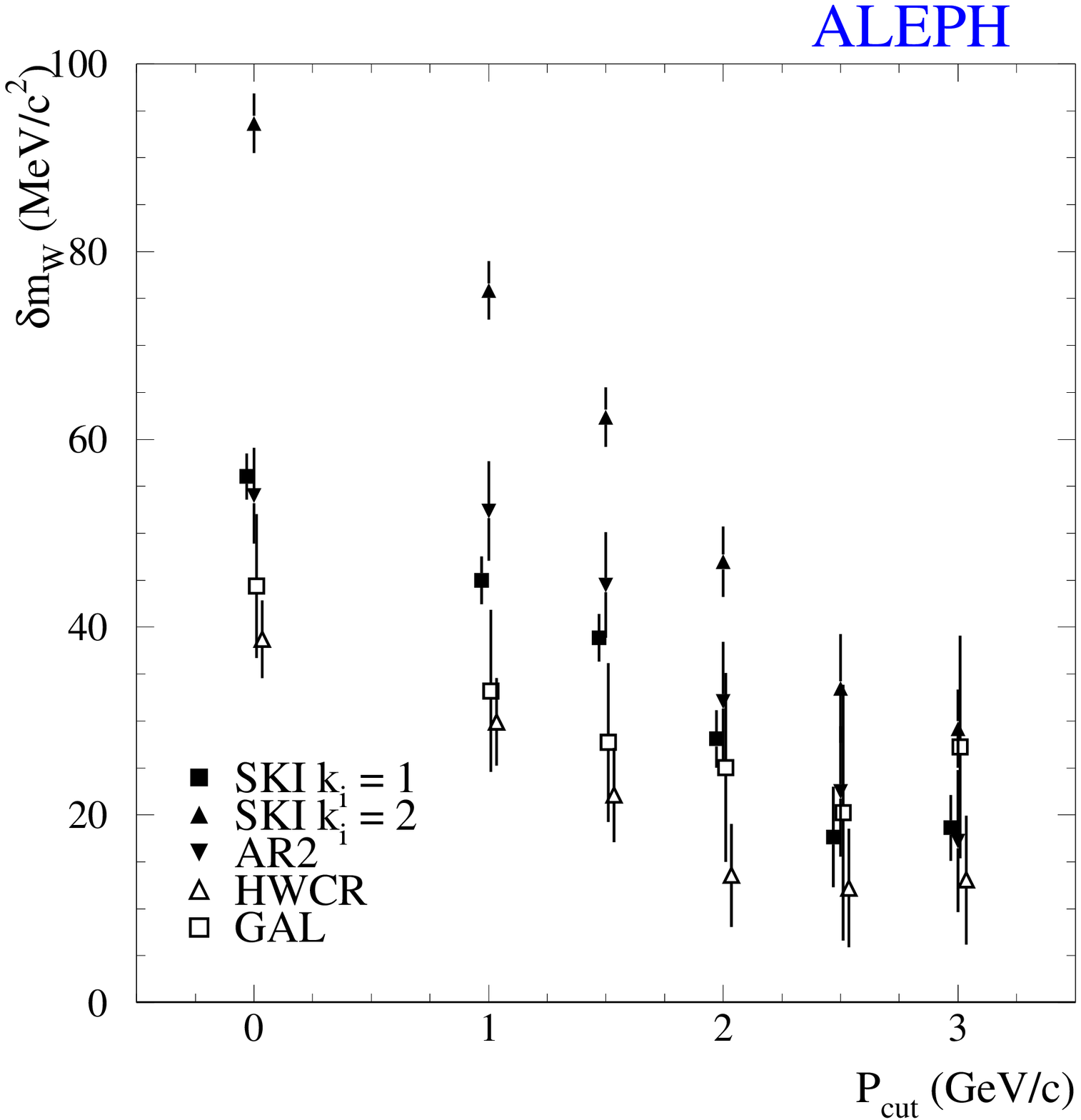,width=8.0cm}}
    {\epsfig{file=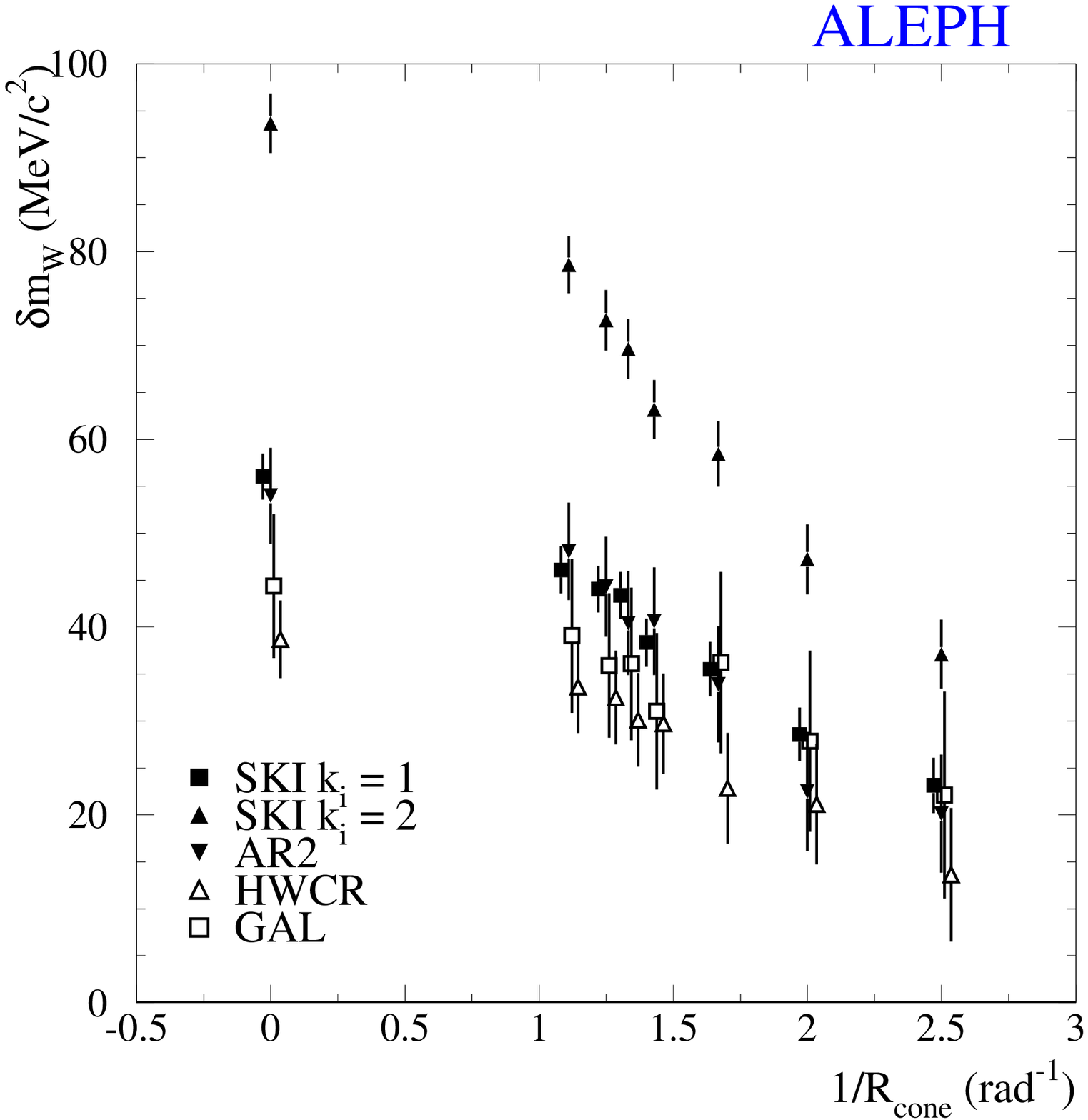,width=8.0cm}}
        }
    \caption{\footnotesize 
      $\delta\PMW$ versus (a) PCUT in \GeVc\ and (b) inverse CONE radius (R) in
rad$^{-1}$ for SKI (2 $k_{\rm i}$ values), AR2, HWCR and GAL models in the 
\qqbar\qqbar\ channel. }
  \label{fig:dmwvscut_mc}
  \end{center}
  \vspace{-7.9cm}
  \begin{center}
  \hspace*{-1cm}
  \begin{picture}(400,10)
  \put(170,30){\large (a)}
  \put(400,30){\large (b)}
  \end{picture}
  \end{center}
  \vspace*{6.5cm}
\end{figure}

Figure~\ref{fig:dmwvscut_mc} shows the expected variation of the
mass due to CR as a function of the cut for the tuned AR2, HWCR and 
GAL models in the 183 to 209 \GeV\ energy range. The SKI predictions for two values 
of $k_{\rm i}$ are also included. The predictions for each of the eight CM 
energies are combined using the relative integrated luminosities of the data.  
Table~\ref{tab:mws_ext} lists the \PMW\ shifts, $\delta\PMW^{\rm PCUT}$ and 
$\delta\PMW^{\rm CONE}$,  for the  PCUT(=3 \GeVc) and CONE (R=0.4 rad) 
reconstructions respectively. The corresponding \PMW\ shifts in the standard 
analysis $\delta\PMW^0$ are shown for comparison. Within errors the shifts for 
each reconstruction are comparable for all tuned models and consistent with SKI 
($k_{\rm i}$=1). 

\begin{table}[htbp]
\begin{center}
\caption{\protect\footnotesize Predicted \PW\ mass shifts ($\delta\PMW$) from 
various models averaged over all CM energies for the CONE (R=0.4 rad) 
and PCUT (=3 \GeVc) reconstructions (units in \MeVcsq). 
\label{tab:mws_ext} 
   } 
\begin{tabular}[h]{|l||c|c|c|} 
\hline
Model     & $\delta\PMW^0$ & $\delta\PMW^{\rm PCUT}$ & $\delta\PMW^{\rm CONE}$  \\
 \hline\hline
SKI ($k_{\rm i}$=1.0) & +56$\pm$2 & +19$\pm$4  & +23$\pm$3  \\
\hline
AR2-AR20          & +54$\pm$5 & +17$\pm$8  & +20$\pm$6  \\ 
HWCR              & +39$\pm$4 & +13$\pm$7  & +14$\pm$7 \\
GAL ($R_0$=0.04)  & +44$\pm$8 & +27$\pm$12  & +22$\pm$11  \\   
\hline
\end{tabular}
\end{center}
\end{table} 

\begin{figure}[htp]
  \begin{center}
    \mbox{
    {\epsfig{file=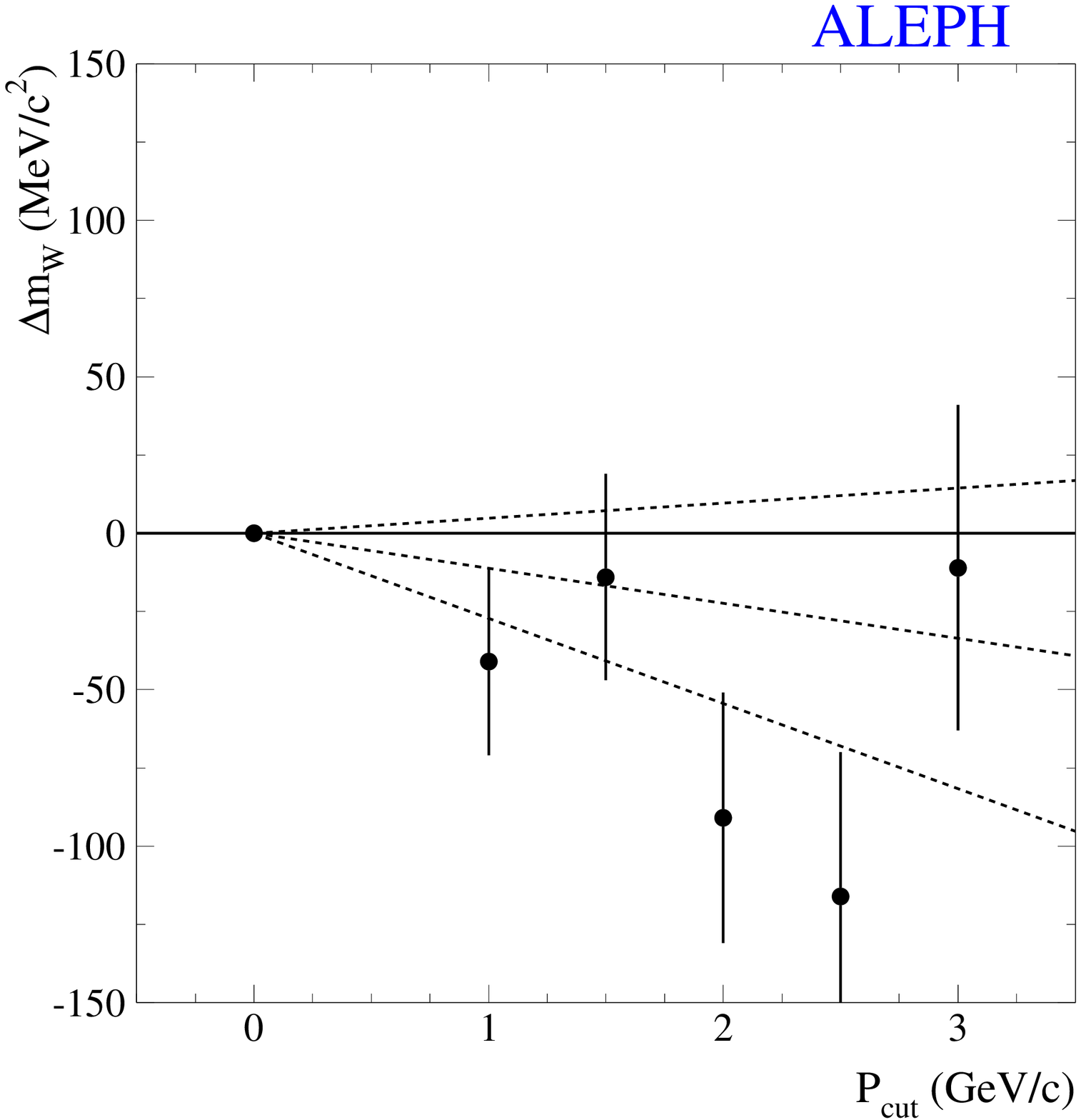,width=8.0cm}}
    {\epsfig{file=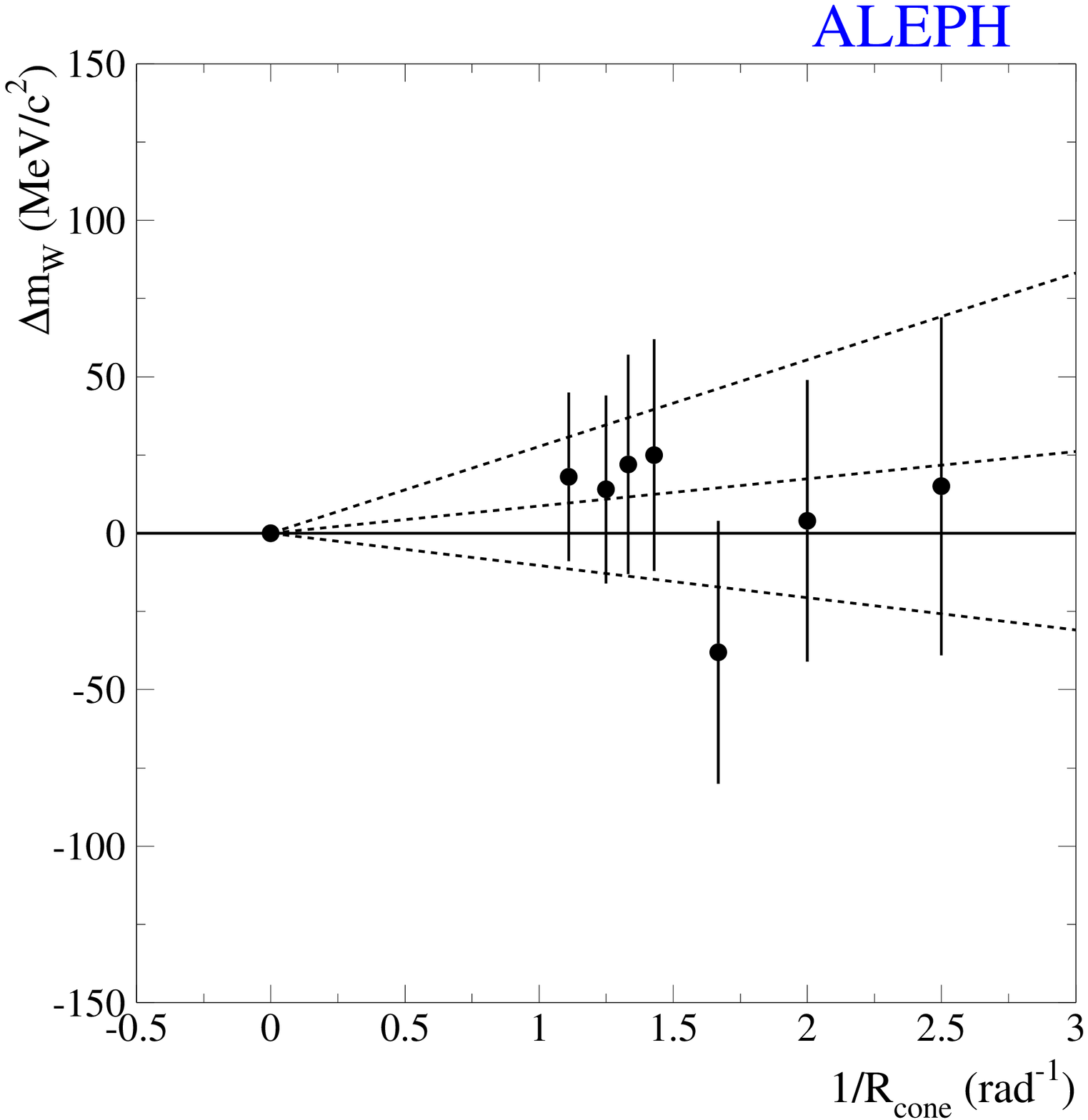,width=8.0cm}}
         }
    \caption{\footnotesize 
      $\Delta\PMW$ versus (a) PCUT in \GeVc\ and (b) inverse CONE radius (R) in
rad$^{-1}$ for \qqbar\qqbar\ data. The dashed lines denote the fitted 
slopes and $\pm 1\sigma$ errors. The correlation with respect to the standard 
analysis is taken into account in the error on the mass difference for each 
reconstruction. 
             }
    \label{fig:dmwvscut_data}
  \end{center}
  \vspace{-7.9cm}
  \begin{center}
  \hspace*{-1cm}
  \begin{picture}(400,10)
  \put(160,30){\large (a)}
  \put(400,30){\large (b)}
  \end{picture}
  \end{center}
  \vspace*{6.5cm}
\end{figure}

For the data collected at all CM energies combined, Fig.~\ref{fig:dmwvscut_data} 
shows the mass difference $\Delta\PMW$ between a PCUT or CONE reconstruction and 
the standard mass analysis. The slopes are fitted with the full correlation matrix 
included and amount to $-11 \pm 16$ (\MeVcsq)/(\GeVc) for
the PCUT analysis and $+9 \pm 19$  (\MeVcsq)/(rad$^{-1}$) for 
the CONE. Both values are compatible with no effect.
 
\begin{figure}[htp]
  \begin{center}
    \mbox{
    {\epsfig{file=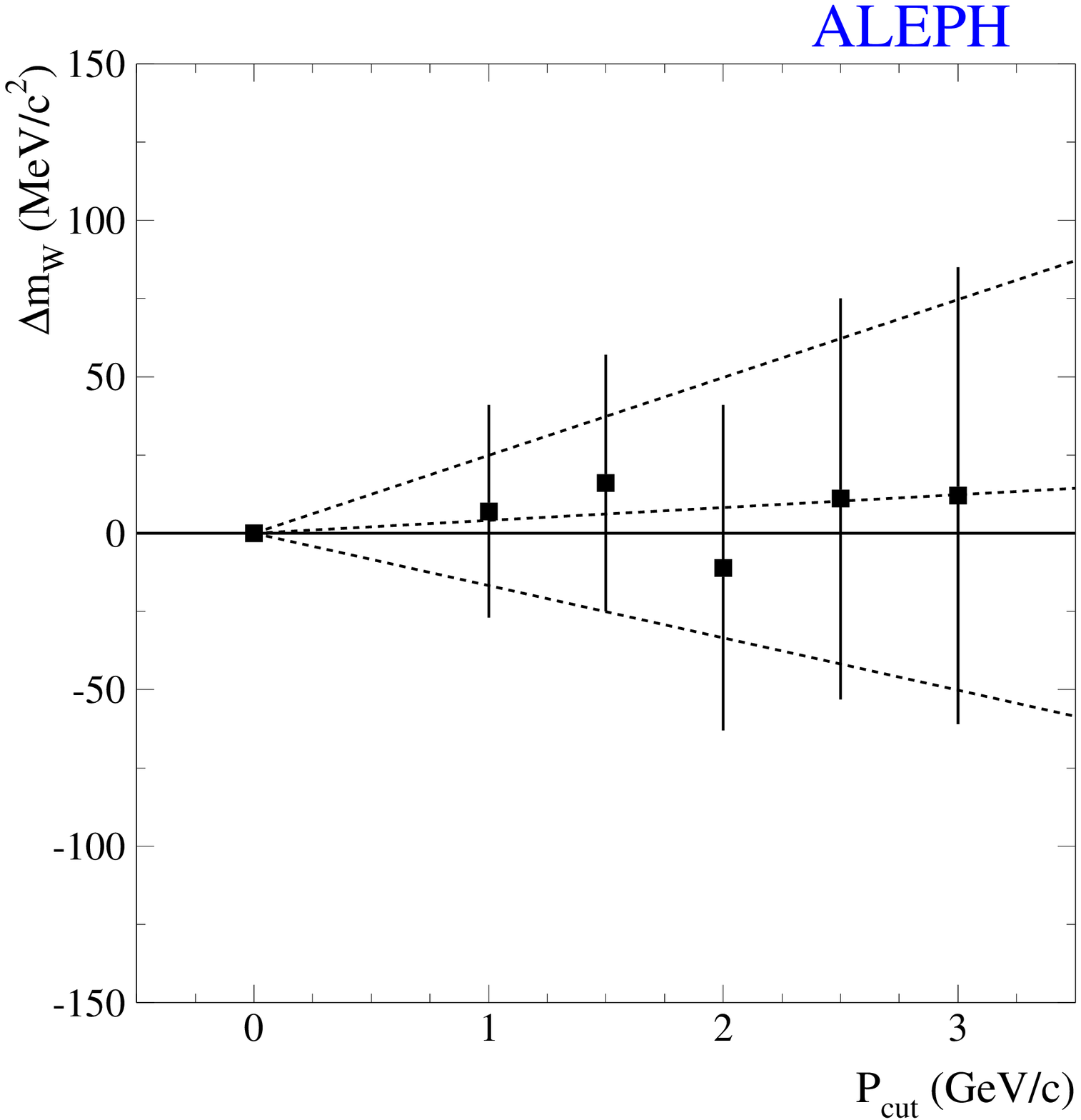,width=8.0cm}}
    {\epsfig{file=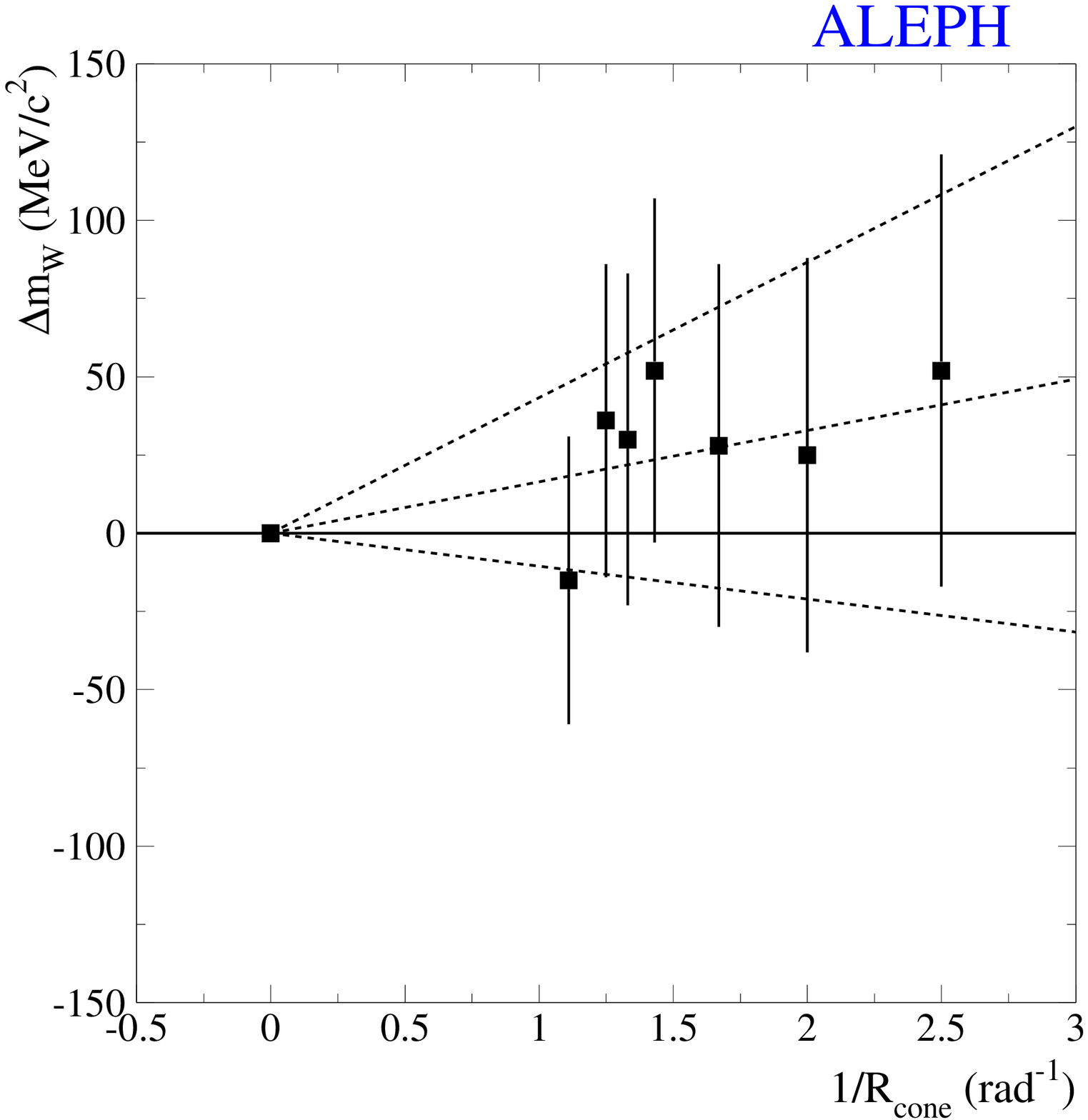,width=8.0cm}}
         }
    \caption{\footnotesize 
      $\Delta \PMW$ versus (a) PCUT in \GeVc\ and (b) inverse CONE radius (R) in
rad$^{-1}$ for data from the $e, \mu, \tau\nu \qqbar$ channels combined, fitting 
with the jets alone in each case}
    \label{fig:dmwlvqq}
  \end{center}
  \vspace{-7.9cm}
  \begin{center}
  \hspace*{-1cm}
  \begin{picture}(400,10)
  \put(70,30){\large (a)}
  \put(300,30){\large (b)}
  \end{picture}
  \end{center}
  \vspace*{6.5cm}
\end{figure}

A cross check was performed on all the semileptonic channels where no CR effect 
between the decay products of the different \PW's can be present. 
The mass analyses in the \evqq\ and \mvqq\ channels were repeated for 
PCUT and CONE following the same kinematic fit procedure as used in the \tvqq\ 
channel where only the hadronic jets are included. Figure~\ref{fig:dmwlvqq} shows 
the corresponding mass differences for each cut value relative to the standard 
analysis after combining the results statistically from the \evqq, \mvqq\ 
and \tvqq\ channels. No significant instability is observed.
The combined $\ell \nu \qqbar$ channels represent a sample of size similar
to the size of the \qqbar\qqbar\ channel and give a slope of 
$+4\pm 21$ (\MeVcsq)/(\GeVc) for the PCUT analysis and 
$+16\pm 27$ (\MeVcsq)/(rad$^{-1}$) for the CONE,
which are not significantly different from zero.

A limit on  $\delta\PMW$ can be inferred from a comparison between
the slopes observed in the data and those from the CR models.
For each model, pseudo-data samples were built, combining 
all the CM energy points weighted by their respective integrated luminosities. 
In the case of the SKI model, 20 different values of the $k_{\rm i}$ parameter are 
chosen, ranging from 0 to 100. 

\begin{figure}[htp]
  \begin{center}
    \mbox{
    {\epsfig{file=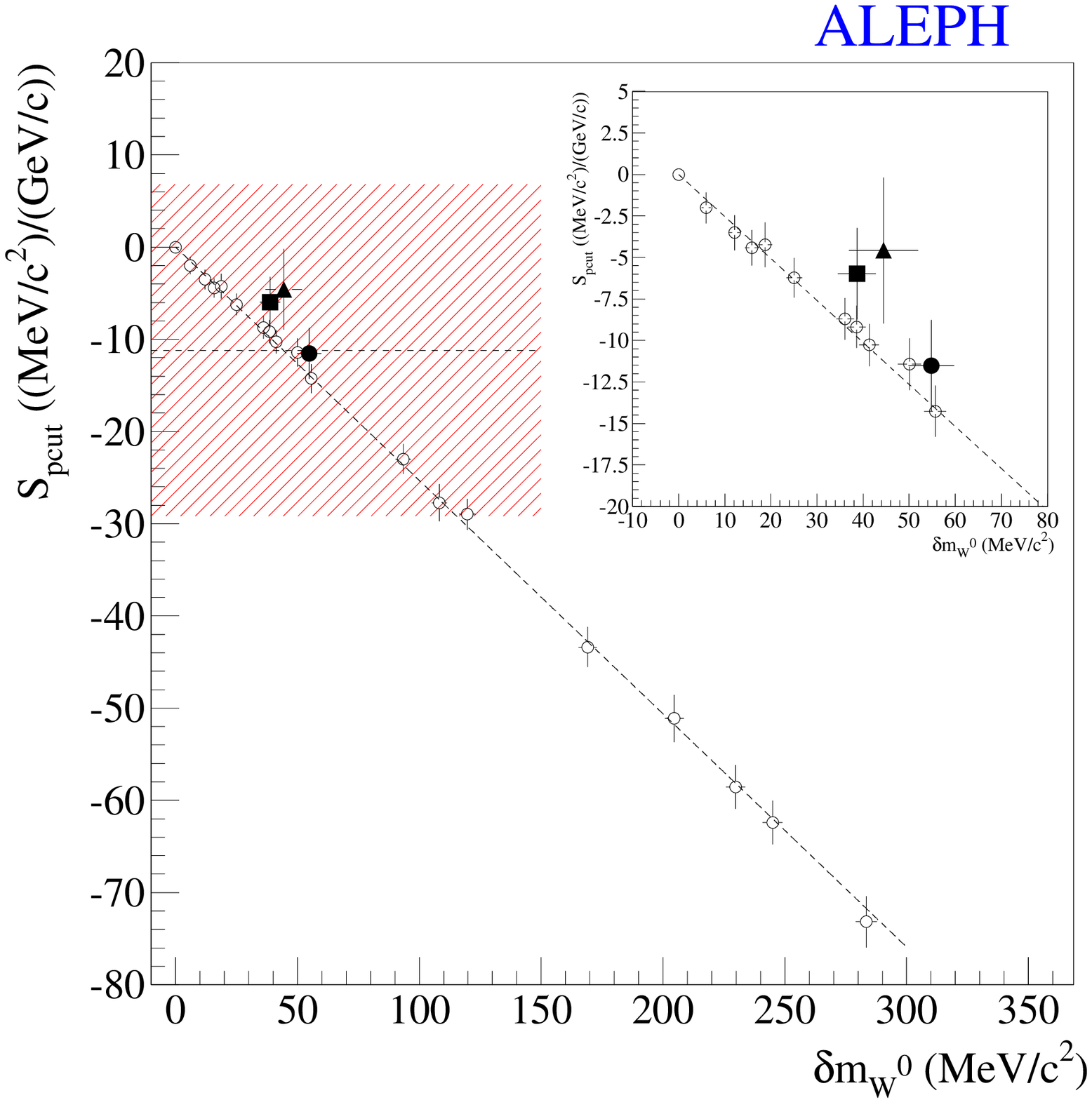,width=8.5cm}}
    {\epsfig{file=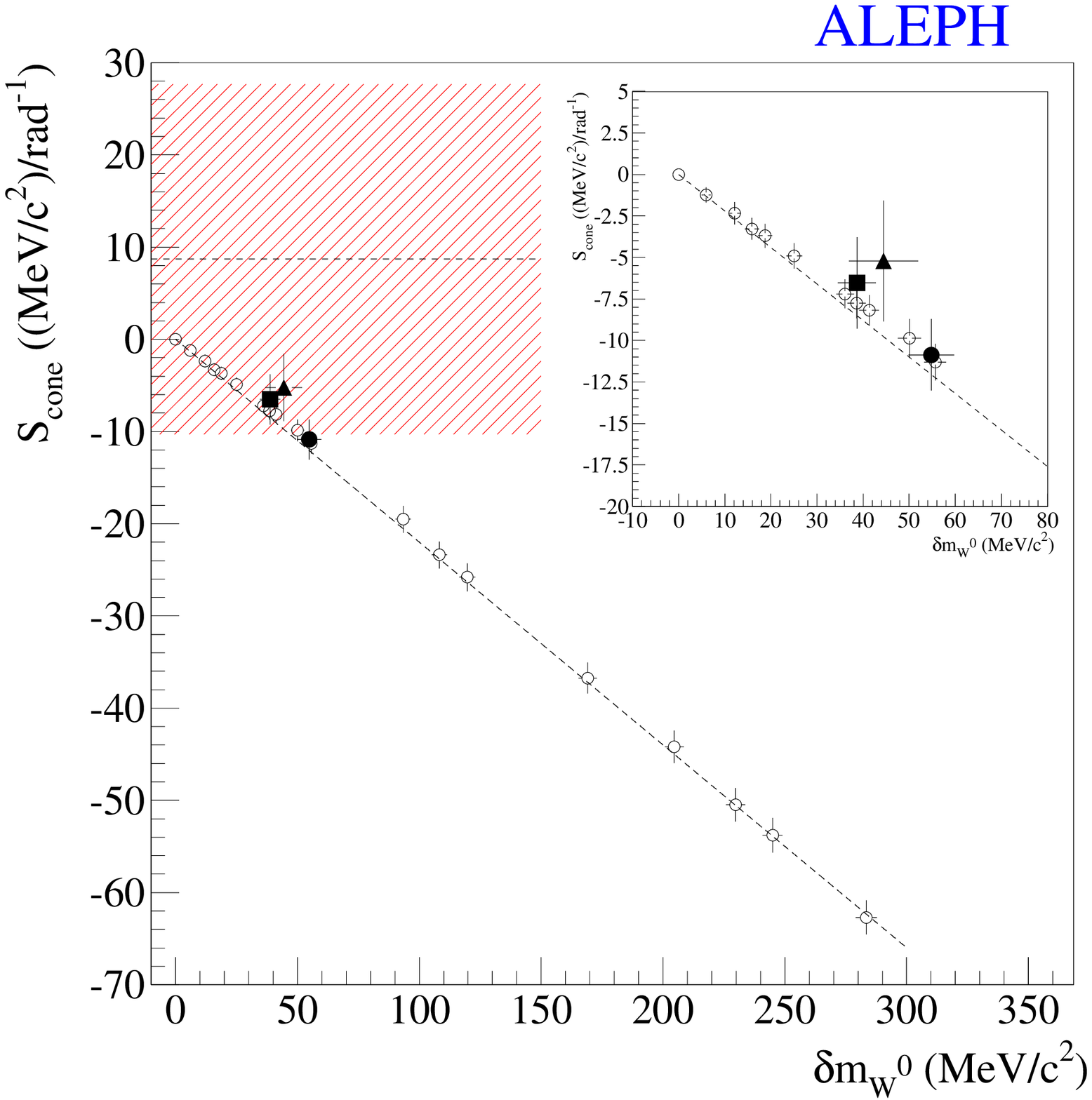,width=8.5cm}}
         }
    \caption{\footnotesize 
      Slope of the mass difference relative to the standard analysis
      as a function of $\delta\PMW^0$ for (a) the PCUT and (b) the CONE 
      reconstructions.
      The dashed line is a straight line fit to the SKI points represented by 
      white circles. The black symbols represent AR2 (circle), 
      HWCR (square) and GAL (triangle) predictions. 
      The horizontal bands represent the measured slopes with their $\pm 1\sigma$ 
      errors.             }
    \label{fig:spcutvsdm}
  \end{center}
  \vspace{-4.9cm}
  \begin{center}
  \hspace*{-1cm}
  \begin{picture}(400,10)
  \put(90,0){\large (a)}
  \put(310,0){\large (b)}
  \end{picture}
  \end{center}
  \vspace*{3.5cm}
\end{figure}

The SKI model predicts a clear correlation between the mass shift for the
standard reconstruction, $\delta\PMW^0$, and the slope of the mass difference
as a function of the PCUT or CONE cuts as shown in Fig.~\ref{fig:spcutvsdm}.
The clustering of the slope values from AR2, 
HWCR and GAL, around $-10 (\MeVcsq)/(\GeVc)$ for PCUT and similarly for CONE 
corresponds to the previously described values of $\delta\PMW$  
quantified in Table 4.

The covariance between the slopes, $S_{\rm{cone}}$ and $S_{\rm{pcut}}$,
is computed as well as the resolution on the slopes from a Gaussian
fit using the pseudo-data samples at each $k_{\rm i}$. The average correlation 
between the PCUT and CONE slopes is 51 \% with little dependence on $k_{\rm i}$. 
The RMS errors on the slopes are 18 (\MeVcsq)/(\GeVc) for the PCUT analysis and 19 
(\MeVcsq)/(\ rad$^{-1}$) for the CONE analysis, in agreement with the values 
obtained in the data sample.
A $\chi^2$ fit, defined as follows:
\[\sum_{\alpha\beta} (S_{\alpha}^{data} - 
S_{\alpha}^{MC}(x))\sigma_{\alpha\beta}^{-1}(S_{\beta}^{data} - 
S_{\beta}^{MC}(x))\]
where $\alpha,\beta$ signify PCUT and CONE respectively and 
$\sigma_{\alpha\beta}$ the covariance matrix, is used to extract the 68\% CL 
Gaussian upper limit on $x$. The parameter $x$ can be either $k_{\rm i}$ or 
$\delta \PMW$. 

\begin{figure}[htp]
  \begin{center}
    \mbox{
    {\epsfig{file=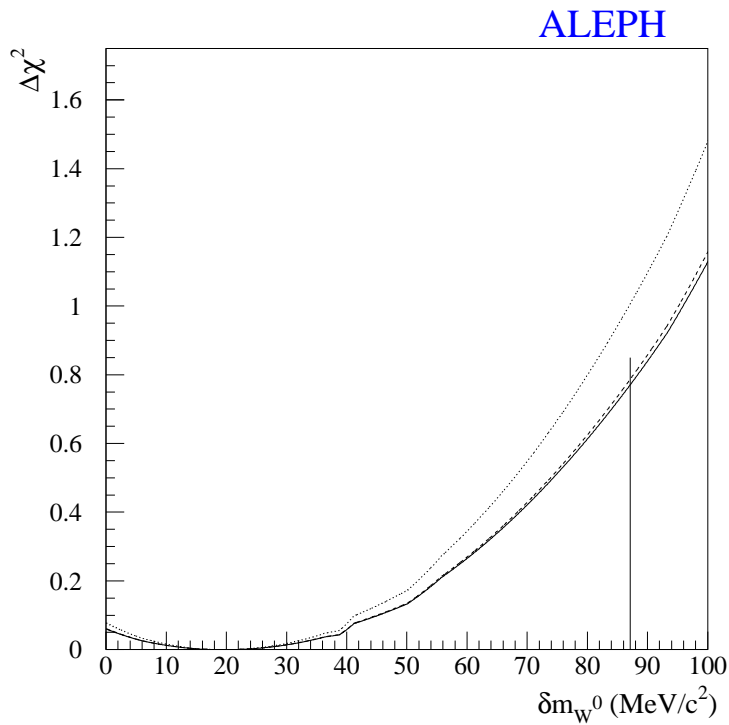,width=10.0cm}}
         }
    \caption{\footnotesize 
      $\Delta\chi^2$ as a function of the mass shift $\delta \PMW^0$ in the standard 
      analysis. The dotted line is with
      statistical errors only, the dashed line includes fragmentation
      errors, and the full line includes all systematic uncertainties. 
    }
    \label{fig:CRchisq}
  \end{center}
\end{figure}

The $\chi^2$ curve is shown in Fig.~\ref{fig:CRchisq} as a function of 
$\delta \PMW^0$.
The upper limit on $\delta \PMW^0$ has been set to +78 \MeVcsq, corresponding to the 
value at which the integral of the Gaussian likelihood from zero is 68\% of the 
full integral over the allowed (positive) range. 

The fragmentation uncertainty on this limit is estimated from the average
bias on the slopes between an {\tt ARIADNE} or {\tt HERWIG} sample and the 
{\tt JETSET} reference
sample, repeated over the pseudo-data samples. The slope biases
from {\tt ARIADNE}, being largest, are used as diagonal terms of a systematic
error matrix with 100\% correlation between the PCUT and CONE slopes.
This matrix is added to the statistical error matrix resulting in an increase in 
the mass limit to +86 \MeVcsq. An estimate of the uncertainty in evaluating the 
statistical error matrix yields a further small increase in the mass limit to  
+87 \MeVcsq (corresponding to $k_{\rm i}$ (68\% U.L.) = 1.88). 
Figure~\ref{fig:CRchisq} shows the progressive effect of adding these systematic 
uncertainties to the $\chi^2$ fits.


The same procedure is used to determine the mass shifts $\delta\PMW^{\rm PCUT}$ 
and $\delta\PMW^{\rm CONE}$. The upper limits on these mass shifts are given in 
Table~\ref{tab:mwshifts_UL} together with the limit from the standard analysis.

Both AR2 and HWCR agree well with the SKI prediction of the slopes as a function 
of  $\delta\PMW$. However, the actual limit from SKI is larger than the AR2, HWCR and 
GAL absolute predictions and therefore is taken as the systematic uncertainty due to 
colour reconnection for each reconstruction.

In practice, these limits on the mass bias depend linearly on the CM
energy within SKI. From 183 to 209 GeV, the limit varies from 45 to 105
\MeVcsq\ when no cut is applied, 12 to 36 \MeVcsq\ for the PCUT and 
18 to 45 \MeVcsq\ for the CONE respectively.

\begin{table}[htbp]
\begin{center}
\caption{\protect\footnotesize CR mass shifts (68 \% C.L. upper limits in \MeVcsq) 
for the three reconstructions: standard, PCUT and CONE using SKI derived from (a) 
the purely statistical analysis and (b) incorporating all systematic uncertainties.
\label{tab:mwshifts_UL} 
   } 
\begin{tabular}[h]{|l||c|c|} 
\hline 
         & (a)SKI(stat.)& (b)SKI(stat.+syst.) \\ 
\hline\hline
$\delta\PMW^0$          &   78  &   87 \\ 
$\delta\PMW^{\rm PCUT}$ &   25  &   27 \\
$\delta\PMW^{\rm CONE}$ &   32  &   35 \\
\hline
\end{tabular}
\end{center}
\end{table}

%% file: systematics.tex
\section{Systematic uncertainties}
\label{sec:systs}
Systematic uncertainties in the measurement of \PMW\ and \PGW\ arise from 
an incomplete description of the $\PWW$ production process, inaccuracies in 
the simulation of event reconstruction in the detector and the 
modelling of the \PW\ decays to di-jets. 
The following subsections describe all the systematic uncertainties 
evaluated for the standard analysis in each of the four event 
categories. They are also determined in the \qqbar\qqbar\ channel  
for the CONE (R=0.4 rad) and PCUT (=3 \GeVc) reconstructions where the 
potential effects of colour reconnection (CR) are minimised.

The LEP energy uncertainties with year-to-year 
correlations are taken from Ref.~\cite{lepewg2004}. 
All other uncertainties in the analysis are 
evaluated at 189 and 207 \GeV, simultaneously in \PMW\ and \PGW\ from the 
two-parameter fits.  When combining all the measurements, any variation over this 
energy range is taken into account using a linear interpolation for the 
intermediate CM energies.     
Table~\ref{tab:syst-4q-198} lists all the systematic uncertainties for the 
standard analysis as well as the optimal PCUT and CONE reconstructions in the 
\qqbar\qqbar\ channel. 
The CR uncertainty in this channel is taken into account at each CM energy.  
 Table~\ref{tab:syst-lvqq-198} 
lists all the systematic uncertainties in the 
standard analyses of the three semileptonic channels.    

\subsection{Detector simulation}
The systematic uncertainties in the detector simulation for the 
\qqbar\qqbar\ events are those arising from the quantitative comparison of the 
reconstructed jet four-momenta with the data as described in 
Sec.~\ref{sec:evrecons}. For the \evqq\ and \mvqq\ channels, the uncertainties in 
the lepton four-momenta are included and combined in quadrature with those from the 
jets.  
Subsidiary studies of particles within the jets have been made by comparing the 
simulation with data for the effect of photon energy miscalibration and charged 
hadron tracking discrepancies. These uncertainties are already taken into account 
in those quoted for the jets.
Each uncertainty is evaluated by first comparing the mean 
fitted parameters from special pseudo-data samples 
with corresponding normal samples each of the size of the 
data. The mean shifts found in \PMW\ and \PGW\ are then 
rescaled to correspond to the residual discrepancies found between data and 
simulation after any corrections have been applied 
(Sec.~\ref{sec:calibrations}). 

\subsubsection{Isolated lepton reconstruction in \boldmath\evqq\ and 
\boldmath\mvqq\ events}
Specific studies (Sec.~\ref{sec:emumom}) have been performed for electrons and muons.
In the \evqq\ channel, the uncertainty is determined from the error (0.04\%) in 
applying the global momentum correction of 0.45\% combined with the percentage error 
of 0.0008\% per \GeV\ in the evaluation of the momentum scale linearity, taking into 
account the correlation (+0.78). The small biases found as a function of polar angle 
have a negligible effect.  

For the \mvqq\ channel, the momentum uncertainty is derived from the full effect of 
the uncorrected global offset of 0.08\%. The percentage error of 0.0025\% per \GeV\ 
in the slope is added in quadrature, taking into account the correlation (-0.22).   
 
Averaged over polar angle, the lepton momentum resolutions in the simulation are 
degraded by 13\% and 8\% for the electrons and muons 
respectively to match the data when averaged over all momenta. 
For \PMW, the effect is relatively insignificant but on \PGW\ it is the 
dominating contribution to the total systematic uncertainty in each channel.

A possible bias in the measurement of the lepton direction
in the \evqq\ and \mvqq\ channels was studied 
by comparing the lepton track $\theta$ and $\phi$ angles as measured by the VDET
and the ITC + TPC separately~\cite{mass_189}.
No difference greater than a fraction of a mrad was observed.
Owing to small offsets in the drift time of the TPC, the z-component of
momentum can be biased for tracks away from $90^\circ$ to the beam axis. 
Conservatively, the effect on the lepton polar angle is parametrised maximally as
$2.0\times\sin 2\theta_{\rm {lepton}}$ mrad with respect to the beam axis. 
Events are generated accordingly, whilst keeping the lepton energy and the total 
momentum of the event conserved. The shift is 
negligible for both \PMW\ and \PGW. Any effect 
from possible lepton $\phi$ angle biases is also negligible.  

Comparing the VDET and ITC + TPC track measurements~\cite{mass_189}, the 
spread of the differences in polar angle measurement for the electrons and muons 
was found to be of order 0.5 mrad. No mean discrepancy greater than 0.3 mrad 
between the data and Monte Carlo distributions was observed.   
Conservatively, an  additional 0.5 mrad smearing has been applied to the simulation 
to compute the uncertainties attributable to the simulation of angular 
resolution. The shifts in \PMW\ are found to be negligible.

\subsubsection{Jet energy corrections before the kinematic fit}
\label{sec:jetcorrs}
As described in Sec.~\ref{sec:jetcorr}, the simulation of jet energies from di-jet 
events produced at the \PZz\ was compared with data in the range 30 to 70 \GeV. 
It was shown that (a) the bias in the relative global energy scale does not exceed 
0.5\% in the central region rising to 2.5\% at low angles and 
(b) the relative slope of the 
data to simulation in the jet energy scale as a function of $E_{\rm{jet}}$ is flat 
setting a limit of $\pm 0.8\times 10^{-4}$ per \GeV. Studies with 
special simulated samples show that the global bias has no significant effect on 
\PMW\ and \PGW\ for all channels. The systematic uncertainties assigned from the 
limit on the slope variation with $E_{\rm{jet}}$ is combined in quadrature with the 
shifts obtained from a mean 1\% global discrepancy between barrel and endcaps. 
Disregarding the presence of b jets in the \PZz\ samples, introduces a shift of 
0.25\% in the relative global energy scale. Since this is only half of the bias 
taken into account, the effect due to b jets is negligible.   
     
\subsubsection{Jet energy resolution}
As described in Sec.~\ref{sec:jetcorr}, the data and Monte Carlo resolutions 
in each $\cos\theta_\mathrm{jet}$ bin as determined from the RMS spread of jet 
energies agree to within $\pm$2\% for di-jet events at the \PZz. Special samples 
were made where the jet four-momenta and energies are smeared degrading the 
resolution by 10\% with respect to the nominal values computed from the kinematic fit 
parametrisations. 
The shifts found are rescaled to correspond with the measured difference.

\subsubsection{Jet angular bias}
\label{sec:jetangbias}
Possible discrepancies in the determination of $\theta_\mathrm{jet}$ were
studied~\cite{mass_189} by comparing, between data and simulation, the direction 
of the charged and neutral jet components in \PZz\ di-jet events. The 
tracking detectors and the ECAL were aligned independently but high statistics 
studies performed at 91.2 GeV show that their relative polar angle alignment is 
about 1 mrad. In order to measure any angular distortions, separately constructed 
charged and neutral components of jets 
are selected and their polar angle directions compared in bins 
of 5 degrees in $(\theta_{\rm {charged}}+\theta_{\rm {neutrals}})$/2. 
The simulation of the jet components is in good agreement with the data except in 
the overlap region between the barrel and endcap calorimeters where the difference 
is up to 2~mrad. The difference
$\Delta(\theta_{\rm {charged}}-\theta_{\rm {neutrals}})$, parametrised as    
0.7(0.2) - 2.4(0.6)$\cos\theta\sin\theta$~mrad (errors in brackets), gives the best 
fit to these discrepancies.  
Further studies confirm that the effect of the global offset of 
0.7~mrad is negligible since the uncertainties effectively cancel in 
$\cos\theta_{\rm{jet}}$ by symmetry. The resulting systematic uncertainties are 
evaluated by applying 
this parametrisation without the offset to special Monte Carlo \PWW\ event samples 
assuming that $\Delta\theta_{\rm{jet}} = 
(\theta_{\rm {charged}}-\theta_{\rm {neutrals}})$/2. 
 
\subsubsection{Jet angular resolution}

Selected di-jet events from the \PZz\ calibration runs have been used to measure
the jet angular resolution for 45 \GeV\ jets from the distribution of the opening 
angles between the two jets. The PCUT and CONE criteria are also applied to the jets 
to measure the variation in jet angle resolution for these reconstructions. Special 
event samples with modified resolutions, which match the data, are used to estimate 
the effect on \PMW\ and \PGW\ for all channels. The resulting uncertainties in \PMW\ 
are very small. 

\subsubsection{Jet boosts}
\label{jetboost}
\begin{figure*}[bth!]
\begin{flushleft}
\hspace*{-0.5cm}
\begin{tabular}[t]{@{}ll@{}}
\mbox{\epsfig{file=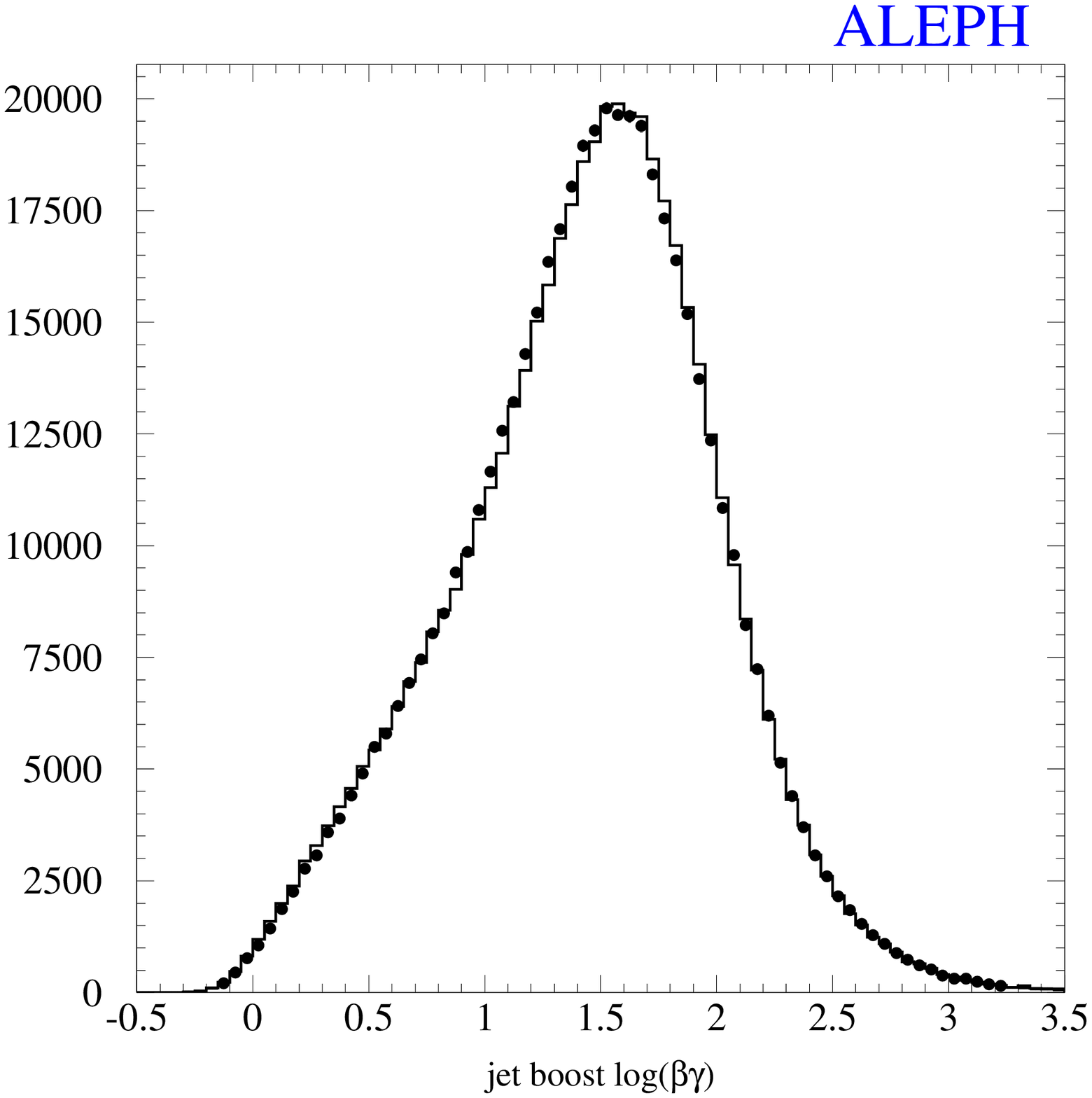,height=8cm}}
&
\mbox{\hspace*{-0.0cm}
      \epsfig{file=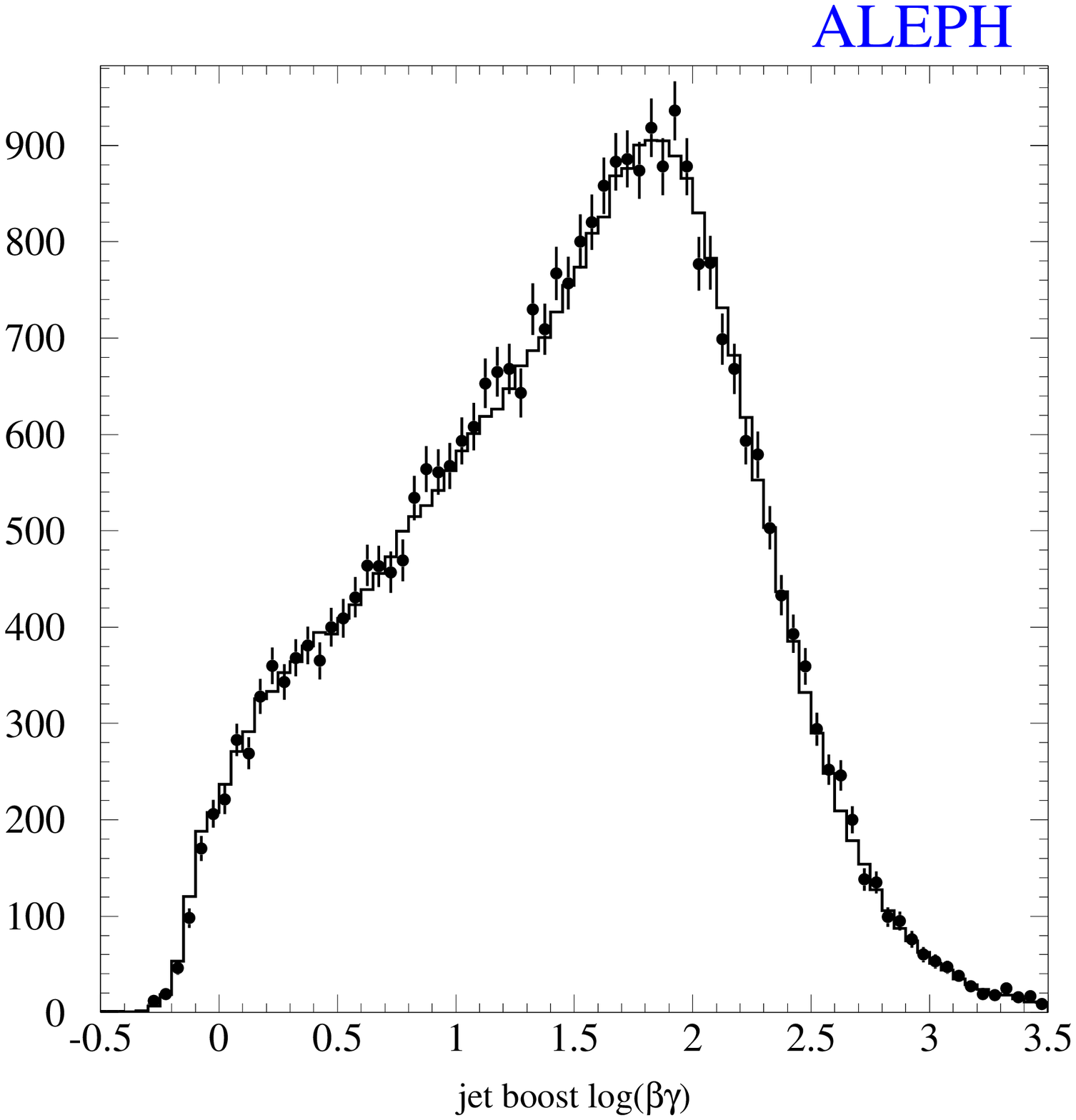,height=8cm}}
\end{tabular}
\end{flushleft}
\vspace{-3.5cm}
\begin{center}
\hspace*{-1cm}
\begin{picture}(400,10)
\put(030,5){\large (a)}
\put(300,5){\large (b)}
\end{picture}
\end{center}
\vspace{1.6cm}
\begin{flushleft}
\hspace*{-0.5cm}
\begin{tabular}[t]{@{}ll@{}}
\mbox{\epsfig{file=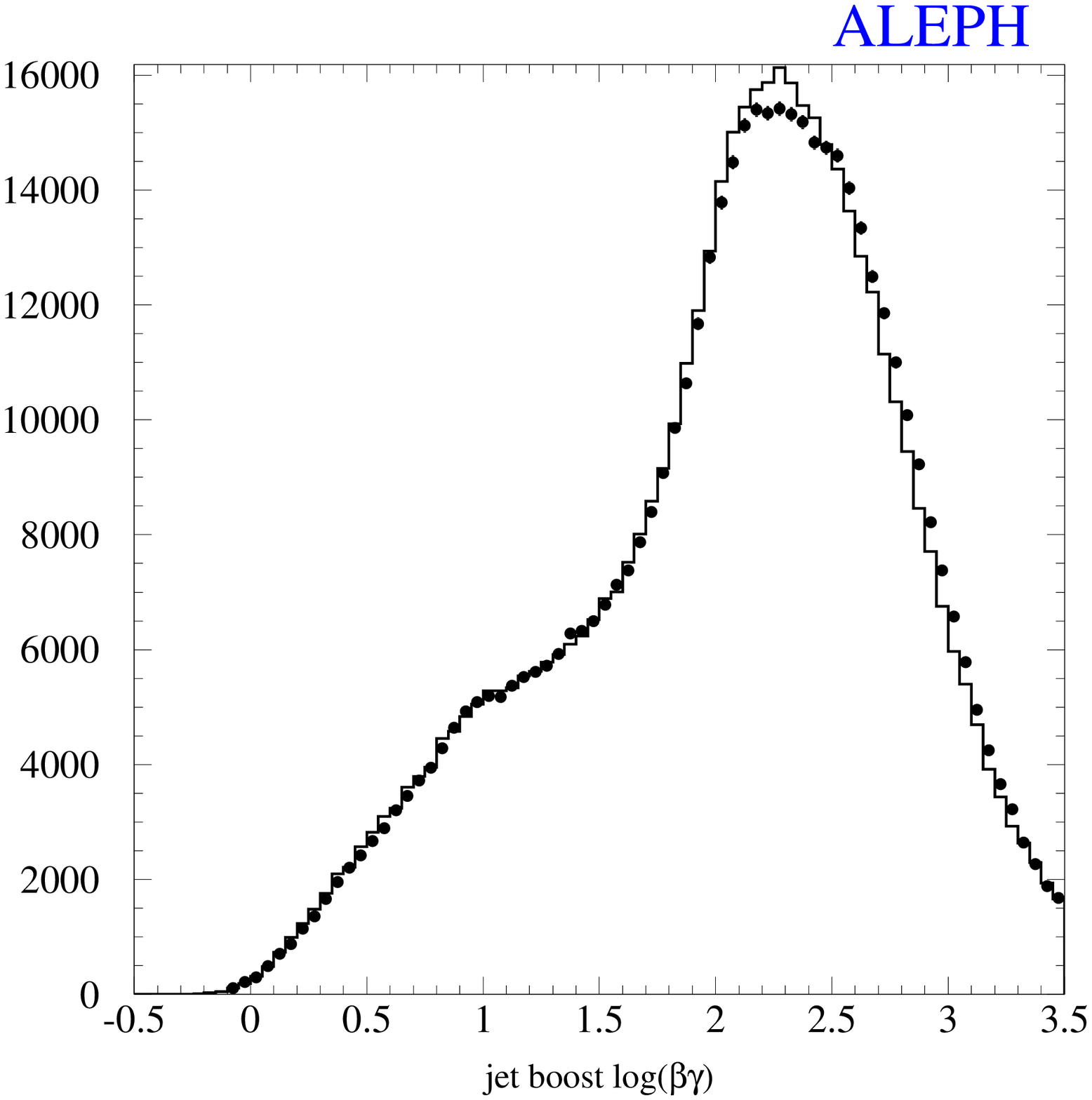,height=8cm}}
&
\mbox{\hspace*{-0.0cm}
      \epsfig{file=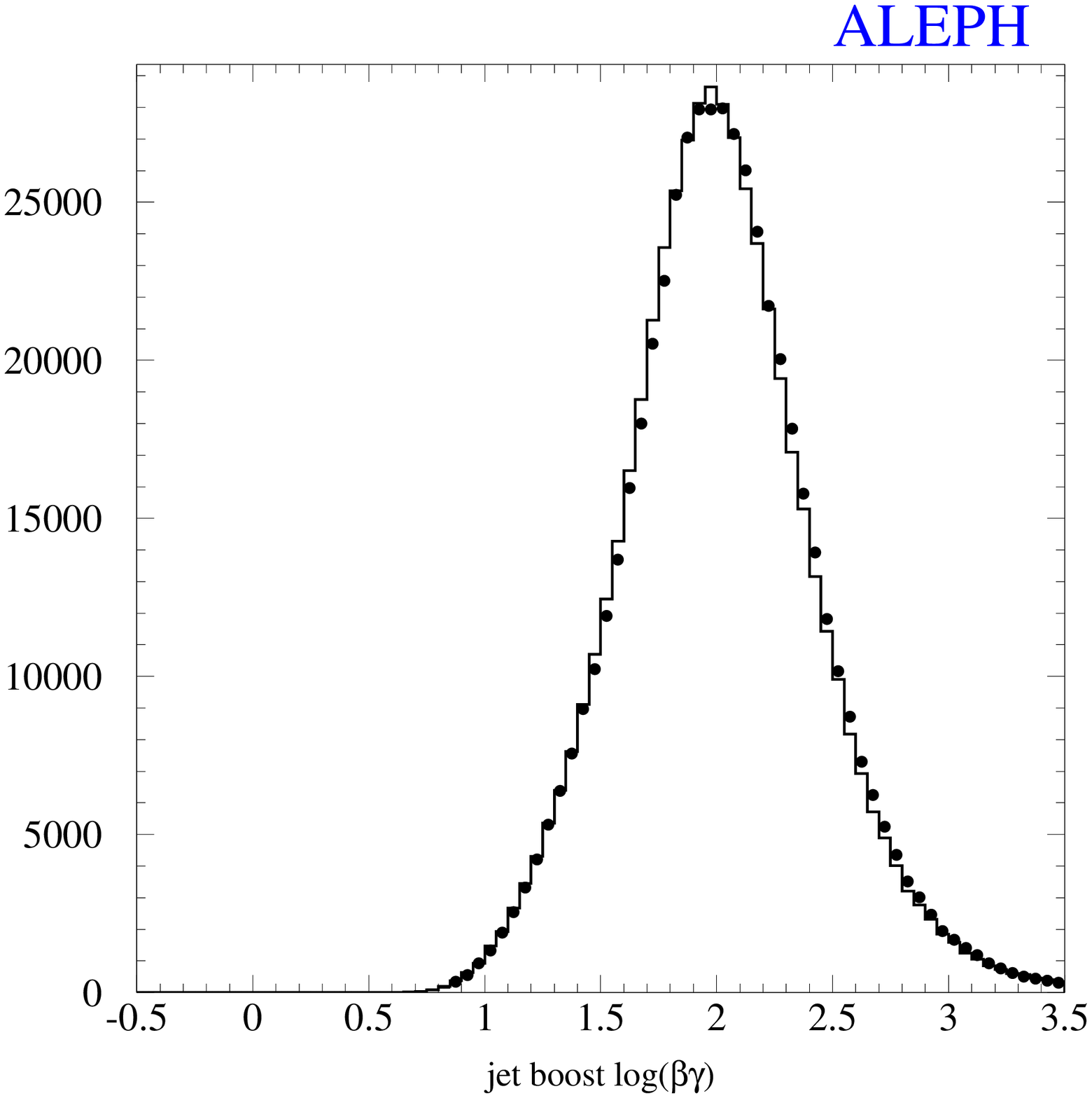,height=8cm}}
\end{tabular}
\end{flushleft}
\vspace{-3.5cm}
\begin{center}
\hspace*{-1cm}
\begin{picture}(400,10)
\put(030,5){\large (c)}
\put(300,5){\large (d)}
\end{picture}
\end{center}
\vspace{1.3cm}
\caption[]
{\protect\footnotesize
Distributions of jet boosts ($\log\beta_{\rm{jet}}\gamma_{\rm{jet}}$) for data 
(circles) and MC (histogram): (a) from $\PZz\ra\qqbar$ events (1998-2000) and (b) from 
high energy di-jet events (183-209 \GeV) using the Durham jet reconstruction in the 
standard analysis, (c) and (d) from $\PZz\ra\qqbar$ events using PCUT and CONE 
reconstructions respectively.  
}
\label{fig:jbt_dijets}
\end{figure*}

The accuracy of the Monte Carlo reconstructed jet masses in each 
channel depends sensitively on the simulation of the charged and neutral particle 
momenta and multiplicity distributions within the jets. Jet boosts, 
$\beta_{\rm {jet}}\gamma_{\rm {jet}}$, are chosen rather than masses to compare data 
with simulation since any momentum discrepancies are factored out and double 
counting minimised. 
Figure~\ref{fig:jbt_dijets} compares the 
data and Monte Carlo distributions of $\log (\beta_{\rm {jet}}\gamma_{\rm {jet}}$) 
for jets built as in the standard analysis, integrated over all polar angles 
from (a) high statistics 
hadronic \PZz\ decays where the average jet momenta are close to those in \PW\ 
decays, (b) higher energy di-jets, (c) hadronic \PZz\ decays for PCUT and (d) 
hadronic \PZz\ decays for CONE. 
These jet samples are 
studied rather than those from the selected \PW\ pairs to avoid the possible 
influence of final state interactions and to benefit from high statistics.
The study includes b-depleted samples and jets from radiative returns to the \PZz\ 
peak (Sec.~\ref{sec:zg}).   
Table~\ref{tab:systbg} gives the largest shifts obtained between data and simulation 
expressed as 
$\Delta\log(\beta_{\rm{jet}}\gamma_{\rm{jet}})$. The small differences between 
central and forward regions of the detector are not statistically significant. 
\begin{table}[htbp]
\begin{center}
\caption{\protect\footnotesize Largest (data-MC) shifts, 
$\Delta\log (\beta_{\rm{jet}}\gamma_{\rm{jet}})$ in 
percent.
The shifts are tabulated for the central region of the detector 
($|\cos\theta_{\rm{jet}}|<0.7$), the forward region ($|\cos\theta_{\rm{jet}}|>0.7$) 
and both combined (errors are shown in brackets). }
\label{tab:systbg}
\begin{tabular}[h]{|l||c|c|c|} 
\hline
Reconstruction  & central & forward & combined \\ \hline\hline
standard  & 0.9 (0.2) & 0.7 (0.2) &  0.8 (0.1)\\
PCUT      & 2.7 (0.2) & 1.9 (0.3) &  2.4 (0.2) \\  
CONE      & 2.1 (0.4) & 2.1 (0.5) &  2.1 (0.3) \\  
\hline
\end{tabular}
\end{center}
\end{table} 
The systematic uncertainties in \PMW\ and \PGW\ are derived using special Monte 
Carlo samples rescaled to match the biases in this table.

\subsection{Fragmentation of the {\boldmath $\PW\ra\qqbar$} decays to hadrons}
\label{sec:mfrag}
In the previous analysis at 189 \GeV~\cite{mass_189}, the uncertainty due to the 
modelling was 
determined mainly from the comparison of \PMW\ and \PGW\ values using event samples 
in which fragmentation is simulated with 
{\tt HERWIG}~\cite{HERWIG} or {\tt ARIADNE}~\cite{ARIADNE} in place 
of {\tt JETSET}. A large uncertainty of $\sim$35 \MeVcsq, fully correlated between 
channels was assigned. It has been found that the 
variation in baryon content between the models is largely responsible. 
The baryon multiplicities predicted by {\tt JETSET} and {\tt ARIADNE} agree with data 
at the \PZz~\cite{HWbaryons} whereas {\tt HERWIG} generates ($\sim 0.5$) fewer 
baryons per event. 

The uncertainties in \PMW\ for each channel in the standard analysis are 
reassessed after correcting for this effect. 
In the \qqbar\qqbar\ channel, the bias in \PMW\ is found to depend linearly 
on the number of protons and neutrons per event. Taking samples with 
0, 2, 4, 6 and 8 nucleons per event, the slope of the bias for all three models is 
statistically equivalent and found to be 20.1$\pm$0.8 \MeVcsq\ per nucleon. A similar 
linear behaviour is seen in the \evqq, \mvqq, and \tvqq\ channels. 
The \PW\ mass differences between the models due 
to the variation in their baryon content is evaluated from the linear 
dependences in each channel assuming that they apply over the entire range of baryon
multiplicities. For {\tt HERWIG}-{\tt JETSET} and {\tt ARIADNE}-{\tt JETSET} , the 
mass shifts before and after 
correcting for the differences in baryon content are given in 
Table~\ref{tab:JTHWfrag}. 
\begin{table}[htbp]
\begin{center}
\caption{\protect\footnotesize For the standard analysis in the 183-209 \GeV\ range, 
the mean \PW\ mass differences between MC samples of {\tt HERWIG} and 
{\tt ARIADNE} relative to {\tt JETSET} are tabulated for each channel before and 
after correcting for the difference in baryon content.
\label{tab:JTHWfrag}    }
  \begin{tabular}[h]{|l||c|c|c|c|}
\hline
  & \multicolumn{2}{c|}{HW-JT}        
  & \multicolumn{2}{c|}{AR-JT}  \\        
                & Uncorrected & Corrected & Uncorrected & Corrected \\ 
\hline\hline
\qqbar\qqbar\   & +12$\pm$8 & -7$\pm$8 & +3$\pm$9 & +5$\pm$9  \\
\evqq\          & +25$\pm$8 & +3$\pm$8 & +1$\pm$9 & +6$\pm$9 \\
\mvqq\          & +10$\pm$8 & -8$\pm$7 & -11$\pm$8 & -7$\pm$8 \\ 
\tvqq\          & +40$\pm$11 & +15$\pm$11 & +5$\pm$13 & +6$\pm$12 \\ 
\hline
\end{tabular}
\end{center}
\end{table}
After correction, the differences between {\tt HERWIG} and {\tt JETSET} become 
insignificant. All three fragmentation models agree within statistical error 
for all channels. The systematic uncertainty is set to 10 \MeVcsq\ for the 
standard analysis, coherent in all channels. The variation in baryon content between 
the models has no significant effect on the values fitted for \PGW. 

For the PCUT and CONE reconstructions in the \qqbar\qqbar\ channel, the 
uncertainties are determined from comparing the same event samples simulated with 
{\tt ARIADNE} and {\tt JETSET} where the variation in baryon content is not 
significant. Any residual differences (AR-JT) are due to other effects unrelated to 
baryon multiplicities and are taken to represent the uncertainties for these 
reconstructions. The differences are larger than the standard analysis but comparable 
for both reconstructions.  

\subsection{Radiative corrections}
The uncertainties in the theoretical treatment of QED initial state 
radiation (ISR) and Coulomb corrections  in {\tt KORALW} as well as 
Next-to-Leading ${\cal{O}}(\alpha)$ 
corrections in {\tt YFSWW3} are determined for each channel and reconstruction by 
comparing Monte Carlo samples with appropriate event weighting. The estimated shifts 
from each study are combined in quadrature.
 
\subsubsection{Missing ISR corrections}
Initial state radiation is simulated in {\tt KORALW} up to 
${\cal{O}}(\alpha^3 L^3)$, i.e. up to third order in the leading-log approximation. 
  The effect of missing higher order ISR terms beyond ${\cal{O}}(\alpha^3 L^3)$ 
  on the measurement of \PMW\ and \PGW\ is estimated by measuring the respective 
  shifts when this QED computation is downgraded to 
  ${\cal{O}}(\alpha^2 L^2)$ as originally suggested in Ref.~\cite{zalewski}.
  Each event in a specially generated {\tt KORALW} 
  sample is weighted according to the calculated ratio of second to third order 
  squared matrix elements: ${\cal{O}}(\alpha^2 L^2)/{\cal{O}}(\alpha^3 L^3)$. 
  Treated as data, fits are made to the weighted events selected in each channel and 
  compared with those from the corresponding unweighted 
  events to evaluate the shifts. The shifts in \PMW\ and \PGW\ are less than 
  1 \MeV\ in all channels. 

\subsubsection{Coulomb corrections}
The unweighted events from {\tt KORALW} include non-factorizable QED 
corrections~\cite{ck_nf} which effectively ``screen'' the Coulomb 
interaction~\cite{mwshiftQED} 
between the two \PW's. 
It is suggested~\cite{Wiesiek} that the difference 
between this ``screened'' Coulomb correction and no Coulomb correction can be 
used to assess an uncertainty. The differences are found to be less than 3 \MeV\ in 
all channels for both \PMW\ and \PGW.

\subsubsection{Next-to-Leading ${\cal{O}}(\alpha)$ corrections}
These corrections are large, ranging for \PMW\ from $\sim$10 \MeVcsq\ in the 
\qqbar\qqbar\ channel to $\sim$20 \MeVcsq\ in the \evqq\ channel. Studies have 
shown~\cite{Cossutti} that the theoretical implementation of these corrections in 
{\tt RacoonWW}~\cite{RacoonWW} are in good agreement with {\tt YFSWW3}. The 
following two possible contributions to the uncertainties in these corrections 
using {\tt YFSWW3} are considered.

(i) The main effect of the NL ${\cal{O}}(\alpha)$ corrections is to 
modify the \PW\ final state radiation (WSR) pattern of photons. In {\tt YFSWW3}, the 
infra-red (IR) contributions to WSR and WSR-ISR interference are exponentiated to 
infinite order including non-IR Next-to-Leading contributions. The uncertainty in 
this calculation is estimated as suggested in Ref.~\cite{Wiesiek} by removing the 
additional non-IR contributions. The effect is found to be 
small, less than 2 \MeVcsq\ in \PMW\ for all channels and reconstructions. The 
shifts in \PGW\ are similar. 
  
(ii) In calculating the weight per event from the {\tt YFSWW3} program, the 
recommended recipe by the 
authors~\cite{YFSWW} is an additive correction in which the Double-Pole 
approximation (DPA) for doubly resonant \PW's is applied only to the \CCC\ part of 
the event weight. An alternative recipe would be to apply the NL correction also to 
the difference between the 4f and \CCC\ contributions - the so-called 
multiplicative NL correction. The additive correction is chosen as 
the default. The systematic uncertainties in \PMW\ and 
\PGW\ are estimated by measuring the difference between the 
additive and multiplicative implementations. These differences are less than 1.5 
\MeVcsq\ in all channels.  
 
\subsection{Calibration curves}
\label{sec:CCCME}
As stated in Sec.~\ref{sec:extraction}, the reweighting procedure was tested by 
comparing the fitted with the input \PW\ masses and widths in 
each channel individually. No deviations were observed in the fitted slopes or 
intercepts of the produced calibration curves except in the \evqq\ channel.
Combining statistically the fitted masses at five points between 79.85 and 
80.85 \GeVcsq\ from 189 and 207 \GeV\ pseudo-data in this channel, the 
calibration curve for \PMW\ is found to be linear but with a slope of 
0.954$\pm$0.023. At the 
measured mass, this deviation from unity corresponds to an uncertainty of 
10 \MeVcsq\ in \PMW. A similar analysis for \PGW\ 
found no significant effect in the \evqq\ channel. An upper limit of 20 \MeV\ is 
assigned as the systematic uncertainty from the statistical precision of the test.


\subsection{Background contamination}
The expected numbers of events in each channel included in the reweighting fits from 
non-\PWW\ background processes are shown in Table~\ref{tab:selec-xsec}.  

The dominant background in the \qqbar\qqbar\ channel is $\qqbar(\gamma)$ (14\% of 
all selected events) followed by  $\PZz\PZz$ (2\%). The normalisations of the 
these contributions are varied conservatively by 5\% and 10\% respectively and the 
consequent shifts added in quadrature. In addition, the uncertainty in the 
fragmentation modelling of the $\qqbar(\gamma)$ events is estimated by 
replacing {\tt JETSET} with {\tt ARIADNE}; its impact is significant only for 
\PMW.  The effect of any \qqbar\ hadronisation uncertainty in the $\PZz\PZz$ 
contribution is very small and has been ignored.
 
In all \lvqq\ channels the contamination is relatively small but also dominated 
by events from the $\qqbar(\gamma)$ and $\PZz\PZz$ processes. Their rates are also 
varied by 5\% and 10\% respectively to produce the quoted uncertainties in 
\PMW\ and \PGW. Any effect from hadronisation is found to be insignificant.  

For all channels, the 
$\PZz\mathrm{ee}$ contributions are flat in the defined mass windows and their 
effects on \PMW\ and \PGW\ are negligible. 
 
\subsection{Final State Interactions in the 4q channel}
\subsubsection{Colour reconnection}
\label{systs:CR}
The studies on the mass shift coming from possible colour reconnection 
between decay products of the W pairs have been discussed in 
section~\ref{sec:crlimit}.

\begin{figure}[htp]
  \begin{center}
    \mbox{
    {\epsfig{file=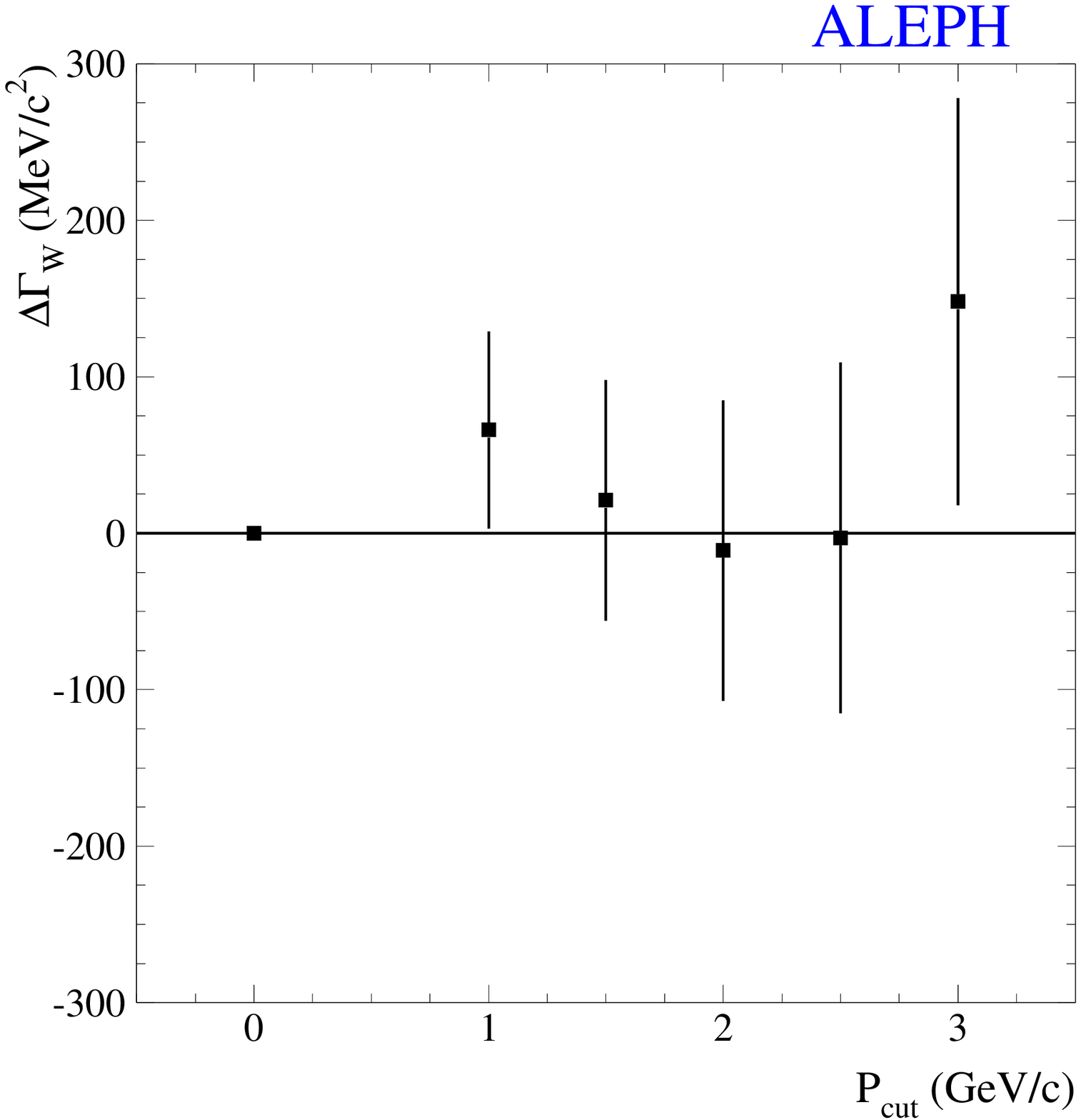,width=8.0cm}}
    {\epsfig{file=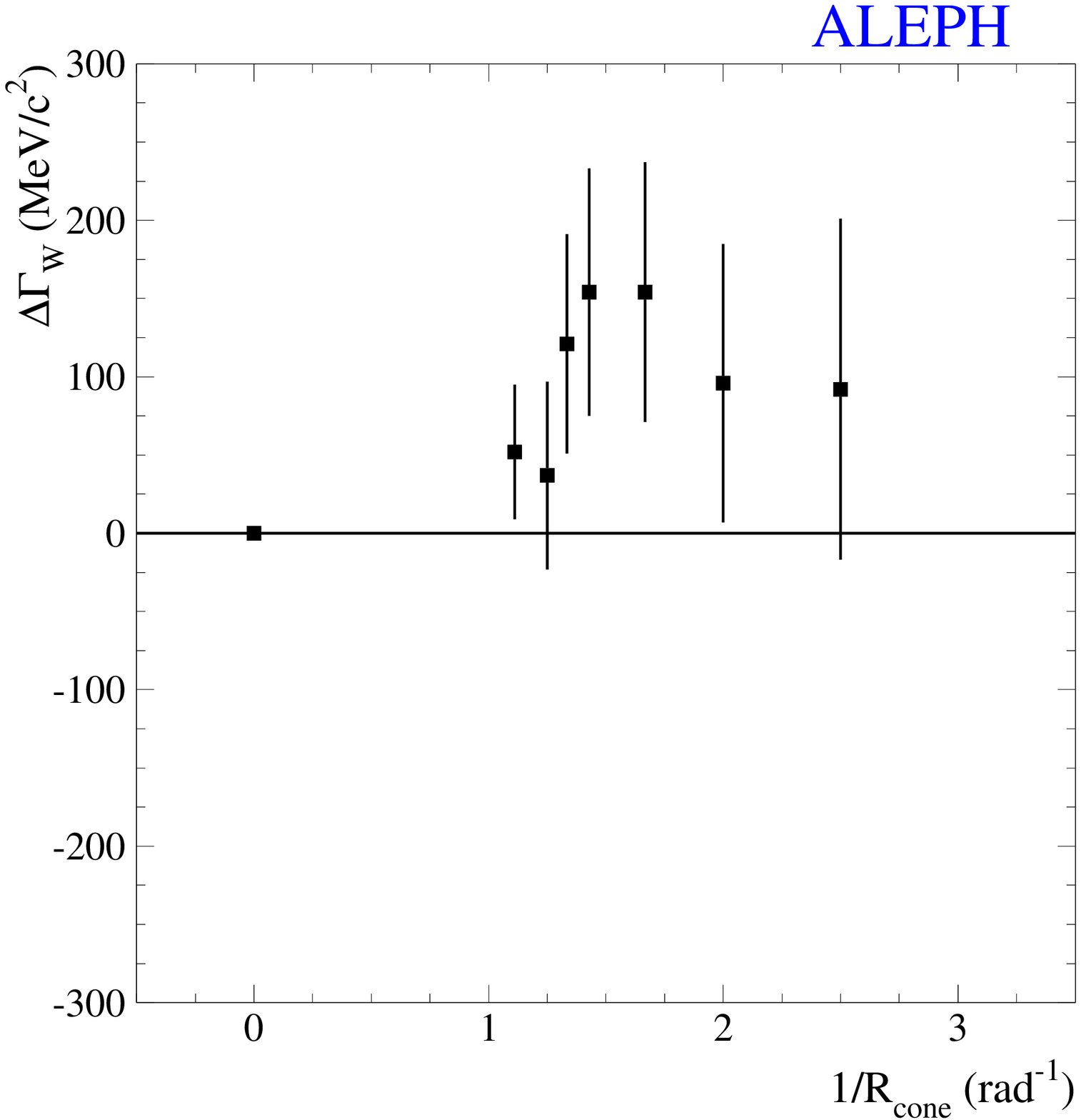,width=8.0cm}}
         }
    \caption{\protect\footnotesize 
      \PW\ width differences $\Delta$ \PGW\ versus (a) PCUT  and (b) inverse CONE
      radius for data from the \qqbar\qqbar\ channel. The errors take into account the
      correlation at each point with the standard analysis.    
             }
    \label{fig:width_CR}
\vspace{-3.5cm}
\begin{center}
\hspace*{-1cm}
\begin{picture}(400,10)
\put(060,5){\large (a)}
\put(300,5){\large (b)}
\end{picture}
\end{center}
\vspace{1.3cm}
  \end{center}
\end{figure}

The mass differences obtained when comparing PCUT or CONE analyses
with the standard analysis give no indication of an effect
within our data statistics; nor do the differences in \PGW\ 
(Fig.~\ref{fig:width_CR}). The upper limit
derived with SKI exceeds the predictions of the HWCR, AR2 and GAL models. The 
predictions of AR2 and GAL on particle distributions in three jet events at 
the \PZz\ have been studied in Ref.\cite{cr_ygap_lep}. They are disfavoured by the 
data. Disregarding the SKII and SKII$^\prime$ models which predict a very small 
effect, the 68\% upper limits on $\delta \PMW$ and $\delta \PGW$ obtained with
the SK1 model are taken as conservative estimates for the systematic uncertainties 
from CR. The averaged values of $\delta\PMW$ in Table 5 are determined using 
only the 
statistical errors in \PMW\ at each CM energy. In the final optimising 
combination 
procedure, the systematic uncertainties for CR quoted in Table 8 are computed
taking into account both the statistical and systematic uncertainties from all
sources. For the standard analysis where the CR uncertainty dominates and 
varies significantly with CM energy, the value quoted in Table 8 is consequently reduced 
from 87 to 79 \MeVcsq.

\subsubsection{Bose Einstein correlations}
The presence of Bose-Einstein correlations between the decay products 
of the two \PW's in the $\mathrm{WW \rightarrow q\bar{q}q\bar{q}\,}$ 
selected events could influence the \PW\ mass
measurement \cite{LoSj,zalewski}. 
When simulated events are modified according to the {\tt JETSET-LUBOEI} model 
\cite{Lonnblad} of Bose-Einstein correlations between the \PW's, tuned on hadronic 
\PZz\ decay data, a shift on \PMW\ of $-32\pm$5 \MeVcsq\ is predicted in the 
standard analysis. This shift is reduced in the optimal CONE or PCUT analysis 
by a factor of two. 
The ALEPH dedicated analysis of Bose-Einstein correlations based on
the comparison of like-sign and unlike-sign pion pairs using the
so-called ``mixed'' method, is described 
in Refs.~\cite{ALEPHstandard,ALEPHmixed}.
The data are in agreement with the hypothesis where
Bose-Einstein correlations are present only for pions coming from the same \PW.
The {\tt JETSET-LUBOEI} model with Bose-Einstein correlations applied also on
pions from different \PW\ bosons is disfavoured by up to 4.7$\sigma$ using the 
different variables studied. 
The systematic uncertainty on \PMW\ is determined from the fraction of the full 
prediction of this model which is consistent with these experimental results, 
using the 
value predicted with and without Bose-Einstein correlations between 
pions from different \PW's. This fraction 
is $-5$\%$\pm22$\%, giving an uncertainty on \PMW\
of 6 \MeVcsq, if a linear dependence between the \PMW\ 
shift and the value of this fraction is assumed.

\subsection{LEP energy}
The LEP beam energies were recorded every 15 minutes, or more frequently if 
required by the machine conditions. 
The instantaneous values recorded nearest in time to the selected events are used 
in the analysis. For the year 2000, as the CM energy was continuously increased, the 
dataset is split into two samples, the first integrating data at energies from 
202.5 \GeV\ to 205.5 \GeV\ centred at 204.86 \GeV\ and the second including all data 
above 205.5 \GeV\ centred at 206.53 \GeV. The effect on \PMW\ of any discrepancy 
between the data and generated reference beam energies was investigated 
and found to range from 8 \MeVcsq\ per \GeV\ difference at 189 \GeV\ to 
16 \MeVcsq\ per \GeV\ at 207 \GeV. The resulting uncertainties at each CM energy are 
small compared with the LEP energy uncertainties and have been ignored.     

The year-on-year correlated uncertainties in the LEP beam energy $E_{\rm LEP}$ taken 
from Ref.~\cite{lepewg2004}, are used to determine the quoted systematic 
uncertainties in \PMW\ and \PGW. 
The relative uncertainty on \PMW\ for the \tvqq\ and 
\qqbar\qqbar\ channels is obtained directly from the relative error in $E_{\rm LEP}$ 
whereas for the \evqq\ and \mvqq\ channels, the relative uncertainty is 
0.9$\times\Delta E_{\rm LEP}/E_{\rm LEP}$.
The effect of smearing in the event-by-event collision energy~\cite{lepewg2004}, 
which also introduces a longitudinal boost in the CM frame 
of ALEPH, are both taken into account in the evaluation of the uncertainty in \PGW. 

\begin{table}[htbp]
\caption{\protect\footnotesize Summary of the systematic errors on \PMW\ and \PGW\ 
averaged over 183-209 \GeV\ in the \qqbar\qqbar\ channel for the standard, 
PCUT (= 3.0 \GeVc) and CONE (R=0.4) reconstructions. }
\begin{center}
\begin{tabular}{|l|c|c|c|c|c|c|}\hline
  & \multicolumn{3}{c|}{$\Delta\PMW$ (\MeVcsq)}        
  & \multicolumn{3}{c|}{$\Delta\PGW$ (\MeV)}       \\
\hline
 Source   & standard & PCUT & CONE 
          & standard & PCUT & CONE \\
\hline
\hline
 Jet energy scale/linearity  &   2  &   2  &  3  & 2    & 12  & 4 \\
 Jet energy resoln      &   0  &   1  &  0     & 7      & 9  & 10 \\
 Jet angle              &   6  &   6  &  6     & 1      & 3 & 3 \\
 Jet angle resoln       &   1  &   3  &  2     & 15     & 18  & 9 \\
 Jet boost              &  14  &  15  & 11     & 5      & 5  & 4 \\
 Fragmentation          &  10  &  20  & 20     & 20     & 40 & 40  \\
 Radiative Corrections  &   2  &   2  &  2     & 5      & 7  & 7 \\
 LEP energy             &  9   &  10  & 10     &  7  & 7  &  7 \\
 Ref MC Statistics      &  2   &  3   &  3     &  5     & 7  & 7   \\
 Bkgnd contamination    &  8   &  5   &  5     & 29     & 31 & 32 \\
 Colour reconnection    & 79   &  28  &  36    & 104    & 24 & 45  \\
 Bose-Einstein effects  &  6   &  2   &  3     & 20     & 10 & 10 \\
\hline
\end{tabular}
\end{center}
\label{tab:syst-4q-198}
\end{table}

\begin{table}[htbp]
\begin{center}
\caption{\protect\footnotesize Summary of the systematic errors on \PMW\ 
and \PGW\ in the standard analysis averaged over 183-209 \GeV\ for all 
semileptonic channels. The column labelled \lvqq\ lists the uncertainties in \PMW\ 
used in combining the semileptonic channels.
\label{tab:syst-lvqq-198}
   }
\begin{tabular}{|l|c|c|c|c|c|c|c|c|}\hline
  & \multicolumn{4}{c|}{$\Delta\PMW$ (\MeVcsq)} 
  & \multicolumn{4}{c|}{$\Delta\PGW$ (\MeV)}       \\
\hline
 Source   & \evqq\ & \mvqq\ & \tvqq\ & \lvqq\
          & \evqq\ & \mvqq\ & \tvqq\ & \lvqq\  \\
\hline
\hline
 e+$\mu$ momentum         &  3  &  8  &   -  & 4 &  5  &  4 & - & 4  \\
 e+$\mu$ momentum resoln
                          &   7  &  4   &   -  & 4 &  65  & 55 & - & 50 \\
 Jet energy scale/linearity
                          &   5  &   5  &  9  & 6  &  4  &  4  & 16 & 6  \\
 Jet energy resoln    
                          &  4   &  2   &  8  & 4 & 20  & 18 & 36 & 22  \\
 Jet angle                &  5  & 5  & 4  & 5 & 2 & 2 & 3 & 2  \\
 Jet angle resoln
                          &  3  &  2  &  3  & 3 &  6 &  7  & 8 & 7 \\
 Jet boost                &  17  &  17  &  20  & 17 & 3  & 3  & 3 & 3 \\
 Fragmentation 
                          & 10 & 10 & 15 & 11 & 22 & 23 & 37 & 25  \\
 Radiative corrections    & 3   &  2   &  3    & 3 &  3  &  2  & 2 & 2 \\
 LEP energy               &  9   &  9  & 10  & 9 &  7  & 7  &  10 & 8  \\
Calibration (\evqq\ only) 
                          & 10 & -  & -  & 4  & 20  &  - & - & 9 \\
Ref MC Statistics
                          & 3  &   3  &  5  & 2 &  7  &  7 & 10  & 5 \\
 Bkgnd contamination      &  3  &   1  &  6  & 2 &  5  &  4  & 19 & 7  \\
\hline
\end{tabular}
\end{center}
\end{table}

%% file: rad_returns.tex
\newpage
\section{Radiative returns to the \PZz\ peak}
\label{sec:zg}
Radiative events $\epem\ra\mathrm{f\bar{f}}\gamma$ where the invariant mass of 
the $\mathrm{f\bar{f}}$ system is in the vicinity of the \PZz\ mass are selected 
over the full 
CM energy range $\sqrt{s}$ = 183-209 \GeV. The hadronic final states 
producing two jets are analysed using the same jet reconstruction methods 
as applied to the $\PW\ra\qqbar$ decays, providing a cross check of the \PW\ mass 
reconstruction. Furthermore, the analysis of the \mm$\gamma$ 
channel provides a direct measurement of the LEP energy~\cite{alephzg} reaching an 
interesting precision when combined with the other LEP experiments~\cite{refzg}.  

\begin{figure}[htp]
  \begin{center}
    \mbox{
    {\epsfig{file=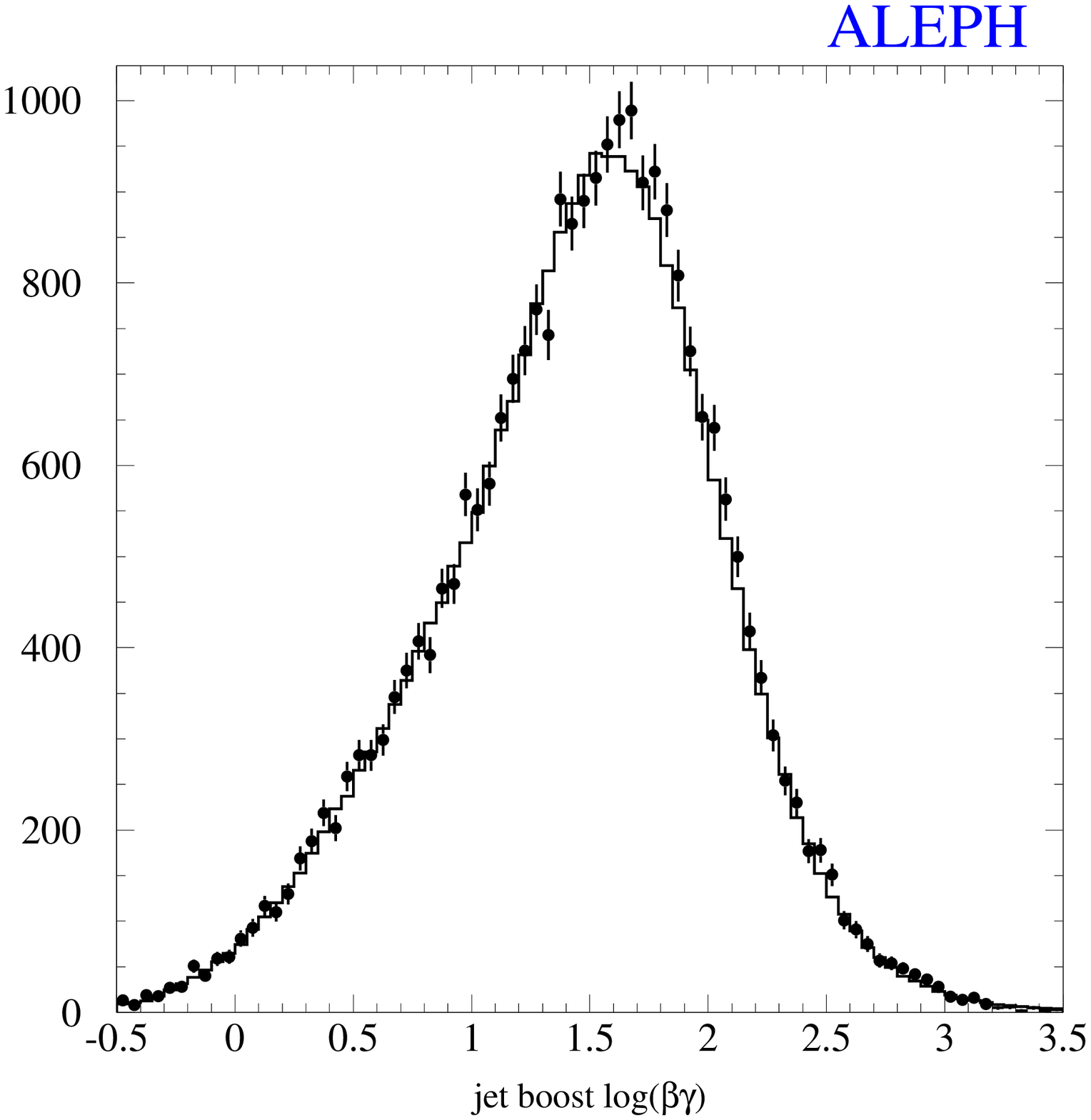,width=8.0cm}}
    {\epsfig{file=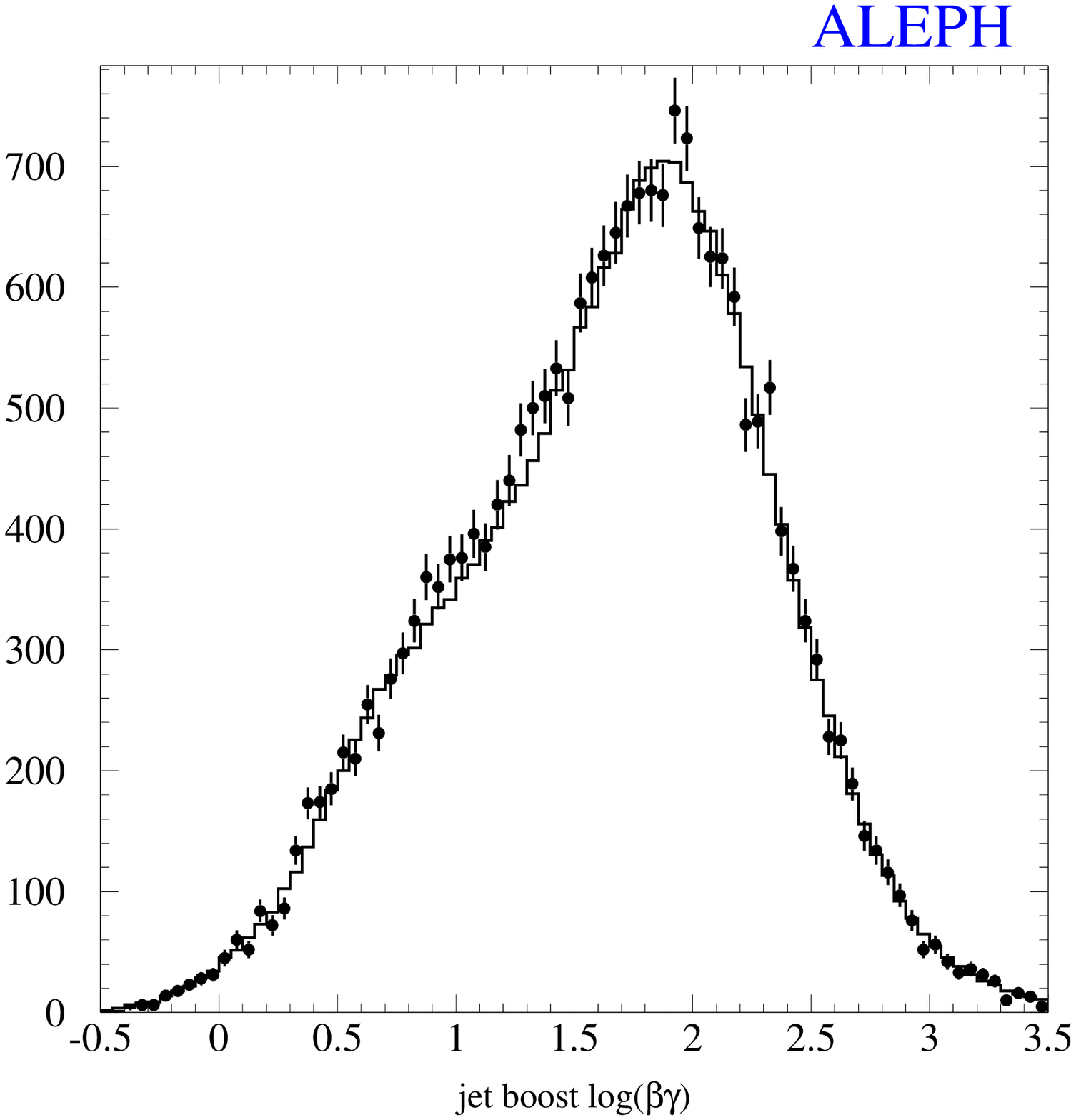,width=8.0cm}}
         }
    \caption{\footnotesize 
Distributions of jet boosts ($\log\beta_{\rm {jet}}\gamma_{\rm {jet}}$) for 
hadronic events at 183-209 \GeV\ comparing Data and simulation (a) in the central 
region ($\cos(\theta_{\rm {jet}}) < 0.7$) and (b) in the forward 
region ($0.7 < \cos(\theta_{\rm {jet}}) < 1.0$).    
             }
    \label{fig:jbt_zgam}
  \end{center}
  \vspace{-7.9cm}
  \begin{center}
  \hspace*{-1cm}
  \begin{picture}(400,10)
  \put(170,30){\large (a)}
  \put(400,30){\large (b)}
  \end{picture}
  \end{center}
  \vspace*{6.5cm}
\end{figure}

Candidate \qqbar$\gamma$ events are required to have at least
eight good tracks with total energy exceeding 20\%
of the nominal CM energy. 
The scalar sum of the transverse momentum components of the good 
tracks must further exceed 12\% of the nominal CM energy.
Identified photons with energy exceeding 5\%
of the nominal CM energy, and isolated 
from the good tracks, are rejected and ignored 
in the analysis. As described in Section~\ref{sec:eflow}
for the W mass analysis, all energy flow objects below 
15$^{\circ}$ to the beams are rejected, and the same thresholds applied to ECAL  
and HCAL neutral objects.
Reconstruction efficiencies for good tracks in data 
have been compared with the simulation at $\sqrt{s}$=91.2 \GeV\ 
revealing lower efficiencies for 
soft tracks from data in the forward direction.
Correction factors have been applied to the simulation for tracks with 
$|\cos\theta|>0.6$ and $p_{\rm T}<5$ \GeV.
Both this last correction for forward good tracks and the removal of 
neutral objects near the beam line are of crucial importance
for the correct simulation of the forward region and the 
following reconstruction of the hadronic \PZz\ mass.
The forward tracks correction was not applied to the generated \PWW\ events. 
However, its effect was found to be closely correlated to the jet angular bias 
discussed in Sec.~\ref{sec:jetangbias} and covered by the corresponding systematic 
uncertainty. 
Figure~\ref{fig:jbt_zgam} shows the distributions of the jet boosts in 
$\log(\beta_{\rm {jet}}\gamma_{\rm {jet}})$ comparing data and simulation separately 
for central and forward reconstructed jets. 

As in previous ALEPH studies of hadronic
radiative returns~\cite{alephzg}, and similarly to the W mass reconstruction, 
the \PZz\ mass is obtained by clustering the hadronic system into two jets with 
the {\tt DURHAM-PE} algorithm, and performing a kinematic reconstruction 
based on fixing the jet velocities to their measured values but 
rescaling their energies to conserve four-momentum. 
It is assumed that the ISR photon is emitted along the beam line, and thus 
the boost of the produced \qqbar\ system is in the opposite direction.
The di-jet rescaled \PZz\ mass can be expressed in terms of the polar 
angles ($\theta_1,\theta_2$) and velocities ($\beta_1,\beta_2$) of the two jets as
$$
M_{\rm Z}^2= s 
\frac{\beta_1\sin\theta_1 + \beta_2\sin\theta_2 - 
              \beta_1\beta_2 | \sin (\theta_1+\theta_2 )|}
     {\beta_1\sin\theta_1 + \beta_2\sin\theta_2 + 
              \beta_1\beta_2 | \sin (\theta_1+\theta_2 )|}.
$$
Requiring a di-jet rescaled mass in the window 75$<M_{\rm Z}<$115 \GeVcsq, 
a total of 25908 events are selected from the data compared with 25904 events from 
the simulation. The expected signal purity is 93.8\%.

The shift of the \PZz\ mass peak is measured by means of 
an unbinned likelihood fit to a p.d.f. built from  reference 
distributions.
The calibration of the fit is done with 
 pseudo-data samples and the small bias is corrected.

Various sources of systematic uncertainties on the $\Delta M_{\rm Z}$
measurement have been considered.
Background uncertainties have been evaluated by varying
the expected contribution of the background events as
Zee ($\pm$50\%), We$\nu$ ($\pm$25\%) and $\gamma\gamma\qqbar$
($\pm$100\%), leading to a combined effect on $\Delta M_{\rm Z}$ of 
16 \MeVcsq.
Fragmentation systematics have been evaluated by comparing 
results obtained with different models 
{\tt JETSET}, {\tt ARIADNE} and {\tt HERWIG}
leading to an uncertainty on $\Delta M_{\rm Z}$ of 19 \MeVcsq.
For the calorimeter systematics that have an impact on the jet boost, different 
shower simulations have been
used (Section~\ref{ECAL:FULLSIM}) and in particular the use of FULLSIM leads 
to a difference in $\Delta M_{\rm Z}$ of 30 \MeVcsq.
For the tracking affecting the jet angles, half of the full effect due to 
reconstruction inefficiencies in 
the forward direction is taken as the systematic uncertainty of 16 \MeVcsq.
The uncertainty related to the ISR model is estimated to be 7 \MeVcsq.
The uncertainty coming from limited Monte~Carlo statistics is dominated 
by the fit calibration uncertainty and is 12 \MeVcsq.
Using different fit methods the uncertainty due to the 
fit method is estimated to be 20 \MeVcsq.
Possible global biases of 0.2 mrad on the polar angle measurements of the good 
tracks lead to an uncertainty of 24 \MeVcsq.
The combined systematic uncertainty on $\Delta M_{\rm Z}$ due to all the above 
sources is then 54 \MeVcsq. 

The resulting shift in the di-jet \PZz\ mass peak reconstruction
in radiative events is
\[  \Delta \PMZ = +40 \pm 30 \mathrm{(stat.)} \pm 54 \mathrm{(syst.)}~\MeVcsq, \]
which is consistent with zero. 
This conclusion remains unchanged when the di-jet rescaled \PZz\ mass
is evaluated using jets built with CONE, PCUT or with
good tracks only.
The jet reconstruction methods studied here are 
applied to the determination of the \PW\ mass and similar uncertainties are used.
Thus, these conclusions give further confidence in the \PW\ mass analysis.

The previous result was obtained using the beam energies supplied by LEP.
If the measurement is in turn interpreted as a shift of the nominal LEP2 CM 
energy, where \PMZ\ is kept fixed to the published value,
\[  \Delta \sqrt{s} = -86\pm 64 \mathrm{(stat.)} \pm 116 \mathrm{(syst.)}~\MeV \] 
is obtained, which again is in good agreement with zero.

The analysis of muon pairs from the process $\epem\ra\PZz\gamma\ra\mm\gamma$ 
provides an additional check on the reconstruction of the LEP beam energies.
Two variables are used 
(i) the plain invariant mass $M_{12}$, defined as 
$M_{12}^2=2P_1 P_2 (1-\cos\theta_{12})$, where $P_1, P_2$ are the 
momenta of the two muons, $\theta_{12}$ is the 
angle between them and (ii) 
the angular mass $m_{12}$ given by 
\[m_{12}^2= s \frac{\sin\theta_1 +\sin\theta_2 - |\sin\theta_{12}|}
                   {\sin\theta_1 +\sin\theta_2 + |\sin\theta_{12}|}.\]

Selected di-muon events are required to be in the range  
$M_{12}>60~\GeVcsq$ and 
$80<m_{12}<100~\GeVcsq$. 
A total of 976 events are selected from the data and 
971.2 are expected from the simulation, with an 
expected signal purity of 93.4\%.

Any discrepancy between data and simulation in the 
$M_{12}$ and $m_{12}$ distributions are evaluated as a shift of the 
data distribution with respect to  reference
distributions, and are measured with an unbinned likelihood fit 
calibrated with  pseudo-data samples.  
Results from the two di-muon masses are combined
in terms of a mean LEP CM energy shift giving
\begin{eqnarray}
  \Delta \sqrt{s} = -334\pm 190 \mathrm{(stat.)} \pm 76 \mathrm{(syst.)}~\MeV 
\end{eqnarray}
\noindent where the main sources of systematic errors come from (i) possible biases 
in the muon polar angle measurement up to $\delta\theta=0.2$~mrad (52~\MeV), 
(ii) the uncertainty from the shift fitting method (48~\MeV) and (iii) 
from the absolute calibration of the muon momenta at the 0.5\% level (24~\MeV).
This shift is 1.6$\sigma$ from the nominal LEP CM average energy, consistent with 
no significant effect.

%% file: results.tex
\boldmath
\section{Combined results}
\label{sec:results}
\unboldmath

\subsection{{\boldmath $\qqbar\qqbar$} channel}
\subsubsection*{\boldmath\PW\ mass}
For each of the standard, optimal CONE and PCUT reconstructions, the 
individual measurements of \PMW\ and \PGW\ at each CM energy are combined 
weighted by their statistical errors and systematic uncertainties as shown  
in Table~\ref{tab:syst-4q-198}. Correlations 
in these uncertainties with CM energy are taken into account. 
The \PW\ masses found from the one-parameter maximum likelihood fits to the data 
are given in Table~\ref{tab:4q-masses}.
\begin{table}[htbp]
\begin{center}
\caption{\protect\footnotesize \PW\ masses in the \qqbar\qqbar\ channel from all 
data for the standard, optimal PCUT and CONE reconstructions with 
corresponding statistical errors and systematic uncertainties (units are \GeVcsq).
}
\label{tab:4q-masses} 
\begin{tabular}[h]{|l||c|c|c|} 
\hline
Reconstruction           & standard & PCUT (3 \GeVc) & CONE (R=0.4) \\ 
\hline\hline
Number of events         & 4861   & 4484   & 4641   \\
\PMW\                    & 80.481 & 80.475 & 80.502 \\
$\chi^2$/dof		 & 7.4/7  & 4.9/7  & 5.1/7  \\
\hline
Statistical error        & 0.058  & 0.070  & 0.070  \\ 
Experimental Uncertainty & 0.022  & 0.028  & 0.026   \\  
FSI Uncertainty          & 0.079  & 0.028  & 0.036  \\  
\hline
Total error              & 0.100  & 0.081  & 0.082  \\
\hline
\end{tabular}
\end{center}
\end{table} 
The experimental systematic uncertainties are derived from all sources in quadrature 
including the LEP CM energy. The FSI uncertainties are the limits from colour 
reconnection and Bose-Einstein effects added in quadrature. 
Taking into account correlations, the \PW\ mass values are in good 
agreement for all three reconstructions. The total errors are also closely comparable 
but in the standard analysis, the systematic uncertainty due to FSI exceeds the 
statistical error. To suppress the dominant non-Gaussian contribution from colour 
reconnection, the result from the PCUT reconstruction with the smallest FSI 
uncertainty is selected to produce the final result to combine with the mass from the 
semileptonic channels. 

\begin{eqnarray}
m^\mathrm{4q}_\mathrm{W}  &=&\ \ 80.475 \pm 0.070
{\mathrm{ (stat.)}} 
\pm 0.028{\mathrm{ (syst.)}}
\pm 0.028{\mathrm{ (FSI)}}
~\GeVcsq. \nonumber
\end{eqnarray}
The corresponding expected statistical error is 0.069 \GeVcsq.

The \PW\ masses with measured and expected statistical errors determined at each CM 
energy from the standard, CONE and PCUT analyses are given in Appendix B.  
  
\subsubsection*{\boldmath\PW\ width}
The \PW\ widths found from the two-parameter maximum likelihood fits to the data 
are given in Table~\ref{tab:4q-widths} for each reconstruction.
\begin{table}[htbp]
\begin{center}
\caption{\protect\footnotesize \PW\ widths in the \qqbar\qqbar\ channel from all 
data for the standard, optimal PCUT and CONE reconstructions with 
corresponding statistical errors and systematic uncertainties (units in \GeV).
}
\label{tab:4q-widths} 
\begin{tabular}[h]{|l||c|c|c|} 
\hline
Reconstruction             & standard & PCUT (3 \GeVc) & CONE (R=0.4)  \\ 
\hline\hline
\PGW\                      & 2.31    & 2.48  & 2.34 \\
$\chi^2$/dof		   & 9/7     & 5/7   & 16/7 \\
\hline
Statistical error          & 0.12   & 0.16   & 0.15 \\ 
Experimental uncertainties & 0.04   & 0.06  & 0.05 \\  
FSI uncertainties          & 0.11   & 0.03   & 0.05 \\  
\hline
Total error                & 0.16   & 0.17 & 0.17  \\
\hline
\end{tabular}
\end{center}
\end{table} 
The statistical error dominates the total systematic uncertainty in all three 
reconstructions. Therefore, the measurement from the standard analysis with the 
smallest total error is used in combination with the semileptonics to derive the most 
precise value for \PGW. Thus, the \PW\ width from the \qqbar\qqbar\ channel is taken 
to be  
\begin{eqnarray}
\Gamma^\mathrm{4q}_\mathrm{W}  &=&\ \ 2.31 \pm 0.12
{\mathrm{ (stat.)}} 
\pm 0.04{\mathrm{ (syst.)}}
\pm 0.11{\mathrm{ (FSI)}}
~\GeV. \nonumber
\end{eqnarray}
The corresponding expected statistical error is 0.11 \GeV.

The \PW\ widths with measured and expected statistical errors determined at each CM 
energy from the standard analysis are given in Appendix C.  

\subsection{\boldmath\evqq, \boldmath\mvqq\ and \boldmath\tvqq\ channels}
\label{sec:lvqqmass}
\subsubsection*{\boldmath\PW\ mass}
The mass values in the standard analysis from the one-parameter fits to the 
data, with the statistical and systematic errors including the LEP energy, are
\begin{eqnarray}
m^{\mathrm{e}\nu\mathrm{q\bar{q}}}_\mathrm{W} &=& 80.536
\pm 0.087{\mathrm{ (stat.)}} 
\pm 0.027{\mathrm{ (syst.)}}
~\GeVcsq, \nonumber \\
m^{\mu\nu\mathrm{q\bar{q}}}_\mathrm{W} &=& 80.353
\pm 0.082{\mathrm{ (stat.)}} 
\pm 0.025{\mathrm{ (syst.)}}
~\GeVcsq, \nonumber \\
m^{\tau\nu\mathrm{q\bar{q}}}_\mathrm{W} &=& 80.394
\pm 0.121{\mathrm{ (stat.)}} 
\pm 0.031{\mathrm{ (syst.)}}
~\GeVcsq. \nonumber 
\end{eqnarray}
The expected statistical errors are $\pm$0.087, 
$\pm$0.082 and $\pm$0.122 \GeVcsq\ for the $\mathrm{e}$, $\mu$ and $\tau$ 
semileptonic channels, respectively.  

 The individual measurements of \PMW\ for each channel  
are combined statistically at each CM energy. The combined semileptonic \PMW\ over 
all CM energies is determined by minimising a 
$\chi^2$ built from the full covariance matrix, taking into account all systematic 
uncertainties derived at each CM energy with the 
appropriate correlation and the statistical errors.
The systematic uncertainties listed in 
Table~\ref{tab:syst-lvqq-198} are found 
to be 89\% correlated between \evqq\ and \mvqq, 85\% between \evqq\ and \tvqq\ and 
89\% between \mvqq\ and \tvqq\ channels. 

The resulting combined mass for the semileptonic channels from the 
one-parameter fits is
\begin{eqnarray}
 m^\mathrm{\ell\nu{q}\bar{q}}_\mathrm{W} &=& 80.429 
\pm 0.054{\mathrm{ (stat.)}} 
\pm 0.025{\mathrm{ (syst.)}}~\GeVcsq, \nonumber
\end{eqnarray}
with a $\chi^2$/dof of 38/23. The expected statistical error is $\pm$ 0.054 \GeVcsq. 

The \PW\ masses with measured and expected statistical errors determined at each CM 
energy are given in Appendix B.  

\subsubsection*{\boldmath\PW\ width}
A two-parameter fit to the data gives the following results for the \PW\  
width in the standard analysis for each channel:
\begin{eqnarray}
\Gamma^{\mathrm{e}\nu\mathrm{q\bar{q}}}_\mathrm{W} &=& 1.84 
\pm 0.20{\mathrm{ (stat.)}} 
\pm 0.08{\mathrm{ (syst.)}}
~\GeV, \nonumber \\
\Gamma^{\mu\nu\mathrm{q\bar{q}}}_\mathrm{W} &=&2.17
\pm 0.20{\mathrm{ (stat.)}} 
\pm 0.06{\mathrm{ (syst.)}}
~\GeV, \nonumber \\
\Gamma^{\tau\nu\mathrm{q\bar{q}}}_\mathrm{W} &=&2.01
\pm 0.32{\mathrm{ (stat.)}} 
\pm 0.06{\mathrm{ (syst.)}}
~\GeV, \nonumber \\
\end{eqnarray}
where the expected errors are determined 
to be $\pm$0.21, $\pm$0.20 and  $\pm$0.31 \GeV\
for the $\mathrm{e}$, $\mu$ and $\tau$ channels respectively. 
The systematic uncertainties listed in Table~\ref{tab:syst-lvqq-198} are found 
to be 100\% correlated between \evqq\ and \mvqq, 43\% between \evqq\ and \tvqq\ and 
48\% between \mvqq\ and \tvqq\ channels.

The \PW\ widths with measured and expected statistical errors determined at each CM 
energy are given in Appendix C.  

The combined total width from the two-parameter fits in all \lvqq\ channels is 
\begin{eqnarray}
 \Gamma^\mathrm{\ell\nu{q}\bar{q}}_\mathrm{W} &=& 2.01
\pm 0.13{\mathrm{ (stat.)}} 
\pm 0.06{\mathrm{ (syst.)}}~\GeV, \nonumber 
\end{eqnarray}
with a $\chi^2$/dof of 15/21. The expected statistical error is $\pm$ 0.13 \GeV.

\subsection{All channels}
The combined results from all channels using the PCUT results in the 
\qqbar\qqbar\ channel for the mass and standard results for the width are:
\begin{eqnarray}
 m_{\rm W} &=& 80.444 
\pm 0.043{\mathrm{ (stat.)}} 
\pm 0.024{\mathrm{ (syst.)}}
\pm 0.009{\mathrm{ (FSI)}}
\pm 0.009{\mathrm{ (LEP)}}
~\GeVcsq,\nonumber \\
 \Gamma_\mathrm{W} &=& 2.140 
\pm 0.090{\mathrm{ (stat.)}} 
\pm 0.045{\mathrm{ (syst.)}}
\pm 0.046{\mathrm{ (FSI)}}
\pm 0.007{\mathrm{ (LEP)}}
~\GeV. \nonumber 
\end{eqnarray}
The combinations are performed in the same way as described in 
section~\ref{sec:lvqqmass}. The $\chi^2$/dof  
is 43/31 and 26/29 for the mass and width combinations, respectively.
Alternatively, if the \PW\ mass from the CONE analysis in the \qqbar\qqbar\ channel 
is combined with those from the \lvqq\ channels, the mass is:
\begin{eqnarray}
 m_{\rm W} &=& 80.453
\pm 0.043{\mathrm{ (stat.)}} 
\pm 0.023{\mathrm{ (syst.)}}
\pm 0.011{\mathrm{ (FSI)}}
\pm 0.009{\mathrm{ (LEP)}}
~\GeVcsq.\nonumber 
\end{eqnarray}

Similarly, combining the \PW\ mass from the standard analysis in the 
\qqbar\qqbar\ channel with those from the \lvqq\ channels gives 
\begin{eqnarray}
 m_{\rm W} &=& 80.440 
\pm 0.043{\mathrm{ (stat.)}} 
\pm 0.022{\mathrm{ (syst.)}}
\pm 0.019{\mathrm{ (FSI)}}
\pm 0.009{\mathrm{ (LEP)}}
~\GeVcsq.\nonumber 
\end{eqnarray}

To assess the effect of any unexpected correlation between the measured \PW\ mass and 
width on the one-parameter fits where the width is fixed to standard model values, 
the mass from each channel is compared with the corresponding two-parameter fit 
results. Combining all channels in the standard analysis, the difference is found 
to be 8 \MeVcsq\ indicating no significant effect.    

To investigate whether there is a significant difference between the masses from 
the \qqbar\qqbar\ and combined \lvqq\ channels due to final state interactions, 
a fit is performed to extract this difference retaining all 
systematic uncertainties from Tables~\ref{tab:syst-4q-198} and 
\ref{tab:syst-lvqq-198} except those from 
Bose-Einstein correlations and colour reconnection. The standard analysis in 
the \qqbar\qqbar\ channel is used to enhance any effects. The result is 
$${\langle m_\mathrm{W}^\mathrm{\qqbar\qqbar}\rangle - 
\langle m_\mathrm{W}^\mathrm{\lvqq}\rangle = 
+62 \pm 76~\mathrm{(stat.+syst.)}~\MeVcsq \ ,}$$
to be compared with the $+79$~\MeVcsq\ FSI uncertainty. 


%% file: conclusions.tex
\section{Conclusions}

  The mass and width of the \PW\ boson have been measured from \PW\ pair 
events using the direct reconstruction of the invariant mass of their decay 
products in fully hadronic and semileptonic final states. 
Following constrained kinematic fits to each event, the \PW\ 
parameters were extracted by reweighting fully simulated invariant mass 
spectra to the measured distributions, employing an unbinned maximum 
likelihood procedure to find the best fits. To produce the most precise value 
of \PMW, one-parameter fits are performed where \PGW\ 
varies with \PMW\ according to the Standard Model. Two-parameter fits, where 
\PMW\ and \PGW\ are allowed to vary independently, are used to measure the 
\PW\ width.

All data collected at centre-of-mass energies between 183 and 209 \GeV\ are 
fully reprocessed and analysed homogeneously to produce the final values with 
statistical errors. The systematic uncertainties are determined taking into 
account correlations between all channels and CM energies. For the \PW\ mass, 
these measurements are combined with the earlier published ALEPH results 
obtained from the total \PW\ pair cross sections at 161~\cite{ALEPH-THRESHOLD} 
and 172 \GeV~\cite{wwxsec_172} to produce the final result as follows:  

$${ m_{\rm W} = 80.440 \pm 0.043{\mathrm{\small (stat.)}} 
\pm 0.024{\mathrm{\small (syst.)}}
\pm 0.009{\mathrm{\small (FSI)}} 
\pm 0.009{\mathrm{\small (LEP)}}~\GeVcsq,} $$
\noindent where the first error is statistical, the second derived from all 
ALEPH experimental systematic uncertainties, the third from 
the final state Bose-Einstein and colour reconnection uncertainties in the 
\qqbar\qqbar\ channel and the last is the LEP energy uncertainty. The L3 and 
OPAL collaborations have recently published their 
results~\cite{OPAL2005,L32005} using all their available data. 
Also, earlier results have been published by DELPHI~\cite{DLO_189} as well as 
final results from the Tevatron Run I $\ppbar$ collider experiments using 
large samples of single W's decaying into electrons and muons~\cite{Wmass_pp}.

No evidence is found for final state interactions between the \PW\ hadronic 
decay products in the \qqbar\qqbar\ channel. The limit on colour reconnection 
is derived from the search for any  
significant variation in the value of \PMW\ when low momentum particles or 
those between jets are progressively excluded in the invariant mass 
reconstructions. To minimise any colour reconnection effects, the \PW\ mass in 
the \qqbar\qqbar\ channel used in the final combination is taken from the 
reconstruction where all particles with momenta lower than 3 \GeVc\ are 
removed.         

This measurement of the \PW\ mass agrees with the earlier ALEPH 
measurement~\cite{mass_189} and other 
measurements~\cite{OPAL2005,L32005,DLO_189,Wmass_pp} as well as 
with the indirect prediction from the Standard Model fit to electroweak 
observables~\cite{lepew2004}. The consistency with the electroweak fit is only 
possible if the Standard Model Higgs boson is light. 

Finally, from the 183-209 \GeV\ data in all channels, the \PW\ width is 
determined to be
$${ \PGW = 2.14 \pm 0.09 {\mathrm{\small (stat.)}} 
\pm 0.04{\mathrm{\small (syst.)}}
\pm 0.05{\mathrm{\small (FSI)}} 
\pm 0.01{\mathrm{\small (LEP)}}~\GeV,} $$
\noindent consistent with the other LEP 
measurements~\cite{OPAL2005,L32005,DLO_189}.
  

%% file: appendices.tex
\clearpage


\boldmath
\section*{Appendix A: Generator setup tunings used for CR studies}
\unboldmath

Modified values of hadronisation and fragmentation parameters are tabulated for
the model variants of {\tt GAL} and {\tt ARIADNE} with CR using the 
{\tt JETSET} framework.  The modified parameters with and without CR for 
the {\tt HERWIG} model are also tabulated.   
The internal name of the parameters is given in each case.
The unmodified parameters used in the generation of events for all models 
without CR are given in Ref.~\cite{Barate}.

\begin{table} [ht]
\begin{center}
\begin{tabular} {| c || c || c || c  | c  | }
\hline
\multicolumn{1}{|c||}{} & \multicolumn{1}{c||}{{\tt JETSET}} 
&\multicolumn{1}{c||}{{\tt GAL}}& \multicolumn{2}{c|}{{\tt ARIADNE}} \\
Parameter       & standard and &    & no CR   &  intra-W only(AR21)/   \\
                & SK models &    & (AR20) &  intra and inter-W(AR2)   \\
\hline
\hline
azimuthal distribution &  3 &  0  & -  & - \\
 in PS MSTJ(46)& & & &   \\
\hline
momentum transverse & & & &  \\
 width for hadron $\sigma_{\rm qt}$ & 0.371 & 0.364 & 0.358 & 0.352 \\
PARJ(21) ($\GEV$)& & & &  \\
\hline
LUND fragmentation & & & & \\
parameter b PARJ(42) &0.805  & 0.815 & 0.823  & 0.762 \\
\hline
\hline
$\Lambda_{\rm QCD}$  ($\GEV$) PARJ(81) &0.291 & 0.307  &  &  \\
\hline
$Q_0$ cut-off in PS ($\GEV$) & &  & & \\
PARJ(82)   & 1.52 & 1.57 &  & \\
\hline
strength parameter $R_0$  & - & 0.039 & - & - \\
\hline
\hline
Colour Reconnection & &  & & \\
switch MSTA(35) & -  & - & 0  & 1 (AR21) or 2 (AR2) \\
\hline
$\Lambda_{\rm QCD}$ ($\GEV$) PARA(1)& - & - & 0.230 & 0.231 \\
\hline
 ptmin cut ($\GEV$)& & & &  \\
PARA(3) & - & - & 0.791 &  0.781 \\
\hline
$E_{\rm gluon}$ cut  ($\GEV$)& - & - & 0. & 2.\\
PARA(28) & & & &  \\
\hline
\end{tabular}
\end{center}
\end{table}

\begin{table} [h]
\begin{center}
\begin{tabular} {| l l || c | c  | }
\hline
\multicolumn{2}{|c||}{} & \multicolumn{2}{c|}{{\tt HERWIG}}  \\
\multicolumn{2}{|c||}{Parameter}      & no CR   &  with CR       \\
\hline
\hline
$\Lambda_{\rm QCD}$ ($\GEV$) & QCDLAM   & 0.190  &  0.187  \\
\hline
Maximum cluster mass &  CLMAX &3.39  & 3.40  \\
\hline
\multicolumn{2}{|c||}{Split cluster spectrum parameter}  & &  \\
 light flavour clusters & PSPLT(1)  &  0.945 &  0.886\\
 heavy flavour clusters & PSPLT(2)  &  0.330 &  0.320 \\
\hline
Width of gaussian angle smearing &  CLSMR(1)  &  0.58 &  0.66  \\
\hline
Decuplet baryon weight & DECWT &0.71 &   0.70\\
\hline
CR probability &  PRECO  & 0.&  1/9\\
\hline
Gluon mass     &  RMASS(13)  & 0.774 & 0.793 \\
\hline
\end{tabular}
\end{center}
\end{table}

\boldmath
\section*{Appendix B: \PW\ masses from the standard, optimal CONE and 
PCUT  analyses}
\unboldmath
\begin{table}[tbhp]
\caption{
\protect\footnotesize
Individual fitted \PMW\ values from the standard analysis for each 
channel and 
CM energy, together with the number of selected events and expected 
statistical errors. 
}
\begin{center}
\begin{tabular}{|l|c|c|c|c|}
\hline
Channel  & CM energy & $N_{evts}$ & \PMW\     & Expected error  \\
         &   (\GeV)  &            & (\GeVcsq)    &  (\GeVcsq)         \\
\hline
\hline
\qqbar\qqbar\ & 183 &  435 & 80.525$\pm$0.168 &  0.171 \\
              & 189 & 1169 & 80.518$\pm$0.105 &  0.102 \\
              & 192 &  234 & 79.995$\pm$0.235 &  0.235 \\
              & 196 &  556 & 80.506$\pm$0.151 &  0.151 \\
              & 200 &  627 & 80.303$\pm$0.151 &  0.146 \\
              & 202 &  283 & 80.635$\pm$0.206 &  0.218 \\
              & 205 &  589 & 80.583$\pm$0.149 &  0.146 \\
              & 207 &  968 & 80.573$\pm$0.119 &  0.118 \\
\hline
\evqq\        & 183 &  112 & 80.440$\pm$0.265 &  0.276 \\
              & 189 &  317 & 80.437$\pm$0.170 &  0.162 \\
              & 192 &   52 & 80.621$\pm$0.502 &  0.404 \\
              & 196 &  148 & 80.420$\pm$0.244 &  0.250 \\
              & 200 &  160 & 80.607$\pm$0.247 &  0.251 \\
              & 202 &   96 & 80.203$\pm$0.303 &  0.363 \\
              & 205 &  140 & 81.089$\pm$0.276 &  0.269 \\
              & 207 &  234 & 80.620$\pm$0.218 &  0.212 \\
\hline
\mvqq\        & 183 &   98 & 79.991$\pm$0.265 &  0.259 \\
              & 189 &  344 & 80.185$\pm$0.160 &  0.153 \\
              & 192 &   60 & 80.483$\pm$0.385 &  0.381 \\
              & 196 &  149 & 81.109$\pm$0.246 &  0.236 \\
              & 200 &  171 & 79.884$\pm$0.237 &  0.233 \\
              & 202 &   86 & 81.210$\pm$0.324 &  0.334 \\
              & 205 &  165 & 80.409$\pm$0.250 &  0.250 \\
              & 207 &  298 & 80.277$\pm$0.186 &  0.197 \\
\hline
\tvqq\        & 183 &   97 & 80.595$\pm$0.414 &  0.396 \\
              & 189 &  306 & 80.277$\pm$0.230 &  0.232 \\
              & 192 &   59 & 80.950$\pm$0.551 &  0.548 \\
              & 196 &  158 & 80.589$\pm$0.343 &  0.347 \\
              & 200 &  163 & 80.210$\pm$0.345 &  0.348 \\
              & 202 &   70 & 80.676$\pm$0.515 &  0.509 \\
              & 205 &  149 & 80.750$\pm$0.352 &  0.370 \\
              & 207 &  224 & 79.959$\pm$0.299 &  0.296 \\
\hline
\end{tabular}
\end{center}
\label{tab:reco1_masses}
\end{table}

\begin{table}[tbhp]
\caption{
\protect\footnotesize
Individual fitted \PMW\ values from the CONE analysis in the 4q channel and 
CM energy, together with the number of selected events and expected 
statistical errors. 
}
\begin{center}
\begin{tabular}{|l|c|c|c|c|}
\hline
Channel  & CM energy & $N_{evts}$ & \PMW\     & Expected error  \\
         &   (\GeV)  &            & (\GeVcsq)    &  (\GeVcsq)         \\
\hline
\hline
\qqbar\qqbar\ & 183 &  420 & 80.607$\pm$0.214 &  0.229 \\
              & 189 & 1116 & 80.528$\pm$0.138 &  0.132 \\
              & 192 &  224 & 80.477$\pm$0.309 &  0.319 \\
              & 196 &  537 & 80.470$\pm$0.195 &  0.194 \\
              & 200 &  589 & 80.230$\pm$0.203 &  0.191 \\
              & 202 &  272 & 81.004$\pm$0.305 &  0.278 \\
              & 205 &  565 & 80.547$\pm$0.202 &  0.197 \\
              & 207 &  918 & 80.428$\pm$0.163 &  0.159 \\
\hline
\end{tabular}
\end{center}
\label{tab:reco9_masses}
\end{table}

\begin{table}[tbhp]
\caption{
\protect\footnotesize
Individual fitted \PMW\ values from the PCUT analysis in the 4q channel and 
CM energy, together with the number of selected events and expected 
statistical errors. 
}
\begin{center}
\begin{tabular}{|l|c|c|c|c|}
\hline
Channel  & CM energy & $N_{evts}$ & \PMW\     & Expected error  \\
         &   (\GeV)  &            & (\GeVcsq)    &  (\GeVcsq)         \\
\hline
\hline
\qqbar\qqbar\ & 183 &  409 & 80.587$\pm$0.219 &  0.230 \\
              & 189 & 1089 & 80.529$\pm$0.138 &  0.134 \\
              & 192 &  214 & 79.935$\pm$0.316 &  0.322 \\
              & 196 &  519 & 80.517$\pm$0.197 &  0.198 \\
              & 200 &  572 & 80.357$\pm$0.203 &  0.197 \\
              & 202 &  256 & 80.614$\pm$0.320 &  0.281 \\
              & 205 &  541 & 80.333$\pm$0.205 &  0.200 \\
              & 207 &  884 & 80.588$\pm$0.165 &  0.159 \\
\hline
\end{tabular}
\end{center}
\label{tab:reco8_masses}
\end{table}

\clearpage
\boldmath
\section*{Appendix C: \PW\ widths from the standard analysis}
\unboldmath
\begin{table}[tbhp]
\caption{
\protect\footnotesize
Individual fitted \PGW\ values from two-parameter fits in the standard 
analysis for each channel and CM energy with expected statistical errors. The 
number of selected events is the same as the corresponding mass analysis in 
Appendix B.
}
\begin{center}
\begin{tabular}{|l|c|c|c|}
\hline
Channel  & CM energy & \PGW\     & Expected error  \\
         &   (\GeV)  & (\GeV)    &  (\GeV)         \\
\hline
\hline
\qqbar\qqbar\ & 183 & 1.894$\pm$0.336 & 0.350 \\
              & 189 & 2.520$\pm$0.242 & 0.214 \\
              & 192 & 2.761$\pm$0.581 & 0.530 \\    
              & 196 & 2.022$\pm$0.331 & 0.319 \\
              & 200 & 2.865$\pm$0.378 & 0.311 \\
              & 202 & 2.134$\pm$0.545 & 0.453 \\    
              & 205 & 1.790$\pm$0.345 & 0.315 \\ 
              & 207 & 2.518$\pm$0.276 & 0.247 \\ 
\hline
\evqq\        & 183 & 2.494$\pm$0.718 & 0.657 \\
              & 189 & 1.662$\pm$0.353 & 0.397 \\
              & 192 & 2.184$\pm$1.172 & 0.811 \\    
              & 196 & 1.791$\pm$0.623 & 0.611 \\
              & 200 & 2.909$\pm$0.715 & 0.606 \\
              & 202 & 3.246$\pm$1.046 & 0.764 \\
              & 205 & 1.210$\pm$0.493 & 0.634 \\ 
              & 207 & 1.694$\pm$0.463 & 0.508 \\ 
\hline
\mvqq\        & 183 & 1.894$\pm$0.630 & 0.632 \\
              & 189 & 1.795$\pm$0.355 & 0.384 \\
              & 192 & 1.955$\pm$0.916 & 0.783 \\ 
              & 196 & 2.667$\pm$0.700 & 0.576 \\ 
              & 200 & 2.187$\pm$0.582 & 0.572 \\
              & 202 & 2.566$\pm$0.957 & 0.750 \\
              & 205 & 3.694$\pm$0.801 & 0.606 \\
              & 207 & 2.186$\pm$0.454 & 0.486 \\
\hline
\tvqq\        & 183 & 2.446$\pm$1.089 & 0.848  \\
              & 189 & 1.720$\pm$0.527 & 0.565 \\    
              & 192 & -               & 0.965 \\
              & 196 & 0.977$\pm$0.501 & 0.776 \\ 
              & 200 & 1.808$\pm$0.761 & 0.776 \\ 
              & 202 & 2.284$\pm$1.226 & 0.925 \\
              & 205 & 2.171$\pm$0.830 & 0.818 \\
              & 207 & 2.429$\pm$0.862 & 0.700 \\ 
\hline
\end{tabular}
\end{center}
\label{tab:reco1_widths}
\end{table}
 In the \tvqq\ channel, the two-parameter fits fail to converge for the width 
at 192 and 196 GeV even though the allowed range is 0.9 to 4.3 GeV. 
At 196 GeV, a one-parameter fit is successfully performed fixing the mass to 
the measured value given in Appendix B.
        

%% file: mass.bbl
\clearpage
\begin{thebibliography}{99}

\bibitem{lepew2004}
The ALEPH, DELPHI, L3, OPAL, SLD Collaborations, the LEP Electroweak Working 
Group, the SLD Electroweak and Heavy Flavour Groups, \ \ {\it Precision 
Electroweak Measurements on the Z Resonance}, CERN-PH-EP/2005-041, SLAC-R-774,
hep-ex/0509008, (2005) to be published in Physics Reports.

\bibitem{mtop}
The CDF and D0 Collaborations and the Tevatron Electroweak Working Group,
{\it Combination of CDF and D0 results on the top-quark mass}, hep-ex/0404010,
 (2004).

\bibitem{mass_172}
ALEPH Collaboration,
{\it Measurement of the W mass by Direct Reconstruction in ${e^+ e^-}$ 
collisions at 172 \GeV}, Phys. Lett. {\bf B422} (1998) 384.

\bibitem{mass_183}
ALEPH Collaboration,
{\it Measurement of the W mass by Direct Reconstruction in ${e^+ e^-}$ 
collisions at 183 \GeV}, Phys. Lett. {\bf B453} (1999) 121.

\bibitem{mass_189}
ALEPH Collaboration,
{\it Measurement of the W Mass and Width in ${e^+ e^-}$ 
Collisions at 189 \GeV}, Eur. Phys. J. {\bf C17} (2000) 241. 

\bibitem{ALEPH-THRESHOLD}
 ALEPH Collaboration, {\it Measurement of the W mass in ${e^+ e^-}$ 
 collisions at production threshold}, Phys. Lett. {\bf B401} (1997) 347.

\bibitem{wwxsec_172}
ALEPH Collaboration, {\it Measurement of the W-pair cross section in 
${e^+ e^-}$ collisions at 172 \GeV}, Phys. Lett. {\bf B415} (1997) 435.

\bibitem{xsec_final}
 ALEPH Collaboration, {\it Measurement of W-pair production in 
 $\mathrm{e^+ e^-}$ collisions at centre-of-mass energies from 183 to 
 209 \GeV}, Eur. Phys.~J. {\bf C38} (2004) 147.

\bibitem{YFSWW}
S.~Jadach et al., Comp. Phys. Commun. {\bf 140} (2001) 432.  

\bibitem{detector}
ALEPH Collaboration, {\it ALEPH: A detector for electron-positron 
annihilations at LEP}, \NIM{{\bf A294}}{1990}{121}.

\bibitem{perf}
ALEPH Collaboration, {\it Performance of the ALEPH detector at LEP}, 
\NIM{{\bf A360}}{1995}{481}.

\bibitem{GEANT}
GEANT3 Manual, CERN Program Library Long Writeup W5013 (1994).

\bibitem{durham}
Yu.L. Dokshitzer, J. Phys. {\bf G17} (1991) 1441.

\bibitem{KORALW}
S.~Jadach et al., Comp. Phys. Commun. {\bf 140} (2001) 475.

\bibitem{JETSET}
T.~Sj\"ostrand, Comp. Phys. Commun. {\bf 82} (1994) 74.

\bibitem{PYTHIA}
T.~Sj\"ostrand et al., Comp. Phys. Commun. {\bf 135} (2001) 238.

\bibitem{HERWIG}
G. Corcella et al., JHEP {\bf 0101} (2001) 010. 

\bibitem{ARIADNE}
L. L\"onnblad, 
Comput. Phys. Commun. {\bf 71} (1992) 15. 

\bibitem{CRSK}
T.~Sj\"ostrand and V.A.~Khoze,
Z. Phys. {\bf C62} (1994) 281;
Phys. Rev. Lett. {\bf 72} (1994) 28.

\bibitem{GAL}
J.~Rathsman, Phys. Lett. {\bf B452} (1999) 364;\\
A.~Edin, G.~Ingelman, J.~Rathsman, Phys. Lett. {\bf B366} (1996) 371; 
Z. Phys. {\bf C75} (1997) 57. 

\bibitem{2-step}
L. L\"onnblad, 
Z. Phys. {\bf C70} (1996) 107 and private communication. 

\bibitem{LoSj} L.~L\"{o}nnblad and T.~Sj\"{o}strand, Phys. Lett. {\bf B351} (1995) 293.

\bibitem{Lonnblad}
  L.~L\"onnblad and T.~Sj\"ostrand, Eur. Phys. J.~{\bf C2} (1998) 165.


\bibitem{KKMC}
S.~Jadach, B.F.L.~Ward and Z.~W\c{a}s,
Comput. Phys. Commun. {\bf 130} (2000) 260.

\bibitem{PHOT02}
J.~A.~M.~Vermaseren,
Proceedings of the IV International Workshop on Gamma Gamma Interactions,
eds G.~Cochard, P.~Kessler (1980).

\bibitem{BHWIDE}
S.~Jadach et al., \PL{{\bf 390}}{1997}{298}.

\bibitem{djamel}
D.Boumediene, {\it Mesure de la masse du boson $W^{\pm}$ dans l'exp\'{e}rience 
ALEPH}, Th\`{e}se de Doctorat de l'Universit\'{e} Paris VI (2002), DAPNIA-2002-03-T, 
CERN-THESIS-2003-008.

\bibitem{jade}
W. Bartel et al.,
Z. Phys. {\bf C 33} (1986) 23; \\
S. Bethke et al.,
\PL{{\bf 213}}{1988}{235}.

\bibitem{conv}
{\it Determination of the mass of the \PW\ boson}, 
in {\it Physics at LEP2}, CERN 96-01,
eds. G.~Altarelli, T.~Sj\"ostrand and F.~Zwirner, 
vol.~1, p.~141.

\bibitem{mwshiftQED} V.A. Khoze and T. Sjostrand, Z. Phys. {\bf C 70} (1996)
  625; \\
W. Beenakker, A.P. Chapovsky and F.A. Berends, Nucl. Phys. {\bf B508}
  (1997) 17; Phys. Lett. {\bf B411} (1997) 203;\\ 
A. Denner, S. Dittmaier and M. Roth, Phys. Lett. {\bf B429} (1998) 145;
Nucl. Phys. {\bf B519} (1998) 39.

\bibitem{CRSKcomm}
T.~Sj\"ostrand, private communication.

\bibitem{cr_ygap_lep}
ALEPH Collaboration,
{\it Test of colour reconnection models using three-jet events in hadronic Z 
decays}, hep-ex/0604042, submitted to Eur. Phys.~J. C.

\bibitem{Hugo_thesis}
H. Ruiz-P\'{e}rez, {\it Measurement of the W mass from the 
$WW \to q\overline{q}q\overline{q}$ channel with the ALEPH detector},
Ph.D. Dissertation Universitat Aut{\`o}noma de Barcelona,
CERN-THESIS-2003-004.

\bibitem{lepewg2004}
R. Assmann et al., LEP Energy Working Group,
{\it Calibration of centre-of-mass energies at LEP2 for a precise measurement 
of the \PW\ boson mass}, Eur. Phys.~J. {\bf C39} (2005) 253.

\bibitem{HWbaryons}
ALEPH Collaboration, {\it A measurement of the semileptonic 
branching ratio ${\rm BR}({\mbox{b-baryon}}\ra{\rm p}l\bar{\nu}X)$ and a study of 
inclusive $\pi^{\pm}, k^{\pm}, (p,\overline{p})$ production in Z decays}, 
Eur. Phys. J. {\bf C5} (1998) 205.

\bibitem{zalewski}
S.~Jadach and K.~Zalewski, Acta Phys. Pol. {\bf B28}
(1997) 1363.

\bibitem{ck_nf}
A.~P.~Chapovsky and V.~A.~Khoze,
Eur. Phys.~J. {\bf C9} (1999)~449. 

\bibitem{Wiesiek}
M.~Skrzypek et al., Phys. Lett. {\bf B523} (2001) 117;\\
W.~Placzek, private communication.

\bibitem{Cossutti}
F.~Cossutti,  Eur. Phys. J. {\bf C44} (2005) 383. 

\bibitem{RacoonWW}
A.~Denner, S.~Dittmaier, M.~Roth and D.~Wackeroth,
Nucl. Phys. {\bf B560} (1999) 33;
Nucl. Phys. {\bf B587} (2000) 67;
Phys. Lett. {\bf B475} (2000) 127;
EPJdirect Vol.2 {\bf C4} (2000) 1.

\bibitem{ALEPHstandard}
  ALEPH Collaboration,
  {\it Bose-Einstein correlations in W-pair decays},
  Phys. Lett. {\bf B478} (2000) 50.

\bibitem{ALEPHmixed}
  ALEPH Collaboration,
  {\it Bose-Einstein correlations in W-pair decays with an 
event-mixing technique},
  Phys. Lett. {\bf B606} (2005) 265.

\bibitem{alephzg}
ALEPH Collaboration, {\it Determination of the LEP centre-of-mass energy from 
$Z\gamma$ events}, Phys. Lett. {\bf B464} (1999) 339.

\bibitem{refzg}
  DELPHI Collaboration, {\it A Determination of the Centre-of-Mass Energy at 
LEP2 using  Radiative 2-fermion Events}, CERN-PH-EP/2005-050, to be published 
in Eur. Phys. J. C. \\
  L3 Collaboration, {\it Measurement of the Z-boson mass using Z-gamma events 
at Centre-of-Mass Energies above the Z pole}, Phys. Lett. {\bf B585} (2004) 42. \\
  OPAL Collaboration, {\it Determination of the LEP Beam Energy using 
Radiative Fermion-pair Events}, Phys. Lett. {\bf B604} (2004) 31.   

\bibitem{OPAL2005}
  OPAL Collaboration,
  {\it Measurement of the mass and width of the \PW\ boson}, \\
  Eur. Phys. J. {\bf C45} (2006) 307.

\bibitem{L32005}
  L3 Collaboration,
  {\it Measurement of the mass and width of the \PW\ boson at LEP}, \\
  Eur. Phys. J. {\bf C45} (2006) 569.

\bibitem{DLO_189}
DELPHI Collaboration, {\it Measurement of the mass of the W boson using direct 
reconstruction at $\sqrt s$ = 189 GeV}, Phys. Lett. {\bf B511} (2001) 159.

\bibitem{Wmass_pp}
CDF Collaboration, {\it Measurement of the W boson mass with the Collider 
Detector at Fermilab}, Phys. Rev. {\bf D64} (2001) 052001. \\
D0 Collaboration, {\it Improved \PW\ boson mass measurement with the D0 
detector}, Phys. Rev. {\bf D66} (2002) 012001.

\bibitem{Barate}
ALEPH Collaboration, {\it Studies of Quantum Chromodynamics with the 
ALEPH detector}, Phys. Rep. {\bf 294} (1998) 1.

\end{thebibliography}
